\documentclass{osa-article}
\journal{oe}
\articletype{Research Article}
\usepackage{subfigure,comment}
\usepackage{textcomp}
\begin{document}
\title{Merit functions and measurement schemes for single parameter depolarization models}\author{Lisa Li\authormark{1} and Meredith Kupinski\authormark{1}}

\address{\authormark{1}Wyant College of Optical Sciences, 1630 E University Blvd, Tucson, AZ 85721}

\email{\authormark{*}lisali@optics.arizona.edu} 
%Potential reviewers: Scott Tyo, Razvigor Ossikovski, Noah Rubin, Baek, someone at Sony, Germer, Francois Goudail,  Igor V Meglinski

\begin{abstract} 
Mueller polarized bi-directional scattering distribution functions (pBSDFs) are $4\times4$ matrix-valued functions which depend on acquisition geometry. The most popular pBSDF is a weighted sum between a Fresnel matrix and an ideal depolarizer. This work's main contribution is relating the relative weight between an ideal depolarizer and Fresnel matrix to a single depolarization parameter. Rather than a 16-dimensional matrix norm, this parameter can form a one-dimensional merit function. Then, instead of a full Mueller matrix measurement, a scheme for pBSDF fitting to only two polarimetric measurements is introduced. Depolarization can be mathematically expressed as the incoherent addition of coherent states\cite{CloudeDepolSynthesis}. This work shows that, for a Mueller matrix to be in the span of a Fresnel matrix and an ideal depolarizer, the weights in the incoherent addition are triply degenerate. This triple degeneracy is observed in five different colored opaque plastics treated with nine different surface textures and measured at varying acquisition geometries and wavebands. 
\end{abstract}

\section{Introduction} 

Bi-directional scattering distribution functions (BSDFs) describe a material's radiometric response when illuminated and observed from different angles. This work proposes a polarized BSDF (pBSDF) model inspired by the spectral decomposition\cite{cloude1986group} of Mueller matrix measurements of opaque, diffuse, plastic materials. A Mueller pBSDF is a 16-element matrix that quantifies the polarization and depolarization effects of a light-matter interaction. Mueller matrices can describe non-depolarizing, partially depolarizing, and completely depolarizing light-matter interactions. A Mueller matrix has sixteen degrees of freedom (DoF) from which 7 are associated with non-depolarizing properties: 3 DoF for retardance, 3 DoF for diattenuation, and 1 DoF for throughput (\emph{e.g.} reflectance, transmittance). The other 9 DoF are associated with depolarization \cite{chipman2005metrics,CloudeDepolSynthesis}. Depolarization can be mathematically expressed by a convex incoherent sum of four or fewer coherent states. The diattenuation and retardance determine the most significant coherent state \cite{CloudeDepolSynthesis,cloudepottier}. The weights in this convex sum are 3 of the 9 DoF for depolarization. When these weights are triply degenerate (\emph{i.e} the smallest three are equal) the depolarization is reduced to 1 DoF. In this work, an approximate triple degeneracy is observed over variations in surface texture and albedo of 5 different colored opaque plastics treated with 9 different surface textures and illuminated by 662, 524, and 451nm wavebands. The albedo, texture, and scattering geometry change the depolarization DoF. Fresnel reflection reasonably approximates the non-depolarizing properties of all measurements. 

The most popular pBSDF model is a weighted sum between a non-depolarizing Mueller matrix (\emph{e.g.} Fresnel relfection) and an ideal depolarizer. Early Mueller pBSDF work by Bickel et al. defined polarizing interactions as arising from perturbations to an idealized smooth surface \cite{bickel1988mueller}. Bickel et al. pointed to an idealized Lambertian surface as the "opposite case" to a perfectly smooth surface and posited that all "real-world surfaces" lie between an ideally smooth and ideally Lambertian surface texture. A Fresnel reflection matrix and ideal depolarizer can be used to represent the ideally specular smooth and ideally diffuse Lambertian surface textures, respectively. A polarimetric interpretation of the original unpolarized microfacet BSDF model uses the ideal depolarizer and Fresnel reflection matrix as component Mueller matrices \cite{torrance1967theory,cooktorrance,Walter2007,MSPIpBRDF}.

Polarized BSDF models of rough surfaces have been explored as both forward models for material rendering \cite{MSPIpBRDF,Baek2018,Baek2020Image,kondo2020accurate} and inverse models for material recognition \cite{TomiYama,polmultiviewstereo,breon2017brdf}. Prior work has shown how applying pBSDF microfacet models to a Fresnel reflection component can improve the realism of a rendered scene\cite{Baek2018,Baek2020Image,kondo2020accurate}. Microfacets are planar structures on a larger macrosurface that only reflect in the specular direction. Recent pBSDF models have experimented with removing the ideally depolarizing component and replacing it with a diffuse or partially-depolarized component to improve measurement agreement\cite{Baek2018,Baek2020Image}. A recent pBSDF model by Kondo et al. experiments with adding a third diffuse component instead of replacing the ideally depolarizing component to further improve model and measurement agreement\cite{kondo2020accurate}. The proposed Mueller pBSDF model in this work uses an alternative approach to potential ray paths when changing the component Mueller matrices in a pBSDF model. The proposed Mueller pBSDF is a two-component model inspired by an observed triple-degeneracy in the eigenspectrum of the measurements' coherency matrices \cite{cloudepottier}.

This work analyzes measurements of an object set consisting of five colors of dyed plastics, nine surface texture treatments, three illumination wavelengths, and 30 measurement geometries. Albedo is the diffuse reflectance of an object which depends on both the illumination waveband and the objects' color. Prior work reported the effects albedo and surface texture have on polarization and depolarization parameters of a measured Mueller matrix\cite{Lietal2020}. The mean square error (MSE) averaged over all albedos, geometries, and textures is used as a figure of merit to compare pBSDF model performance. 

Our proposed $\mathbf{p}^{(0)}$ complementary model (Section \ref{sec:CompModels}) is a normalized Mueller pBSDF model. A normalized Mueller pBSDF model of an ideal depolarizer and Fresnel reflection matrix\cite{cooktorrance,Walter2007,MSPIpBRDF}, the $\mathbf{p}^{(1)}$ base model, and a normalized and modified version of the Kondo et al. model\cite{kondo2020accurate}, the $\mathbf{p}^{(2)}$ bulk model, are used as established model benchmarks. Using normalized Mueller matrices allows polarimetric accuracy to be assessed independently from irradiance described by a scalar-valued BSDF. Unnormalized Mueller pBSDF implementations use many fit parameters, but working with normalized Mueller matrices reduces the number of parameters down to one or two. The normalization of the proposed $\mathbf{p}^{(0)}$ model decouples radiometry from polarimetry by dividing each Mueller matrix by the throughput (\emph{e.g.} average reflectance). This normalization allows the flexibility to use an existing scalar-valued BSDF and only one additional depolarization DoF to create a full radiometric and polarimetric model. 

An important contribution of this work is relating the 1 depolarization DoF to the quotient of the largest coherency eigenvalue and the throughput. This relation offers a new way to quantify error and perform pBSDF fitting between true and estimated normalized eigenvalues. The benefits of a pBSDF parameterized by the largest coherency eigenvalue are model fitting in lower dimensions and more effective measurement strategies that measure the largest normalized coherency eigenvalue rather than the full Mueller matrix.

This paper begins with a background on the applications of pBSDF models. Section \ref{sec:Methods} describes: the object ensemble (\ref{sec:Dataset}, \ref{sec:AcqGeo}), mathematical representations of polarization and polarimetry (\ref{sec:Polarization}, \ref{sec:Polarimetry}), mathematical representations of depolarization (\ref{sec:DepReps}, \ref{sec:3DegEig}). Section \ref{sec:MethodMeasXi0} describes a novel method of directly measuring $\xi_0$ with a minimum of two polarized measurements. The three models assessed in this work, the complementary model $\mathbf{p}^{(0)}$, the base model $\mathbf{p}^{(1)}$, and the bulk model $\mathbf{p}^{(2)}$, are presented in Section \ref{sec:pBSDFModels}. Section \ref{sec:MeasAgree} presents the first merit function in this work, ${\Delta}(\mathbf{m},\mathbf{p}|\mathbf{W})$, which compares a normalized model $\mathbf{p}$ to a normalized measurement $\mathbf{m}$ using simulated irradiances that are computed from the polarimetric measurement matrix $\mathbf{W}$. Section \ref{sec:pBSDFResults} describes the results from fitting $\mathbf{p}^{(0)}$, $\mathbf{p}^{(1)}$, and $\mathbf{p}^{(2)}$ to Mueller matrix measurements. 
Section \ref{sec:EstXi0} presents a second merit function $\Delta\xi_0$ and demonstrates an alternate method of fitting using the largest coherency eigenvalue, $\xi_0$. The conclusion in Sec. \ref{sec:Conclusion} summarizes the novel contributions and findings of this work. 

\section{Background} 
The microfacet BSDF model was first developed to produce more accurate off-specular scattering in computer graphics to improve the appearance of object renderings\cite{Torrance67,cooktorrance}. The pBSDF models by  Baek et al. \cite{Baek2018} in 2018 and Kondo et al. \cite{kondo2020accurate} in 2020 produce well-matched visual renderings for non-polarized and linearly-polarized illumination sources and observers. Many forms of microfacet distribution functions have since been developed for different object types for applications beyond computer graphics, such as remote sensing \cite{Walter2007,Breon,ashikmin2000microfacet,breon2017brdf}. A Mueller pBSDF model is also used in applications such as the realistic rendering of complex scenes, inverse models for material recognition and shape reconstruction, or synthetic training data for neural networks.  

The component Mueller matrices and the microfacet distribution functions in a Mueller pBSDF are changed to improve the agreement between a Mueller pBSDF and measurements. Baek et al. proposed a model\cite{Baek2018} which replaces the ideal depolarizer in a basic Mueller BSDF model, which is a sum of an ideal depolarizer and Fresnel reflection matrix, with a Mueller matrix that traces a hypothetical ray propagation path through a scattering material. This hypothetical ray path consists of Fresnel transmission into the material followed by bulk scattering represented by an ideal depolarizer and then a final Fresnel transmission out of the material. This model was used to improve surface normal estimation\cite{Baek2018} when applied to a $3\times3$ partial Mueller matrix. This $3\times3$ partial Mueller matrix omits the fourth row and fourth column of the full $4\times4$ Mueller matrix. In 2020, Baek et al. tested a full $4\times4$ implementation of this model and observed disagreement in polarimetric accuracy and irradiance when compared to their measurements. Baek et al. 2020 also conclude that rendering a large variety of materials requires the use of look-up tables, which this work does not refute.

The Kondo model\cite{kondo2020accurate} extends the Baek model\cite{Baek2018} as a $4\times4$ model by reintroducing the ideal depolarizer as a third component Mueller matrix. Each component Mueller matrix is normalized, and the weights for each component are constrained to sum to the measured total luminance for an object. The Kondo et al. model is a $3\times3$ partial Mueller matrix model applied only to linear polarization states. Data created using this three-component model is used as part of a training data set for a convolutional neural network to create realistic polarimetric renderings of scenes with elaborate, fine detail. A full $4\times4$ Mueller extension of this model, which assumes sub-surface bulk scattering can be partially isotropically depolarizing, is used as one of the benchmark models in this work.  

\section{Methods}\label{sec:Methods}
\begin{figure}[b]
    \centering
    \begin{subfigure}[Bricks ordered top-to-bottom from smoothest to the roughest texture and labelled T1-T9.]{\includegraphics[height=55mm]{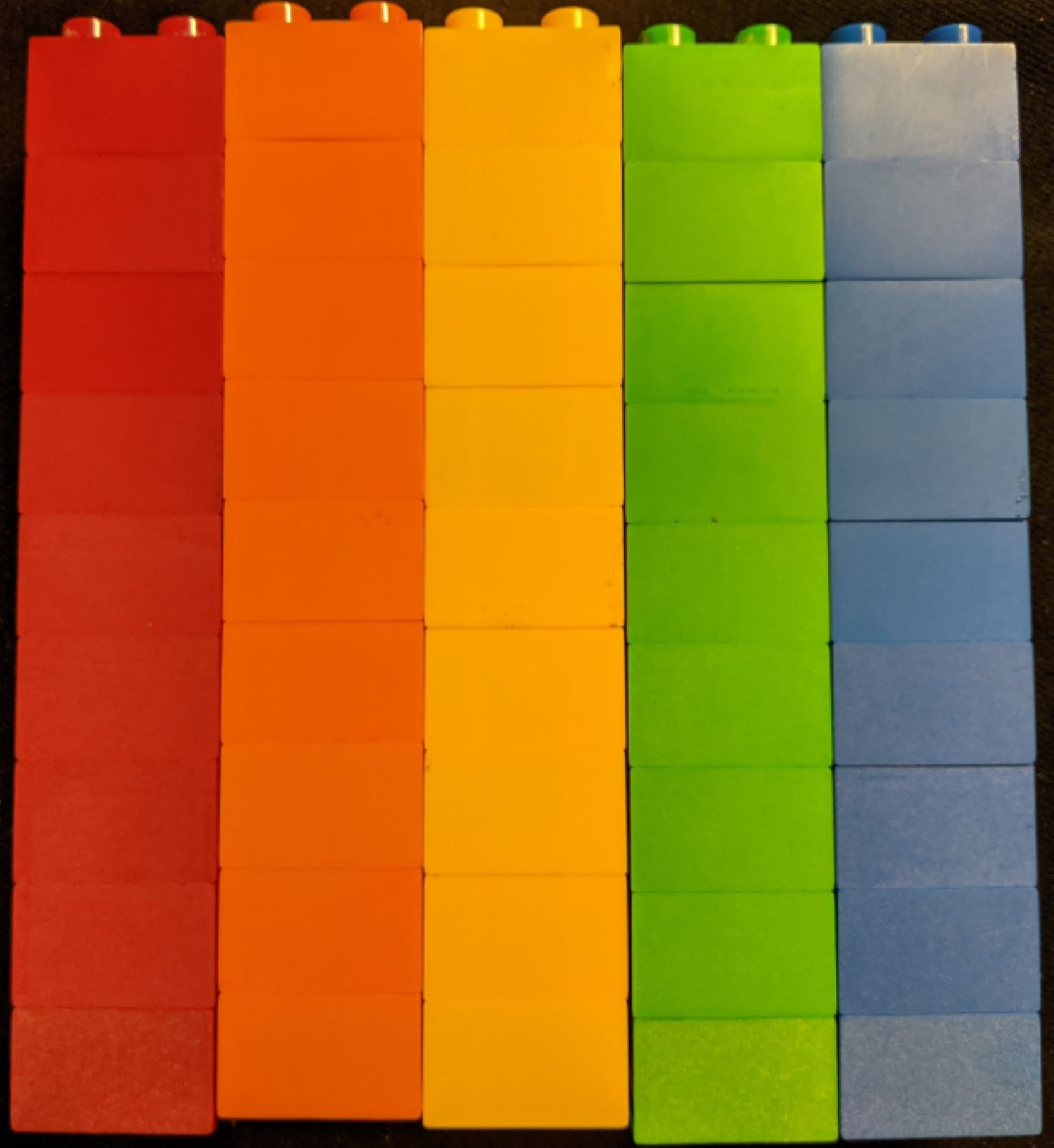}} \end{subfigure}
    \begin{subfigure}[Bricks arranged for measurements and texture location is shuffled.]{\includegraphics[height=55mm]{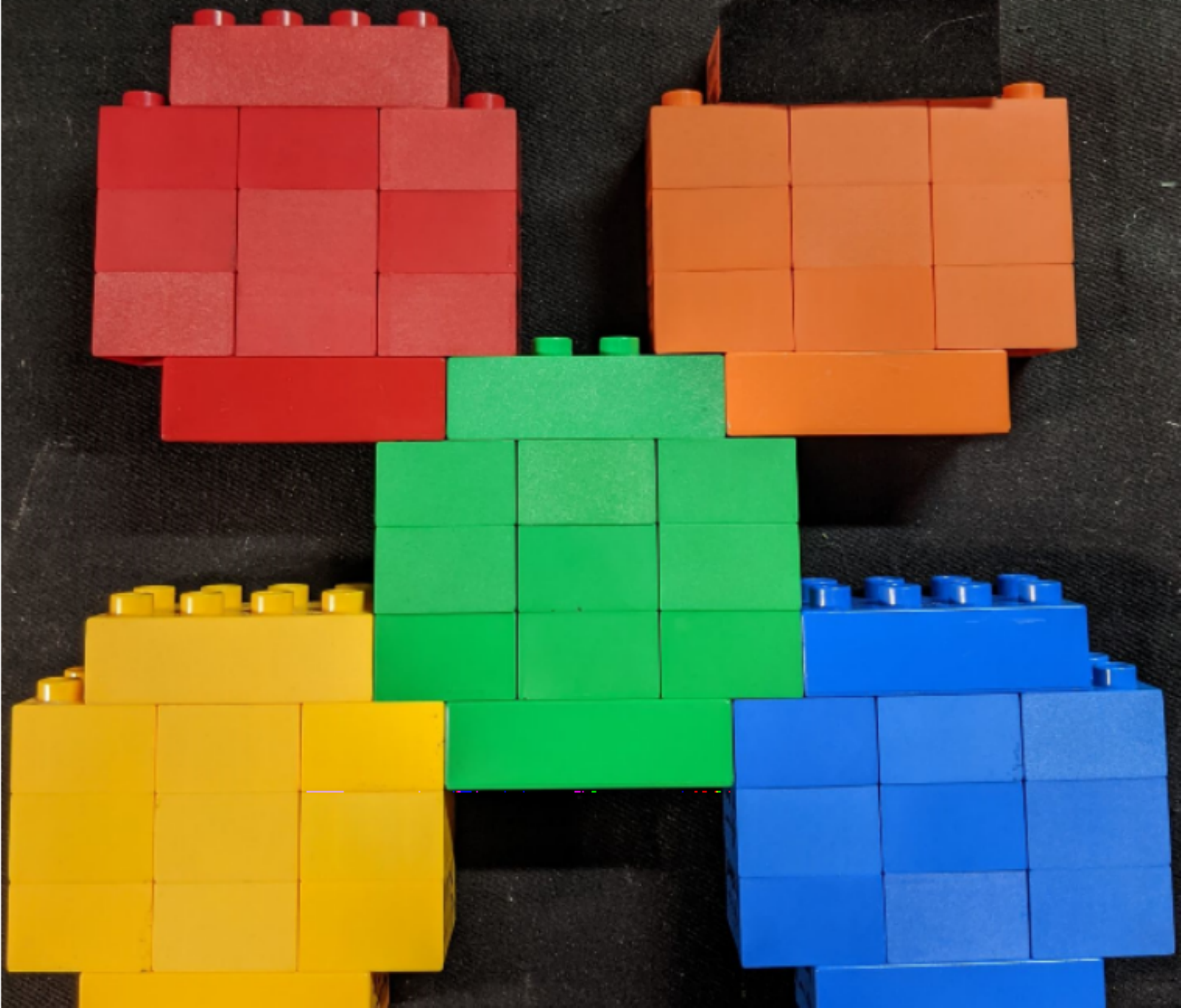}} \end{subfigure}
    \caption{A set of 45 objects with varying albedo and texture is created from five colors of LEGO DUPLO\texttrademark \ bricks sanded by nine grits of sandpaper. (a) Bricks ordered top-to-bottom from smoothest to roughest with texture labels T1-T9. (b) The texture location is shuffled when bricks are imaged; see Table \ref{tab:LegoTextures} in Appendix \ref{app:Dataset} for details.}
    \label{fig:3x3Towers}
\end{figure}  
\subsection{Materials of Varying Albedo and Texture}  \label{sec:Dataset}
The object ensemble is a group of red, orange, yellow, green, and blue LEGO Duplo 
bricks which are roughened using a belt sander equipped with nine different grits of sandpapers. This process creates 45 distinct bricks which differ in color and texture. A white-light interferometer is used to measure each sample's surface profiles and mean surface roughness to quantify surface roughness; see Table \ref{tab:LegoTextures} in the Appendix.

The 45 bricks pictured in Fig. \ref{fig:3x3Towers} are measured using a custom large-aperture Mueller matrix imaging polarimeter called the RGB950 \cite{RGB950}. The RGB950 measures objects under narrow-band illumination at $662\pm11.17$, $524\pm17.31$, and $451\pm9.78$ nm wavelengths. A single Mueller matrix measurement is taken at a given source position, camera position, and illumination waveband. In this study, 30 unique combinations of camera position and sample rotation are selected. These 30 geometries are reported in Tab. \ref{tab:MeasGeos} using angle of incidence onto the macrosurface $\theta_i$ and angle of exitance from the macrosurface $\theta_o$. The $\theta_i$ and $\theta_o$ angles are referenced to the central position in the $3\times3$ brick tower in Fig. \ref{fig:3x3Towers}(b). An $11\times11$ pixel region of interest (ROI) from each brick image is selected for analysis. This corresponds to a $3.2\times3.2$mm projected area on each brick. Each ROI on the $3\times3$ brick towers pictured in Fig \ref{fig:3x3Towers}b has a slightly different incident and exitant propagation vectors. Randomizing the texture positions within an image allows verification of trends with texture that are not dependent on acquisition geometry. 

\subsection{Acquisition Geometry} \label{sec:AcqGeo}
\begin{figure}[t]
    \centering
    \includegraphics[trim=3 1 5 3, clip,height=45mm]{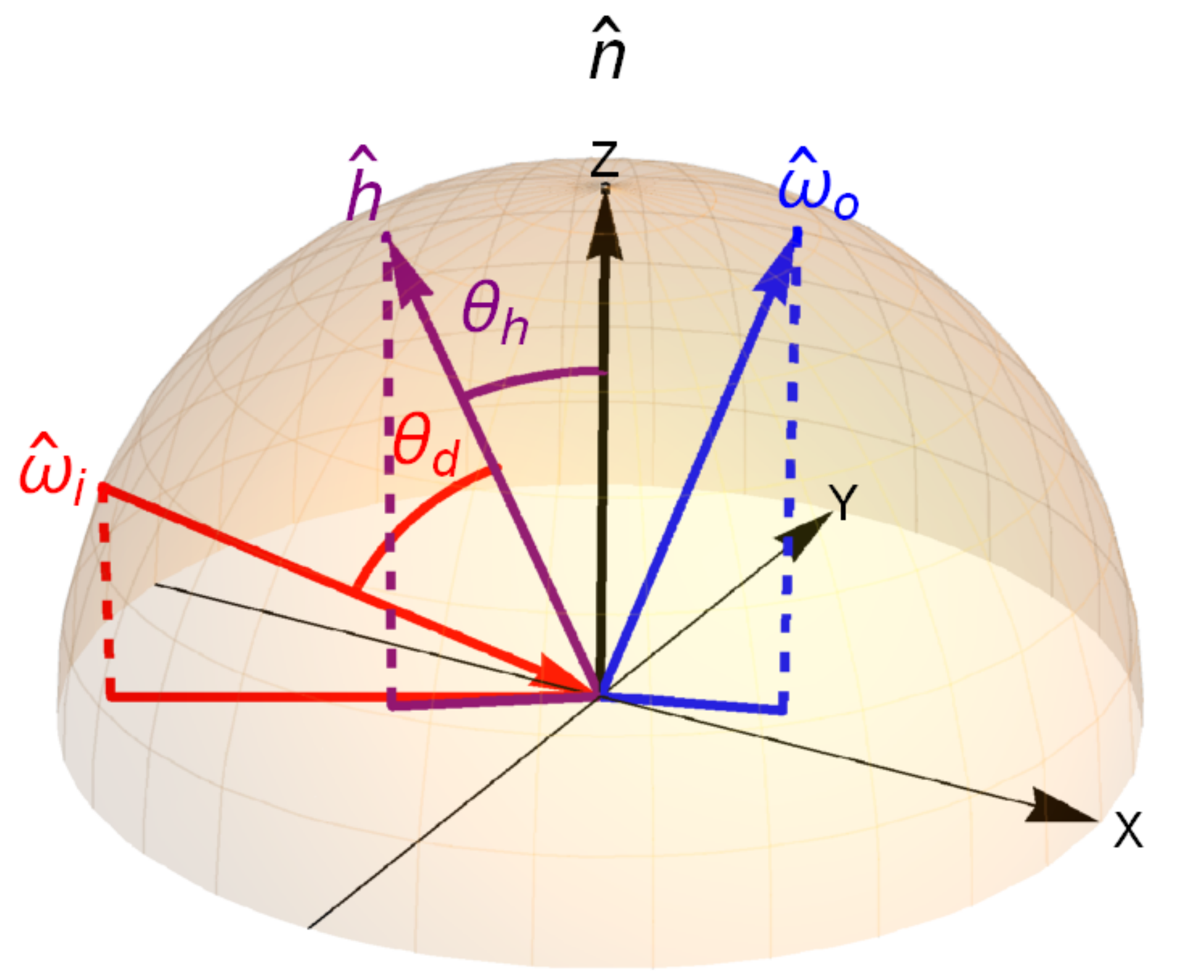} 
    \caption{The measurement geometry is described by the incident and exitant propagation vectors pointing in the direction of light travel: $\widehat{\boldsymbol{\omega}}_i$ and $\widehat{\boldsymbol{\omega}}_o$. For the convention of vectors pointing away from the surface, the incident propagation vector becomes $-\widehat{\boldsymbol{\omega}}_i$. The halfway vector $\widehat{\mathbf{h}}$ is the surface normal of an object which would produce a specular reflection. The surface normal of the measured object is $\widehat{\mathbf{n}}$ (which is parallel to $\widehat{\mathbf{z}}$) and called the macrosurface normal. The halfway angle $\theta_h$ between $\widehat{\mathbf{h}}$ and $\widehat{\mathbf{n}}$ is the deviation of the macrosurface from a specular geometry. The difference angle $\theta_d$ is the angle between $\widehat{\mathbf{h}}$ and $-\widehat{\boldsymbol{\omega}}_i$. The difference angle is $\theta_d = (180^\circ -\Omega)/2$, \emph{i.e.} half of the angle supplementary to the scattering angle $\Omega$. The scattering angle $\Omega$ is the angle between the incident and exitant propagation vectors.}
    \label{fig:Coordinates}
\end{figure}
In this work, unit vectors are denoted by a hat $\widehat{\cdot}$ and all vectors are boldface. All vectors point in the direction light travels (\emph{i.e.} k-vector direction) \cite{Baek2018,Baek2020Image}. The object surface normal is $\widehat{\mathbf{n}}$ which also referred to as a macronormal to distinguish from microfacet normals. The convention $\widehat{\mathbf{n}}=\widehat{\mathbf{z}}=\{0, 0, 1\}$ is adopted. The incident propagation direction of the illumination is $
\widehat{\boldsymbol{\omega}}_i=[
\sin(\theta_i)\cos(\phi_i),
\sin(\theta_i)\sin(\phi_i),
\cos(\theta_i)]$
where $\theta_i$ is the source zenith angle and $\phi_i$ is the source azimuth angle. Therefore backscattering configurations only occur when $\theta_i > 90^\circ$. The exitant propagation direction after a light-matter interaction which reaches the camera is $\widehat{\boldsymbol{\omega}}_o=[
\sin(\theta_o)\cos(\phi_o),
\sin(\theta_o)\sin(\phi_o),
\cos(\theta_o)]$. 
Therefore if $\widehat{\boldsymbol{\omega}}_i = \widehat{\boldsymbol{\omega}}_o$, the observer is looking into the source and the direction of light travel is unchanged. 

Microfacets are planar structures on a larger macrosurface that only reflect in the specular direction. Since the angle of incidence and the angle of specular reflection are equal, the surface orientation produces a specular reflection defined by the incident and exitant propagation vectors. An ensemble of sub-resolution microfacets is used to model a polarized contribution to light scattering in off-specular directions.
Surface texture influences the distribution of these microfacet orientations \cite{germer1999polarization}. For example, the ensemble of microfacet orientations for a perfectly smooth mirror would have no deviation from the macronormal. The micronormal $\widehat{\mathbf{m}}$ is the surface normal of a microfacet on a larger macrosurface. A microfacet distribution, which can be interpreted as a probability density function (pdf) on $\widehat{\mathbf{m}}$, are designed to describe various surface types, see Sec. {\ref{sec:MicroDistFuncs}}. 

The halfway vector $\widehat{\mathbf{h}}$ is the surface normal of an object which would produce a specular reflection for a given source and camera position; see Fig. \ref{fig:Coordinates}. This work adopts the notation for the halfway and difference angles introduced by Rusinkiewicz \cite{Rusinkiewicz:1998:ANC}. The difference angle $\theta_d$ is the angle of incidence onto a specular microfacet defined as the angle between vectors $-\widehat{\boldsymbol{\omega}}_i$ and $\widehat{\mathbf{h}}$. The halfway angle is the angle between $\widehat{\mathbf{n}}$ and $\widehat{\mathbf{h}}$. The cosine of the difference angle is equivalent to $\cos(\theta_d) = -\widehat{\boldsymbol{\omega}}_i \cdot \widehat{\mathbf{h}}$. The difference angle itself is also equivalent to $\theta_d = (180^\circ -\Omega)/2$, or half of the supplementary angle for the scattering angle $\Omega$. The scattering angle $\Omega$ is the angle between the incident and exitant propagation vectors calculated as $\cos{(\Omega)} = \widehat{\boldsymbol{\omega}}_i\cdot\widehat{\boldsymbol{\omega}}_o$.

\subsection{Polarization} \label{sec:Polarization}
Spectrally incoherent light can be fully polarized, partially-polarized, or completely unpolarized. The polarization state and irradiance of reflected, transmitted, and scattered light after a light-matter interaction is, in general, dependent upon the polarization state of the incident light. Mueller calculus describes the polarization transformation of linear light-matter interactions. Mueller calculus is required instead of Jones calculus for describing partial polarization. Both polarizing and depolarizing effects from material interactions are described using the $4\times4$ Mueller matrix, while the polarization state of light is described using the $4\times1$ vector of Stokes parameters. A vector of Stokes parameters $\mathbf{S}_\lambda(\widehat{\boldsymbol{\omega}})$ is defined along a propagation direction, \emph{e.g.} $\widehat{\boldsymbol{\omega}}_i$ or $\widehat{\boldsymbol{\omega}}_o$. 
The four Stokes parameters which describe all possible polarization states of light are
\begin{equation}
\mathbf{S}_\lambda(\widehat{\boldsymbol{\omega}})=
\begin{bmatrix}
 S_0 \\ S_1 \\ S_2 \\ S_3 
\end{bmatrix}
=\begin{bmatrix}
 P_H+P_V \\
 P_H-P_V \\
 P_{45^\circ}-P_{135^\circ} \\ 
 P_R-P_L \\
\end{bmatrix},
\label{eq:stokesintro}
\end{equation}
where $\mathbf{S}_\lambda(\widehat{\boldsymbol{\omega}})$ is called the Stokes vector and P are irradiance measurements in units of [W/$m^2$]. 
The subscripts on P denote transmission through a polarization filter: horizontal linear (H), vertical linear (V), 45$^\circ$ linear, 135$^\circ$ linear, right-circular (R), and left-circular (L). A lowercase $\mathbf{s}_\lambda(\widehat{\boldsymbol{\omega}}) = \mathbf{S}_\lambda(\widehat{\boldsymbol{\omega}})/S_0$ indicates a Stokes vector normalized by the total radiance. The degree of polarization ($DOP$) of a Stokes vector indicates the fraction of radiance that is polarized
\begin{equation}
    DOP(\mathbf{S}_\lambda(\widehat{\mathbf{\omega}})) = \frac{\sqrt{S_1^2+S_2^2+S_3^2}}{S_0}.
    \label{eq:DOP}
\end{equation}
For unpolarized light $DOP=0$ and Stokes parameters in the numerator of Eq. \ref{eq:DOP} are zero and $S_0$ is non-zero. When $DOP=1$ light is fully polarized and $S_0 = \sqrt{S_1^2+S_2^2+S_3^2}$. 

Mueller matrix operations may increase or decrease the $DOP$ of an exitant Stokes vector compared to an incident Stokes vector. The magnitude of this $DOP$ change is, in general, different for each polarization state. The $4\times4$ Mueller matrix $\mathbf{M}$ describes a material's linear interaction with the Stokes parameters
\begin{equation}
\mathbf{S}'_\lambda(\widehat{\boldsymbol{\omega}}_o)=\mathbf{M}_\lambda(\widehat{\boldsymbol{\omega}}_i,\widehat{\boldsymbol{\omega}}_o,\widehat{\mathbf{n}})
\mathbf{S}_\lambda(\widehat{\boldsymbol{\omega}}_i)\label{eq:muellercalc}
\end{equation}
where $\mathbf{S}_\lambda(\widehat{\boldsymbol{\omega}}_i)$ indicates the incident polarization state and $\mathbf{S}'_\lambda(\widehat{\boldsymbol{\omega}}_o)$ indicates the exiting state. Mueller matrices are dependent on the incident propagation vector $\widehat{\boldsymbol{\omega}}_i$, the exitant propagation vector $\widehat{\boldsymbol{\omega}}_o$, surface normal of the material $\widehat{\mathbf{n}}$, and the illumination wavelength $\lambda$. The Stokes vectors for the incident state and the exiting state are defined with respect to different reference planes. The incident Stokes parameters are defined with respect to the incident plane while the exitant Stokes parameters are defined with respect to the meridional plane. The Mueller matrix is dependent on the propagation directions $\widehat{\boldsymbol{\omega}}_i$ and $\widehat{\boldsymbol{\omega}}_o$, so the Mueller matrix is defined in the scattering plane. Appendix \ref{app:RotM} describes Mueller rotation matrices to transform a Stokes vector between these different reference planes.

The individual 16 elements $M_{ij}$ are unitless 
\begin{equation}
\mathbf{M}_\lambda(\widehat{\boldsymbol{\omega}}_i,\widehat{\boldsymbol{\omega}}_o,\widehat{\mathbf{n}})=
\mathbf{M}_\lambda(\theta_h,\theta_d)=
\begin{bmatrix}
 M_{00} & M_{01} & M_{02} & M_{03} \\
 M_{10} & M_{11} & M_{12} & M_{13} \\
 M_{20} & M_{21} & M_{22} & M_{23} \\
 M_{30} & M_{31} & M_{32} & M_{33} \\
\end{bmatrix}.
\label{eq:muellerintro}
\end{equation} 
where the $\mathrm{M}_{00}$ element is the reflectance for unpolarized incident light. A normalized Mueller matrix is divided by this reflectance
\begin{equation}
    \mathbf{m}_\lambda(\widehat{\boldsymbol{\omega}}_i,\widehat{\boldsymbol{\omega}}_o,\widehat{\mathbf{n}}) = \mathbf{M}_\lambda(\widehat{\boldsymbol{\omega}}_i,\widehat{\boldsymbol{\omega}}_o,\widehat{\mathbf{n}})/\left[\mathbf{M}_\lambda(\widehat{\boldsymbol{\omega}}_i,\widehat{\boldsymbol{\omega}}_o,\widehat{\mathbf{n}})\right]_{00}.\label{eq:m_norm}
\end{equation} 
Here the normalized Mueller matrix is lowercase, similar to the normalized Stokes vector, and the notation $[\mathbf{M}]_{00}=M_{00}$ is used. In general, the reflectance, and therefore $\mathrm{M}_{00}$, is a function of incident and exitant propagation directions which is related to the scalar-valued BRDF by \cite{PriestGermer2000}
\begin{equation}
R(\widehat{\boldsymbol{\omega}}_i,\widehat{\boldsymbol{\omega}}_o)=\left[\mathbf{M}(\widehat{\boldsymbol{\omega}}_i,\widehat{\boldsymbol{\omega}}_o)\right]_{00}=\frac{\pi}{\Omega_i\Omega_o}\int_{\widehat{\boldsymbol{\omega}}_i}\int_{\widehat{\boldsymbol{\omega}}_o}f(\widehat{\boldsymbol{\omega}}_i,\widehat{\boldsymbol{\omega}}_o)d\Omega_id\Omega_o.
\label{eq:BRDFReflectance}
\end{equation}
Here $f$ is the BRDF in units of inverse steradians, the reflectance $R$ is unitless, and the differential projected solid angle is $d\Omega_n = \cos\theta_n \sin\theta_n d\theta_n d\phi_n$ where $\theta_n$ and $\phi_n$ are the zenith and azimuth angles of the propagation vector $\boldsymbol{\omega}_n$. Here, the $n$ subscript indicates that the differential projected solid angle definition applies to both $i$ and $o$.

A normalized Mueller matrix operating on an incident Stokes vector $\mathbf{S}_\lambda(\widehat{\boldsymbol{\omega}}_i)$ can be scaled by the reflectance to produce the exitant Stokes vector 
\begin{equation}
    \mathbf{S}'_\lambda(\widehat{\boldsymbol{\omega}}_o) = R(\widehat{\boldsymbol{\omega}}_i,\widehat{\boldsymbol{\omega}}_o)\mathbf{m}(\widehat{\boldsymbol{\omega}}_i,\widehat{\boldsymbol{\omega}}_o,\widehat{\mathbf{n}}) \mathbf{S}_\lambda(\widehat{\boldsymbol{\omega}}_i).\label{eq:norm_n}
\end{equation}
The normalized Mueller matrix is an idealized Lambertian reflector since for all measurement geometries, the element $[\mathbf{m}]_{00}$ equals one. Although the normalization loses absolute radiometry, the polarimetry of the light-matter interaction is preserved. The incident to exitant $DOP$ and the incident to exitant polarization ellipse are identical for both normalized and unnormalized Mueller matrices. 

\begin{figure}[t]
    \centering
    \begin{subfigure}[High albedo: Red brick 662nm illumination]{\includegraphics[width=0.49\textwidth]{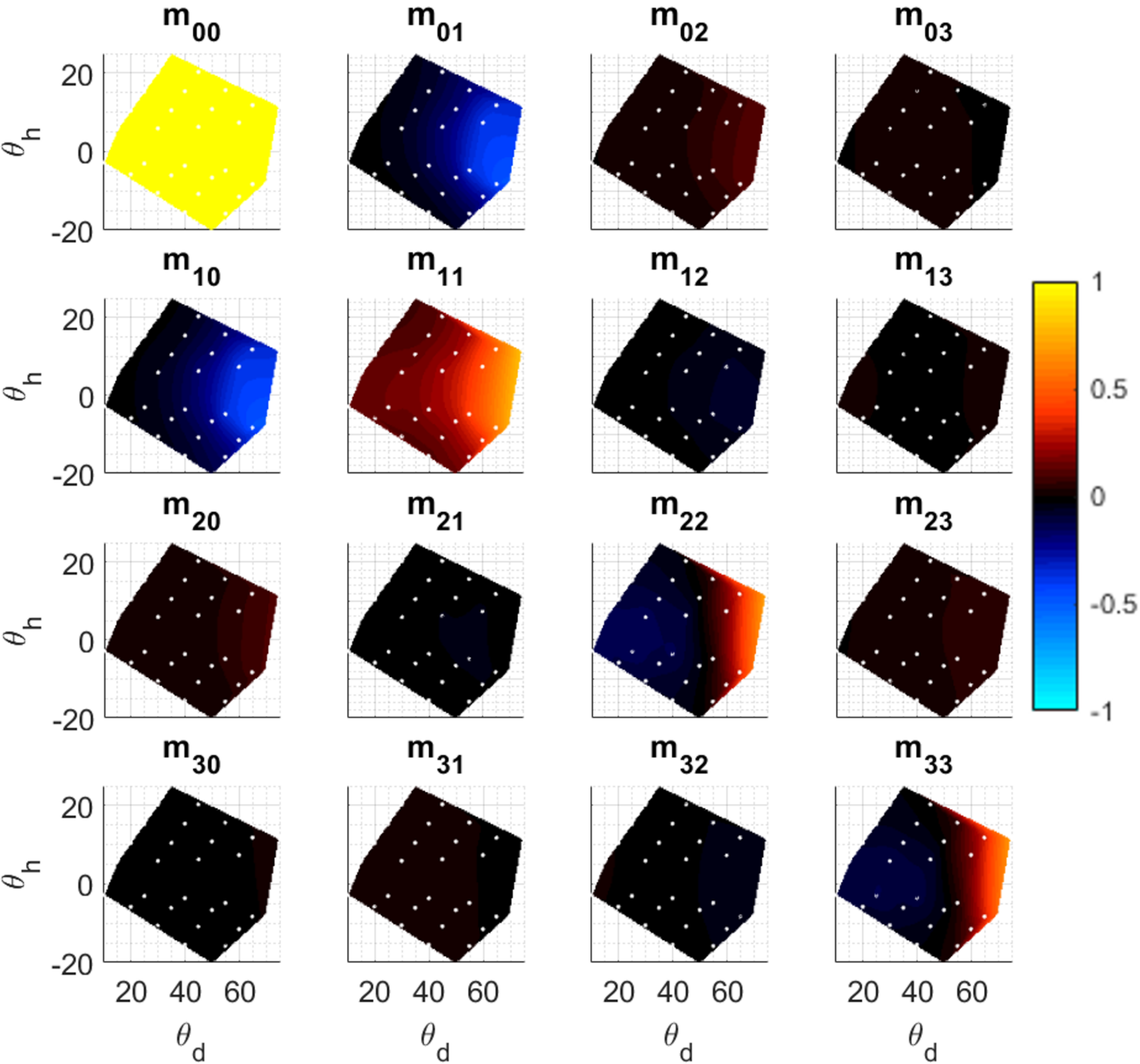}}\end{subfigure}  
    \begin{subfigure}[Low albedo: Red brick 451nm illumination]{\includegraphics[width=0.49\textwidth]{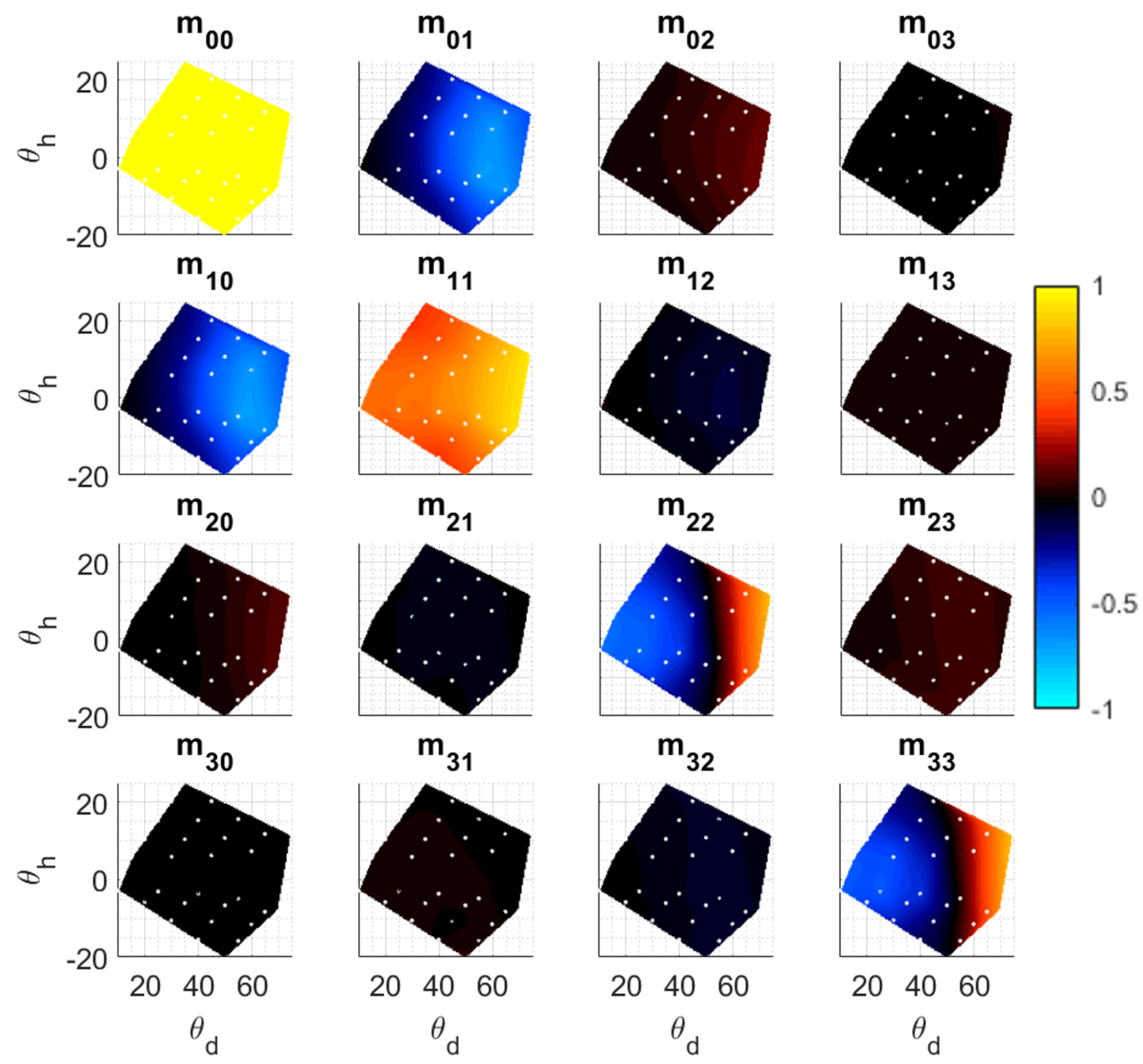}}\end{subfigure}
    \caption{Normalized Mueller matrix measurements $\widehat{\mathbf{m}}_{\lambda}(\widehat{\boldsymbol{\omega}}_i,\widehat{\boldsymbol{\omega}}_o)$ in a signed halfway angle $\theta_h$ and difference angle $\theta_d$ space for a red brick of smoothest texture (T1). In (a), 662nm illumination of the red object is high albedo, and in (b) 451nm illumination is low albedo. The white dots' locations are the 30 capture geometries measured, and MATLAB's natural neighbor interpolation is used to generate images. The most pronounced differences between the Mueller matrices in (a) high albedo and (b) low albedo conditions are the diagonal elements at lower scattering angles. High albedo illumination produces more depolarization than low albedo illumination due to an increase in multiple scattering events. Both high and low albedo are anisotropic depolarizers since for isotropic depolarization, see Eq.\ref{eq:PDMatrix}, the on-diagonal linear $\mathrm{m}_{11},\mathrm{m}_{22}$ and circular $\mathrm{m}_{33}$ elements are equal.}
    \label{fig:HighLowAlbedo}
\end{figure}

This work analyzes normalized Mueller matrices measurements at 30 geometries; see Fig.\ref{fig:HighLowAlbedo} for examples. In this normalized form, the changes to the relative magnitude of the non-$\mathrm{m}_{00}$ elements at varying geometries can be assessed independently from the reflectance changes. Measuring the Mueller matrix at a wide range of incident and exitant geometries requires varying the exposure settings with scattering angle. Varying exposure settings were selected to maximize the detector's dynamic range, thus maximizing the polarimetric accuracy. However, the absolute radiometry required for an unpolarized BSDF profile is lost when the exposure is varied. Any Mueller pBSDF model $\mathbf{P}(\widehat{\boldsymbol{\omega}}_i,\widehat{\boldsymbol{\omega}}_o,\widehat{\mathbf{n}})$ can be factored into a reflectance component $R(\widehat{\boldsymbol{\omega}}_i,\widehat{\boldsymbol{\omega}}_o,\widehat{\mathbf{n}})$ and a normalized Mueller matrix model component $\mathbf{p}(\widehat{\boldsymbol{\omega}}_i,\widehat{\boldsymbol{\omega}}_o,\widehat{\mathbf{n}})$. The normalized pBSDF can be scaled by an unnormalized BSDF to return a polarization-dependent reflectance at every measurement geometry.

Figure \ref{fig:HighLowAlbedo} compares normalized Mueller matrix measurements with 451nm and 662nm illumination for which the smoothest (T1) red brick is low and high albedo, respectively. These low and high albedo measurements differ most in the on-diagonal elements, which correspond to isotropic depolarization, see Eq. \ref{eq:PDMatrix}.
Normalized Mueller matrix elements $m_{11}, m_{22},$ and $m_{33}$ in the low albedo measurement Fig. \ref{fig:HighLowAlbedo}b have magnitudes $\geq0.5$ for the $\theta_h = 0^\circ\pm10^\circ$ and $\theta_d<\theta_B(1,1.54)$ geometries. The same elements for the high albedo Mueller matrix in Fig \ref{fig:HighLowAlbedo}a are $\approx0$. The difference is especially noticeable for measurements $\theta_d<57^\circ$. The low albedo Mueller matrix elements $m_{01},m_{02},m_{10},m_{20}, m_{23},$ and $m_{32}$ also have increased magnitude for geometries which are $\approx0$ for the high albedo Mueller matrix.

For the Mueller matrix visualization in Fig.\ref{fig:HighLowAlbedo}, measurements are represented in a signed halfway angle versus difference angle space. The halfway angle is specified between $0^\circ$ and $90^\circ$, but in signed space, an artificial sign is attached to $\theta_h$. This sign of $\theta_h$ is assigned positive when the azimuth angle of the halfway vector, $\phi_h$, is between $90^\circ$ and $270^\circ$; the sign is assigned negative otherwise. This sign convention is further discussed in Appendix \ref{app:MeasGeo}.

\subsection{Polarimetry}\label{sec:Polarimetry}
A Mueller matrix is estimated from a series of images acquired at varying Polarization State Analyzer (PSA) and Polarization State Generator (PSG) states \cite{Chipman1}. For linear light-matter interactions the relationship between a noise-free scalar-valued irradiance and an object's Mueller matrix is 
\begin{equation}
i=\mathbf{a}^t\mathbf{M}\mathbf{g}.\label{eq1}
\end{equation}
Here $\mathbf{a}$ is the $4$ Stokes parameters describing the PSA, $\mathbf{g}$ is the Stokes parameters of the PSG, $i$ is a noise-free irradiance, and $t$ denotes the transpose of a real-valued vector. Consider forming a single $16\times1$ vector from the PSA and PSG Stokes vectors, $\mathbf{w}=\mathbf{a}\otimes\mathbf{g}$ where $\otimes$ is the Kronecker product. If $\mathbf{A}$ is an $m\times n$ matrix and $\mathbf{B}$ is a $p\times q$ matrix, then the Kronecker product $\mathbf{A}\otimes\mathbf{B}$ is the $mp\times nq$ block matrix
\begin{equation}
\mathbf{A}\otimes\mathbf{B}=\begin{pmatrix}
{a}_{11}\mathbf{B} \hspace{.5cm}$\dots$\hspace{.5cm}{a}_{1n}\mathbf{B} \\ 
\dots \hspace{.5cm}$$\ddots$$\hspace{.5cm} \dots\\
{a}_{m1}\mathbf{B} \hspace{.5cm}$\dots$\hspace{.5cm} {a}_{mn}\mathbf{B} 
\end{pmatrix}.
\end{equation}
Then the irradiance in Eq. \ref{eq1} can be rewritten as the inner-product of vectors
\begin{equation}
i(\mathbf{M})=\mathbf{w}^t\Vec{\mathbf{M}}\label{VectorProduct}
\end{equation}
where the dependence of the irradiance on the Mueller matrix is written explicitly as $i(\mathbf{M})$ and $\Vec{\mathbf{M}}$ is a $16\times1$ vector of the Mueller elements. A series of $L$ irradiance values can be expressed as
\begin{equation}
    \mathbf{i}(\mathbf{M})=\mathbf{W}^t\Vec{\mathbf{M}}\label{eq:measurement}
\end{equation}
where $\mathbf{W}$ is a $16 \times L$ matrix called the polarimetric measurement matrix and each row  can be written as a Kronecker product between the $l^{th}$ PSA/PSG
\begin{equation}
\mathbf{W}=\begin{pmatrix}\mathbf{a}_1\otimes\mathbf{g}_1\\ \mathbf{a}_2\otimes\mathbf{g}_2\\ ...\\ \mathbf{a}_L\otimes\mathbf{g}_L\end{pmatrix}.
\end{equation}

\subsection{Depolarization} \label{sec:DepReps}
Depolarization refers to a reduction in the $DOP$ after a light-matter interaction. A Mueller matrix is depolarizing if the degree of polarization is greater for the incident light than for the exitant light, \emph{i.e.} $DOP(\mathbf{S}_\lambda(\widehat{\boldsymbol{\omega}}_i)) > DOP(\mathbf{S}'_\lambda(\widehat{\boldsymbol{\omega}}_o))$. If $DOP(\mathbf{S}'_\lambda(\widehat{\boldsymbol{\omega}}_o)) = 0$ and $DOP(\mathbf{S}_\lambda(\widehat{\boldsymbol{\omega}}_i)) > 0$, then the Mueller matrix $\mathbf{M}_\lambda(\widehat{\boldsymbol{\omega}}_i,\widehat{\boldsymbol{\omega}}_o,\widehat{\mathbf{n}})$ is completely depolarizing for the incident Stokes vector $\mathbf{S}_\lambda(\widehat{\boldsymbol{\omega}}_i)$. An isotropic partial depolarizer reduces the degree of polarization equally for $S_1, S_2$ and $S_3$ and is given by,
\begin{equation} \label{eq:PDMatrix}
    \mathbf{D}(d) = \begin{bmatrix}
    1&0&0&0\\0&d&0&0\\0&0&d&0\\0&0&0&d
    \end{bmatrix}
\end{equation}
where $d$ is the isotropic depolarization ratio. For an ideal depolarizer $d=0$ and exitant light is always unpolarized, \emph{i.e.} $DOP(\mathbf{S}'_\lambda(\widehat{\boldsymbol{\omega}}_o)) = 0$, for all incident polarization states.
Depolarization characteristics of a Mueller matrix can be described using depolarization parameters such as polarization entropy \cite{cloudepottier}, the Cloude depolarization index \cite{cloude1986group}, the Gil and Bernabeau depolarization index \cite{depind}, and the Ossikovski depolarization indices \cite{ossikovski2010alternative}. 

\begin{figure}[ht!]
    \centering
    \begin{subfigure}[High albedo, rough texture: $\widehat{\mathbf{m}}_0$ for R9, 662nm]{\includegraphics[width = 0.48\textwidth]{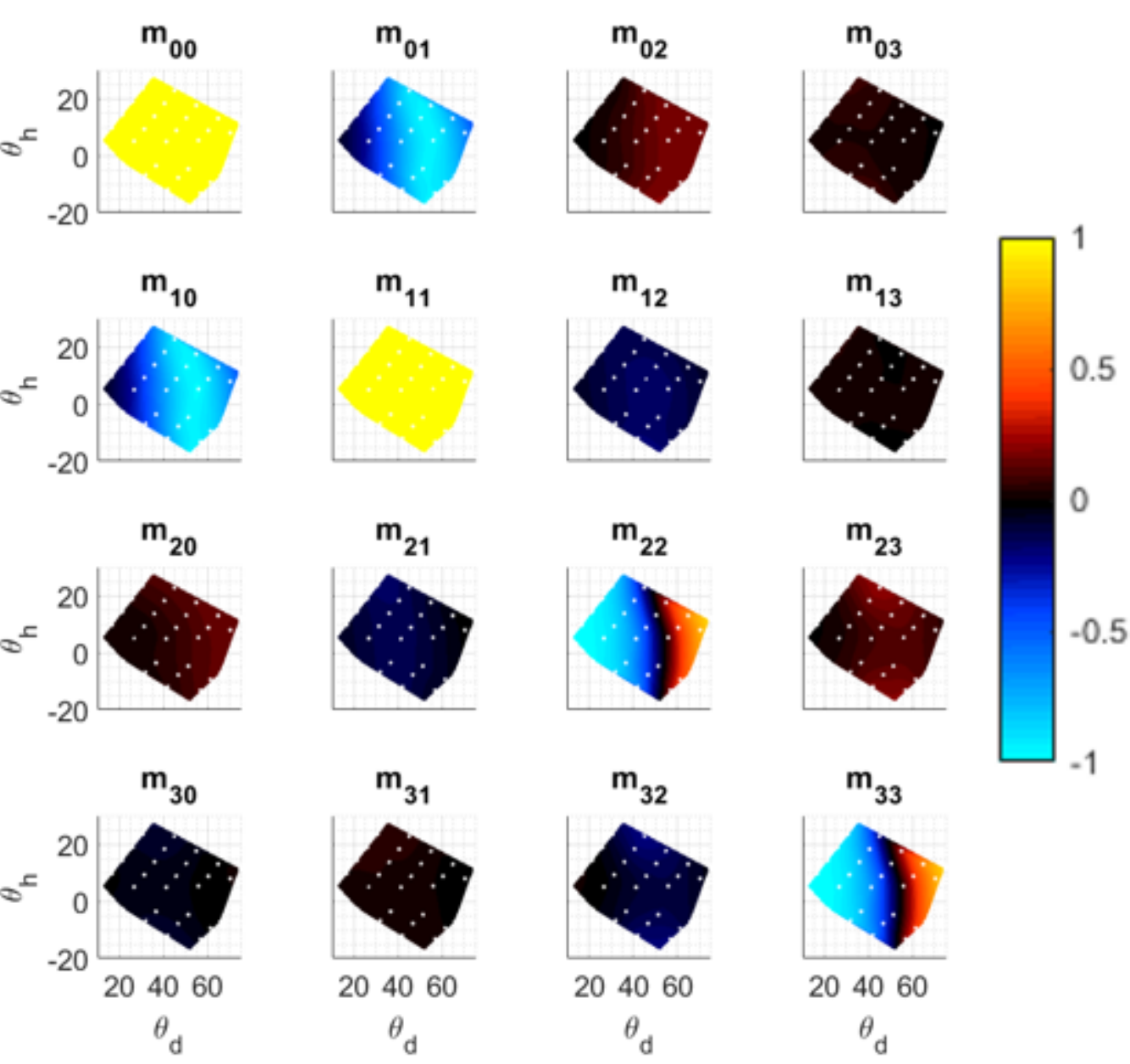}}\end{subfigure}
    \begin{subfigure}[High albedo, smooth texture: $\widehat{\mathbf{m}}_0$ for B1, 451nm ]{\includegraphics[width = 0.48\textwidth]{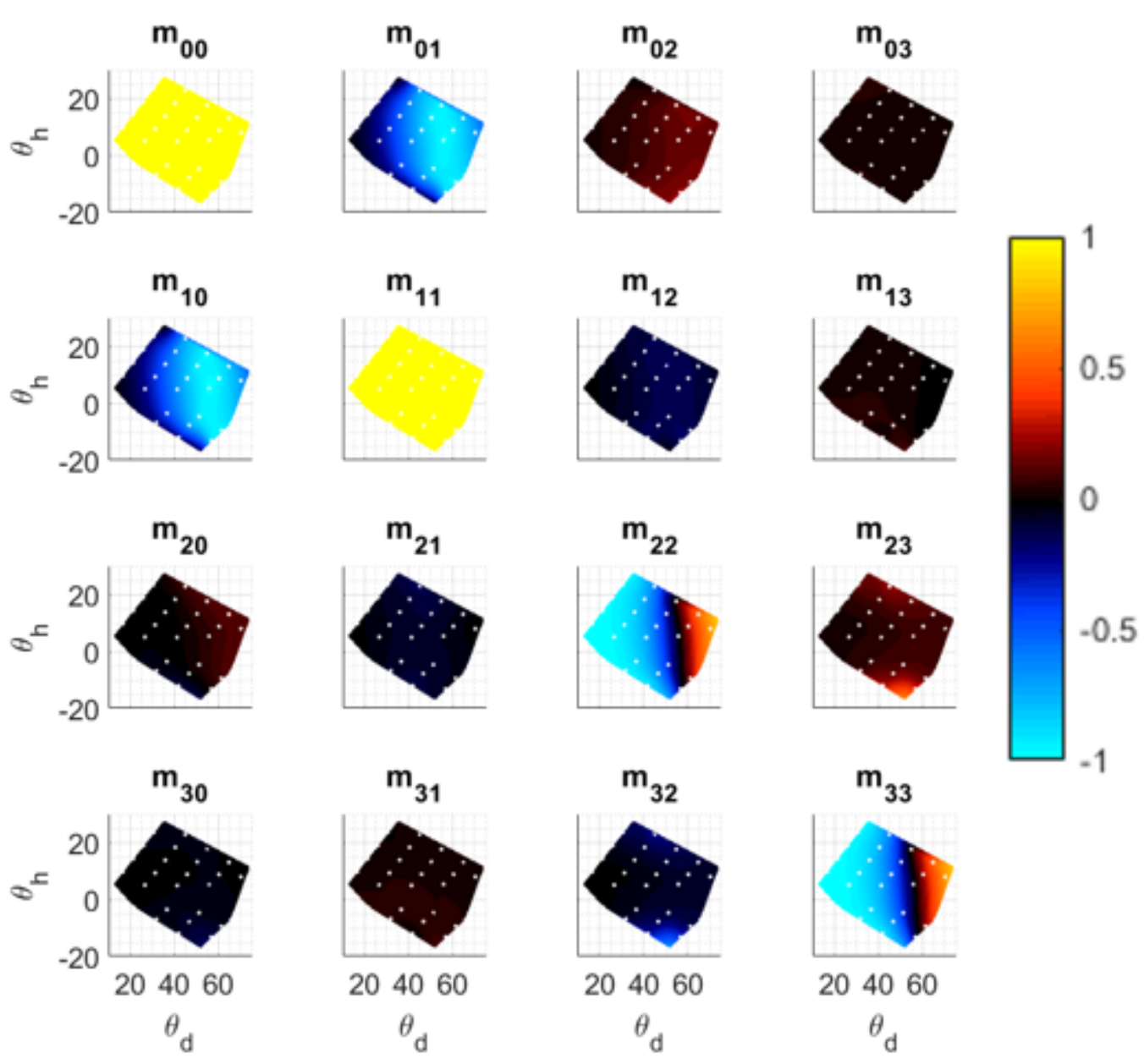}}\end{subfigure} 
    \begin{subfigure}[Low albedo, rough texture: $\widehat{\mathbf{m}}_0$ for R9, 451nm]{\includegraphics[width = 0.48\textwidth]{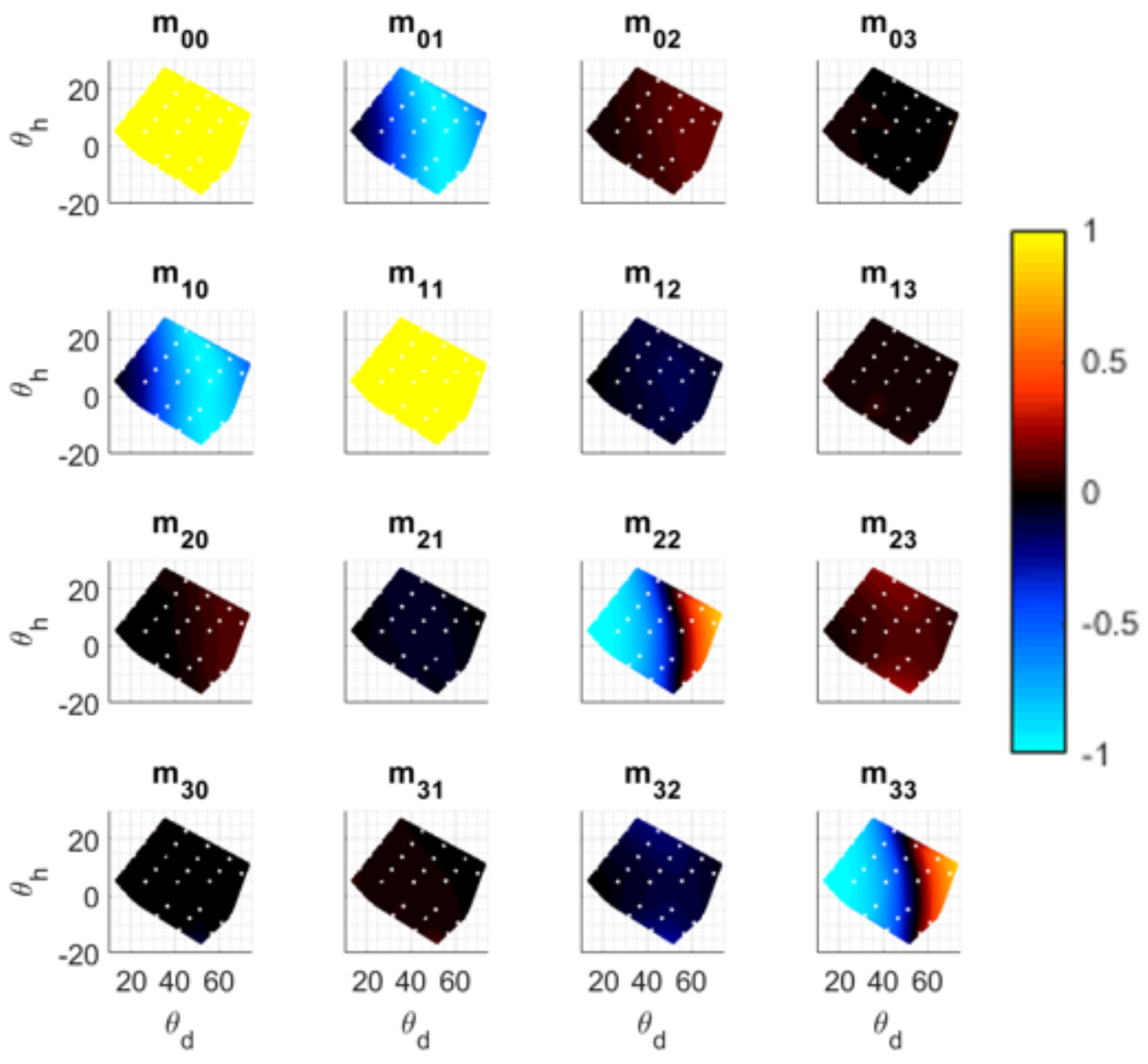}}\end{subfigure}
    \begin{subfigure}[Low albedo, smooth texture: $\widehat{\mathbf{m}}_0$ for B1, 662nm]{\includegraphics[width = 0.48\textwidth]{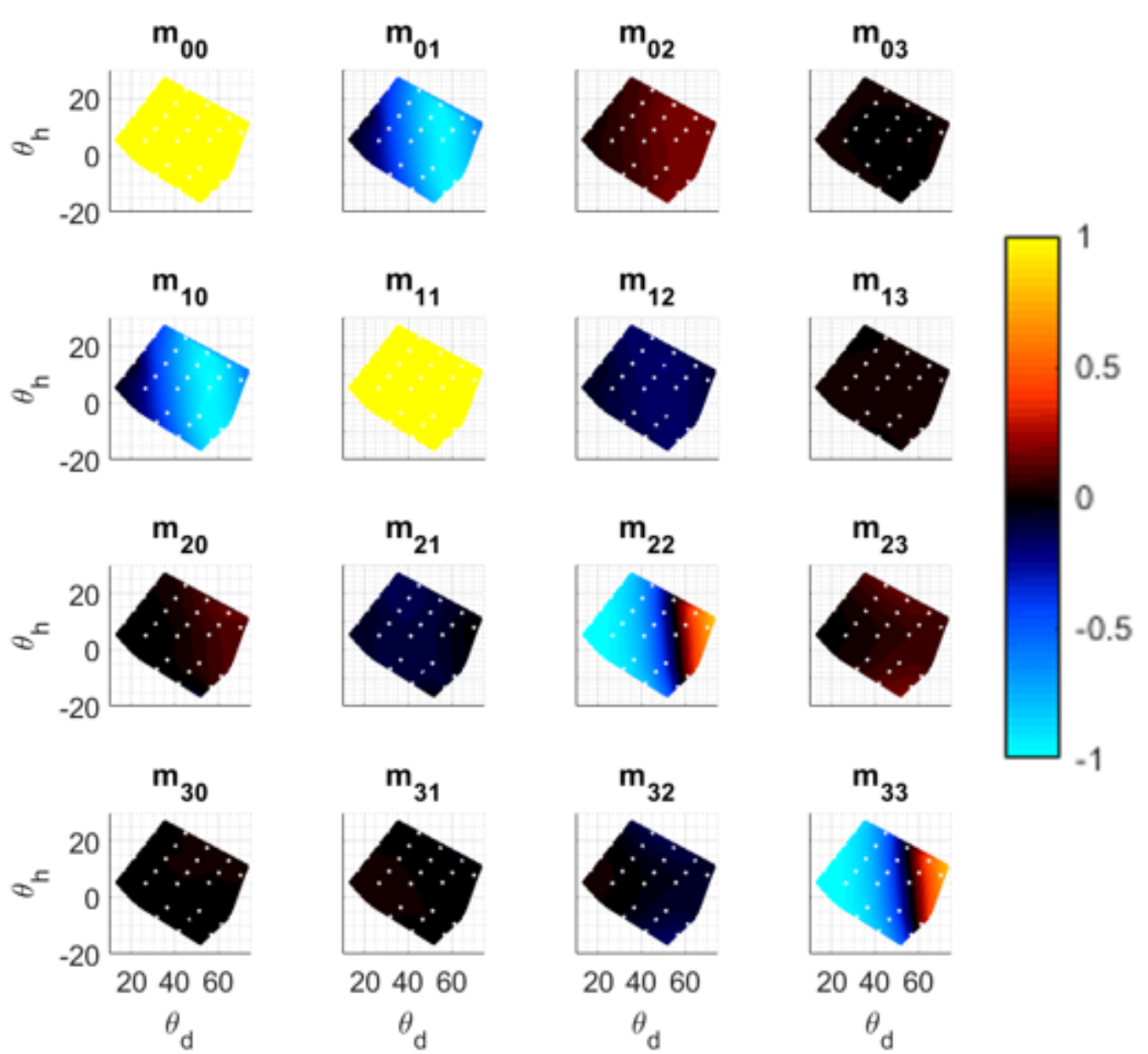}}\end{subfigure} 
    \caption{The most significant Mueller-Jones matrix $\widehat{\mathbf{m}}_0$ for the roughest red brick (R9) in (a,c) and smoothest blue brick (B1) in (b,d) for high and low albedo illumination, respectively. All four of these Mueller matrices resemble the Fresnel reflection matrix in Fig. \ref{fig:ComponentMMs} (b). The location of the 30 white dots are the measurement geometries in the signed halfway angle $\theta_h$ and difference angle $\theta_d$ space. MATLAB's natural neighbor interpolation is used between measurement geometries.}
    \label{fig:BasisMJMs}
\end{figure}  

For every Mueller matrix $\mathbf{M}$ there is a $4 \times 4$ complex-valued Hermitian coherency matrix $\mathbf{C}$ related by \cite{cloudepottier}
\begin{eqnarray}
    \mathbf{C} &=& \frac{1}{4}\sum_{i,j=0}^3 M_{ij}\mathbf{U}\left[\boldsymbol{\sigma}_i\otimes\boldsymbol{\sigma}_j^*\right]\mathbf{U}^\dagger
\end{eqnarray}
where $\boldsymbol{\sigma}$ are Pauli-spin matrices and
\begin{equation} \label{eq:UMatrix}
\mathbf{U}=
\frac{1}{\sqrt{2}}
\begin{bmatrix}
1 & 0 & 0 & 1\\
1 & 0 & 0 &-1\\
0 & 1 & 1 & 0\\
0 & i &-i & 0\\
\end{bmatrix}.
\end{equation}

Here the dagger $\dagger$ indicates a complex conjugate transpose operation and $\otimes$ indicates a Kronecker product. Since the coherency matrix is Hermitian it can be written as
\begin{equation}
    \mathbf{C} =
    \sum^{R-1}_{n=0}\xi_n\mathbf{c}_n\mathbf{c}_n^\dagger
    \label{eq:CohVecs}
\end{equation} 
where $R$ is the rank, $\xi_n$ are the real and non-negative eigenvalues in descending order, and $\mathbf{c}_n$ are the orthonormal eigenvectors of the coherency matrix. 

Each eigenvector of the coherency matrix in Eq. \ref{eq:CohVecs} corresponds to a Jones matrix 
\begin{equation}
    \mathbf{J}_n = [\mathbf{c}_n]_0 \boldsymbol{\sigma}_0 + [\mathbf{c}_{n}]_{1} \boldsymbol{\sigma}_1 +  [\mathbf{c}_{n}]_{2} \boldsymbol{\sigma}_2 +  [\mathbf{c}_{n}]_{3} \boldsymbol{\sigma}_3
\end{equation}
where $n = {0,1,2,3}$ indicates the $n^{th}$ eigenvector and $[\cdot]_i$ indicates the $i^{th}$ element of the vector. 

A Jones matrix is related to a non-depolarizing Mueller matrix (\emph{i.e.} Mueller-Jones matrix), by
\begin{equation}
    \widehat{\mathbf{M}}
    =\mathbf{U}\left(\mathbf{J}\otimes\mathbf{J}^{\ast}\right)\mathbf{U}^{-1}
    \label{eq:JB2M}
\end{equation}
where the $\widehat{\cdot}$ notation indicates that the coherency matrix of any Mueller-Jones matrix is rank one. The Mueller-Jones matrices computed from the orthogonal eigenvectors of the coherency matrix can be incoherently summed to express the original Mueller matrix
\begin{equation}
    \mathbf{M}=\sum_{n=0}^{R-1}\xi_n\widehat{\mathbf{m}}_n. \label{eq:Mn2M}
\end{equation}
This treatment is also called spectral decomposition. Spectral decomposition separates a depolarizing Mueller matrix into non-depolarizing parts. Alternate decompositions of a partially depolarizing Mueller matrix split $\mathbf{M}$ into the combination of a fully depolarizing component and either a fully depolarizing $\mathbf{D}(0)$ component\cite{kostinski1992depolarization} or an isotropically depolarizing component $\mathbf{D}(d)$ \cite{le1996optical}. Each $\widehat{\mathbf{m}}_n$ is the Mueller-Jones matrix associated with the eigenvector $\mathbf{c}_n$ in Eq. \ref{eq:CohVecs} which is constrained to be normalized. The sum of the eigenvalues, $\sum^{R}_{n=0}\xi_n$ equals the $\mathrm{M}_{00}$ element of the original Mueller matrix. Each eigenvalue $\xi_n$ weights the contribution of the normalized Mueller-Jones matrix $\widehat{\mathbf{m}}_n$ to the original Mueller matrix. 

Polarization entropy is a scalar value between 0 and 1 which is related to the depolarization index of a Mueller matrix and is calculated from the coherency matrix eigenspectrum \cite{cloudepottier,aiello2005physical,pires2008statistics}. 

\begin{equation} \label{eq:PolEnt}
    E(\mathbf{m}) = -\sum_{n=0}^{R-1} \xi_n \log_4(\xi_n).
\end{equation}
where a normalized Mueller matrix $\mathbf{m}$ is used to calculate $\xi_n$, which constrains the sum of the eigenvalues to one. A polarization entropy of zero indicates the Mueller matrix is a Mueller-Jones matrix because the coherency matrix is rank one, \emph{i.e.} only one eigenvalue of the coherency matrix is non-zero. A Fresnel reflection matrix is an example of a Mueller-Jones matrix. 

\subsection{Triply Degenerate Eigenspectrum} \label{sec:3DegEig}
\begin{figure}[ht!]
    \centering
    \begin{subfigure}[High albedo, rough texture: entropy and eigenspectrum for R9, 662nm]{\includegraphics[width = \textwidth]{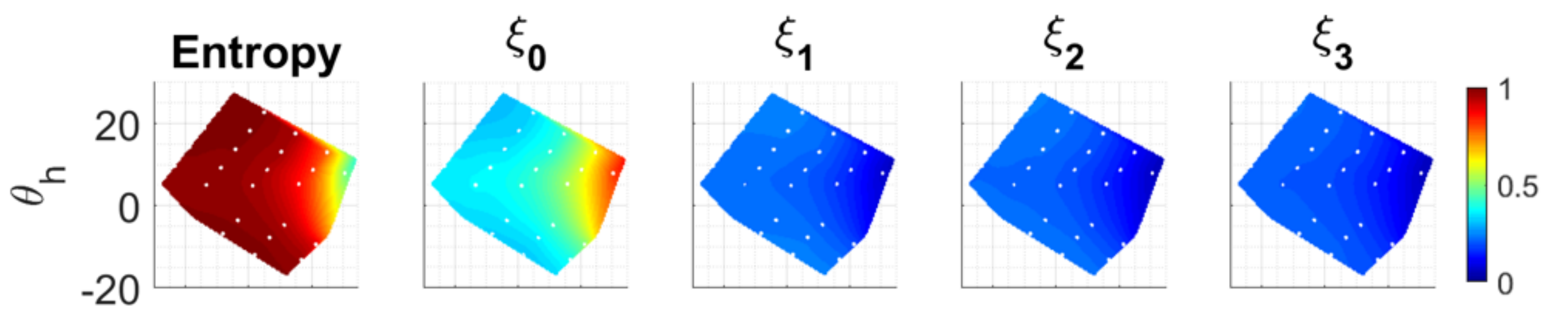}}\end{subfigure}
    \begin{subfigure}[High albedo, smooth texture: entropy and eigenspectrum for B1, 451nm]{\includegraphics[width = \textwidth]{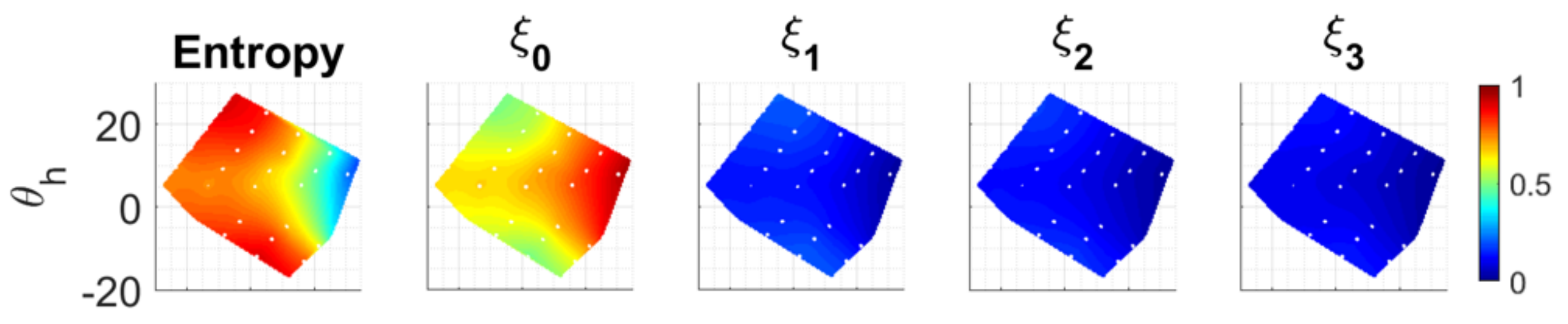}}\end{subfigure} 
    \begin{subfigure}[Low albedo, rough texture: entropy and eigenspectrum for R9
    451nm]{\includegraphics[width = \textwidth]{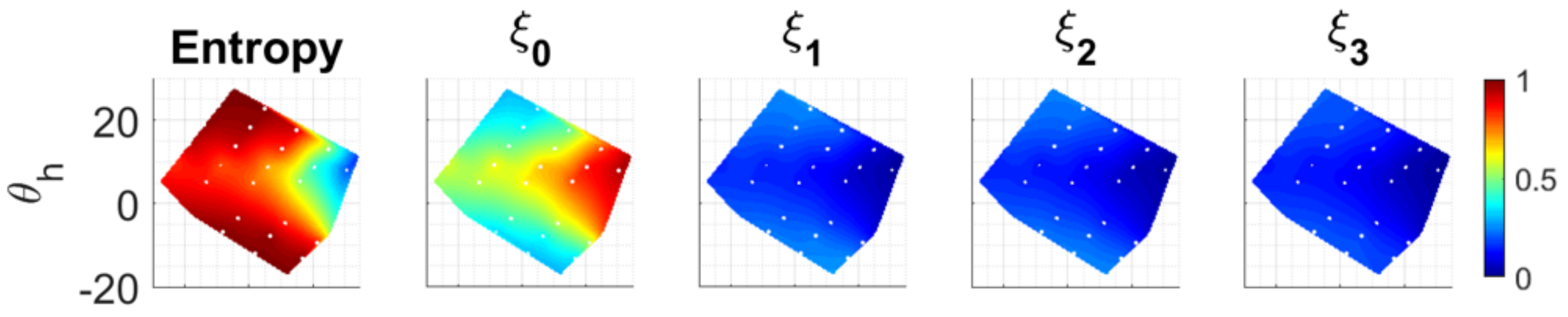}}\end{subfigure}
    \begin{subfigure}[Low albedo, smooth texture: entropy and eigenspectrum for B1, 662nm]{\includegraphics[width = \textwidth]{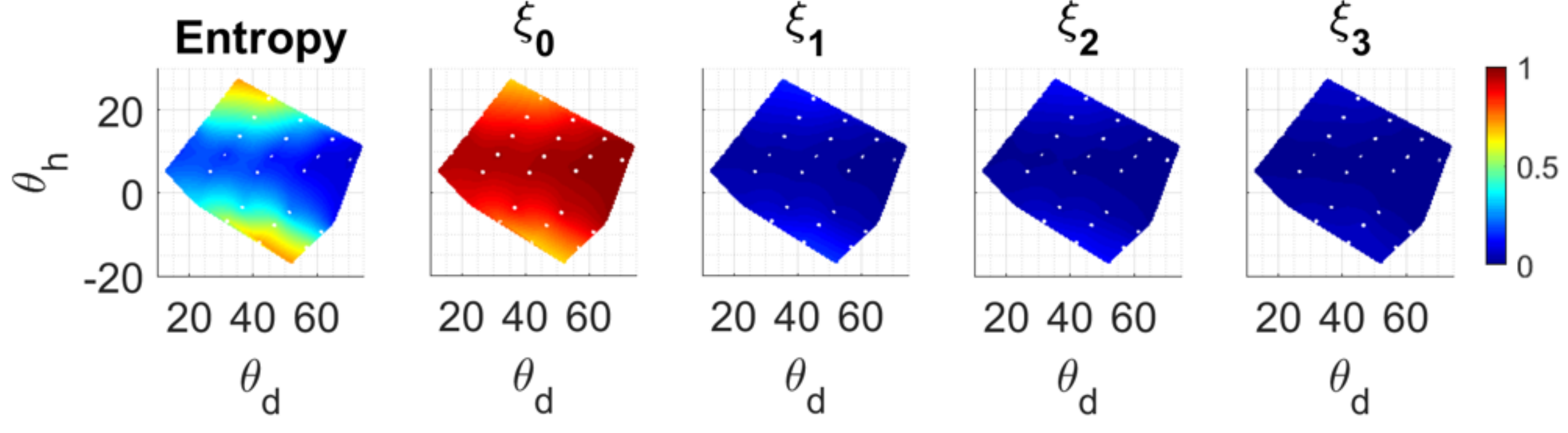}}\end{subfigure}
    \caption{The polarization entropy and normalized eigenspectrum are plotted in signed $\theta_h$ versus $\theta_d$ space for (a) a high albedo, rough texture brick; (b) a high albedo, smooth texture brick; (c) a low albedo rough texture brick; and (d) a low albedo smoother texture brick. Albedo directly affects the magnitude of $\xi_0$: higher albedo trends with increased polarization entropy magnitude for larger $\theta_h$ values as seen in (a) and (c). Surface texture influences the functional form of $\xi_0$ in $\theta_h$ versus $\theta_d$ space; the smoother textures vary faster over $\theta_h$. Polarization entropy increases as the measurement geometry moves away from specular, \emph{i.e.} $\theta_h$ increases. Polarization entropy decreases as the scattering angle is increased, \emph{i.e.} $\theta_d$ increases. At a given measurement geometry, the polarization entropy is greater for high albedo objects due to more bulk scattering. The approximate triple degeneracy relationship $\xi_1 \approx \xi_2 \approx \xi_3$ applies to both albedo objects for all measurement geometries.}
    \label{fig:HiLoT1T9EntEig}
\end{figure}

Figure \ref{fig:BasisMJMs} shows the normalized Mueller-Jones matrices $\widehat{\mathbf{m}}_0$ from Eq. \ref{eq:Mn2M} with the normalized largest eigenvalue for bricks of two different textures and colors under two different wavebands of illumination. For each case, the most significant Mueller-Jones matrix $\widehat{\mathbf{m}}_0$ resembles the  Fresnel reflection matrix; see Fig. \ref{fig:ComponentMMs}(a).  Figure \ref{fig:HiLoT1T9EntEig} shows the entropy and largest normalized eigenvalue $\xi_0$ which corresponds to the largest basis Mueller matrices shown in Fig. \ref{fig:BasisMJMs}. The fractional contribution of $\widehat{\mathbf{m}}_0$ to the measurement is given by the associated eigenvalue $\xi_0$. At large on-specular scattering angles, $\xi_0$ approaches one. For small off-specular scattering geometries, $\xi_0$ is minimized. The eigenvalue $\xi_0$ is smaller for high albedo measurements than low albedo measurements.

\begin{figure}[ht!]
    \centering
    \includegraphics[width=\textwidth]{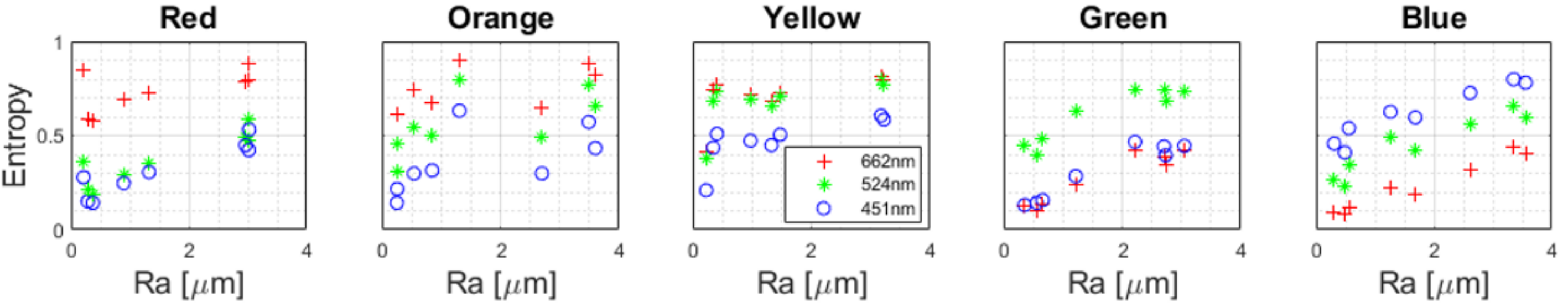}
    \caption{The polarization entropy at a single measurement ($\theta_i=55^\circ, \theta_o=70^\circ$) is plotted for each color brick versus mean surface roughness expressed as arithmetic mean deviation (Ra) in units of [$\mu$m]. The marker color denotes illumination wavelength: 662nm (red), 524nm (green), and 451nm (blue). Polarization entropy (Eq. \ref{eq:PolEnt}) increases as the surface roughness increases. Polarization entropy is higher when the wavelength of illumination is close to the color of the brick.}
    \label{fig:PolEntropyvsRa} 
\end{figure}

Four measurements are compared in Fig.\ref{fig:HiLoT1T9EntEig} to illustrate how polarization entropy is dependent on measurement geometry, albedo, and surface texture. For both albedos, polarization entropy increases as $\theta_d$ decreases or as the measurement geometry moves away from specular (\emph{i.e.} as the halfway angle $\theta_h$ increases). As polarization entropy increases, the largest eigenvalue decreases. The approximate triple degeneracy $\xi_1 \approx \xi_2 \approx \xi_3$ applies to both albedos and textures and for all measurement geometries. For the high albedo cases, the eigenvalues approach equal magnitude as the entropy approaches one. At each measurement geometry, a higher albedo yields a higher polarization entropy due to increased bulk scattering. Figure \ref{fig:PolEntropyvsRa} shows that polarization entropy also generally increases as the surface texture becomes rougher. Other authors have established observations and relations between surface texture and polarized light scattering \cite{germer1999polarization}.

There is a unique Mueller matrix solution for maximum polarization entropy. The polarization entropy equals one when all coherency eigenvalues are equal. If a flat eigenspectrum of $\xi_n=1/4$ for $n=0, 1, 2, 3$ is applied to Eq. \ref{eq:CohVecs}, then the coherency matrix is proportional to the identity matrix and the associated Mueller matrix is proportional to the ideal depolarizer $\mathbf{D}(0)$, given in Eq. \ref{eq:PDMatrix}. Substituting the ideal depolarizer's normalized eigenspectrum [1/4,1/4,1/4,1/4] into Eq. \ref{eq:Mn2M}  yields
\begin{eqnarray}
     \mathbf{D}(0)&=&\frac{1}{4}\left(\widehat{\mathbf{m}}_0+\widehat{\mathbf{m}}_1+\widehat{\mathbf{m}}_2+\widehat{\mathbf{m}}_3\right)\nonumber\\
     &=&\frac{1}{4}\widehat{\mathbf{m}}_0+\frac{1}{4}\sum_{n=1}^3\widehat{\mathbf{m}}_n\nonumber\\
    &=&\frac{1}{4}\widehat{\mathbf{m}}_0+\mathbf{Q}(\widehat{\mathbf{m}}_0).
    \label{eq:sumofMBs}
\end{eqnarray}
Here a Mueller matrix \emph{complementary} to a normalized Mueller-Jones matrix has been defined by
\begin{equation}
    \mathbf{Q}(\widehat{\mathbf{m}})=\mathbf{D}(0)-\frac{1}{4}\widehat{\mathbf{m}}.\label{eq:comp}
\end{equation}
The complementary matrix is neither a normalized matrix nor a Mueller-Jones matrix, but is a physically realizable Mueller matrix \cite{physreal90,physreal93}. This separation between a normalized Mueller-Jones matrix and its complementary matrix is motivated by the triply degenerate eigenspectrum, $\xi_1 \approx \xi_2 \approx \xi_3$, observed for brick measurements in Fig. \ref{fig:HiLoT1T9EntEig}. A normalized Mueller matrix with a triply degenerate coherency matrix eigenspectrum has the form $[\xi_0,(1-\xi_0)/3,(1-\xi_0)/3,(1-\xi_0)/3]$ and entropy

\begin{equation} \label{eq:PolEntFlat}
    E(\mathbf{m}) = -\log_4\left(\frac{\xi_0^{\xi_0}}{(1-\xi_0)^{\xi_0-1}}\right).
\end{equation}
Here $1/4<\xi_0\leq1$ because it is the largest eigenvalue of the coherency matrix of a normalized Mueller matrix. When a triply degenerate eigenspectrum is substituted into Eq. \ref{eq:Mn2M} the normalized Mueller matrix expression is

\begin{eqnarray}
    \mathbf{m}&=&\xi_0\widehat{\mathbf{m}}_0+\left(1-\xi_0\right)\frac{1}{3}\sum_{n=1}^{3}\widehat{\mathbf{m}}_n\nonumber\\
    &=&\frac{4}{3}\left[\left(\xi_0-\frac{1}{4}\right)\widehat{\mathbf{m}}_0+\left(1-\xi_0\right)\mathbf{D}(0)\right] \label{eq:1par_base}\\
    &=&\frac{4}{3}\left[\xi_0\left(\widehat{\mathbf{m}}_0-\mathbf{D}(0)\right)+\mathbf{Q}(\widehat{\mathbf{m}}_0)\right].
\end{eqnarray}
The two terms in these equivalent expressions each have different physical interpretations: 1) a normalized Mueller-Jones matrix which has a dominant contribution, $\xi_0$, and three other normalized Mueller-Jones matrices which are equally weighted, 2) an ideal depolarizer and the dominant Mueller-Jones matrix where both weights involve the dominant eigenvalue $\xi_0$, or 3) the dominant eigenvalue $\xi_0$ isolated to weight the 15 non-$\mathrm{m}_{00}$ elements of the dominant normalized Mueller-Jones matrix and the Mueller matrix complementary to it. The third expansion for a triply degenerate normalized Mueller matrix is the basis for the pBSDF model proposed in this work. Other decompositions of partially depolarizing Mueller matrices exist \cite{kostinski1992depolarization} which separate the Mueller matrix into a non-depolarizing and a completely depolarizing component $\mathbf{D}(0)$.
 
\subsection{Measuring the Largest Normalized Eigenvalue} \label{sec:MethodMeasXi0}
The triple degeneracy of a Mueller matrix is an important constraint which leads to simplified expressions for measuring the normalized largest eigenvalue. Given this triply degenerate assumption, the Mueller matrix is
\begin{equation}
    \mathbf{M}=\frac{4[\mathbf{M}]_{00}}{3}\left[\left(\xi_0-\frac{1}{4}\right)\widehat{\mathbf{m}}_0+\left(1-\xi_0\right)\mathbf{D}(0)\right].\label{eq:1par_a}
\end{equation}
The dominant normalized Mueller-Jones matrix $\mathbf{m}_0$, the normalized largest eigenvalue $\xi_0$, and the unpolarized reflectance $[\mathbf{M}]_{00}$ are all dependent on acquisition geometry. 
The measurement equation, given in Eq. \ref{eq1}, applied to Eq. \ref{eq:1par_a} yields
\begin{eqnarray}
    \mathbf{a}^t\mathbf{M} \mathbf{g}&=&\frac{4[\mathbf{M}]_{00}}{3}\left[\left(\xi_0-\frac{1}{4}\right)\mathbf{a}^t\widehat{\mathbf{m}}_0\mathbf{g}+\left(1-\xi_0\right)\mathbf{a}^t\mathbf{D}(0)\mathbf{g}\right]\nonumber\\
    &=&\frac{4[\mathbf{M}]_{00}}{3}\left[\left(\xi_0-\frac{1}{4}\right)\mathbf{a}^t\widehat{\mathbf{m}}_0\mathbf{g}+\frac{1-\xi_0}{2}\right]
\end{eqnarray}
where $\mathbf{a}^t\mathbf{D}(0)\mathbf{g}=0.5$ assumes that the PSA state $\mathbf{a}$ and the PSG state $\mathbf{g}$ are both fully-polarized. If the dominant Mueller-Jones process is known or assumed then only $\xi_0$ and $\mathrm{M}_{00}$ remain as unknowns. Consider two noise-free measurements $i_1=\mathbf{a}_1^t\mathbf{M}\mathbf{g}_1$ and $i_2=\mathbf{a}_2^t\mathbf{M}\mathbf{g}_2$ using fully-polarized PSA/PSG states. The difference over sum of these two measurements is
\begin{equation}
    i_{\Delta}=\left(\frac{i_1-i_2}{i_1+i_2}\right).
\end{equation}
Here $i_{\Delta}$ will be zero when the measurements are equal, which indicates $\xi_0=1/4$ and the Mueller matrix is an ideal depolarizer. The other extreme is $i_{\Delta}=\pm1$, which only occurs when one of the measurements is zero and the Mueller matrix $\mathbf{M}$ is non-depolarizing. In the general case, the normalized largest eigenvalue is computed from two measurements by
\begin{equation}
    \xi_0(\small{\widehat{\boldsymbol{\omega}}_i},\small{\widehat{\boldsymbol{\omega}}_o},\small{\widehat{\mathbf{n}}})=\frac{1}{4}+\frac{3}{4}i_{\Delta}\left(i_{\Delta}+\left(1-i_{\Delta}\right)\mathbf{a}_1^t\widehat{\mathbf{m}}_0(\small{\widehat{\boldsymbol{\omega}}_i},\small{\widehat{\boldsymbol{\omega}}_o},\small{\widehat{\mathbf{n}}})\mathbf{g}_1-\left(1+i_{\Delta}\right)\mathbf{a}_2^t\widehat{\mathbf{m}}_0(\small{\widehat{\boldsymbol{\omega}}_i},\small{\widehat{\boldsymbol{\omega}}_o,\small{\widehat{\mathbf{n}}}})\mathbf{g}_2\right)^{-1}.\label{eq:2Meas}
\end{equation}
Both the largest normalized eigenvalue $\xi_0$ and the most significant Mueller-Jones matrix $\widehat{\mathbf{m}}_{0}$ depend on measurement geometry $(\widehat{\boldsymbol{\omega}}_i,\widehat{\boldsymbol{\omega}}_o,\widehat{\mathbf{n}})$, as shown in Fig. \ref{fig:BasisMJMs} and Fig. \ref{fig:Estxi0_thhthd} (a). 
Without any assumptions concerning the eigenspectrum, the eigenvalue can be computed from the Mueller matrix using Eq. \ref{eq:CohVecs}. The capability to formulate a pBSDF model from a smaller quantity of measurements than required to formulate a Mueller matrix is a way to utilize a triple degeneracy assumption by computing the largest normalized eigenvalue from Eq. \ref{eq:2Meas}. 
\subsection{Mueller pBSDF Models} \label{sec:pBSDFModels}

\begin{figure}[ht!]
    \centering
    \begin{subfigure}[$\mathbf{D}(0)$]{\includegraphics[width=0.49\textwidth]{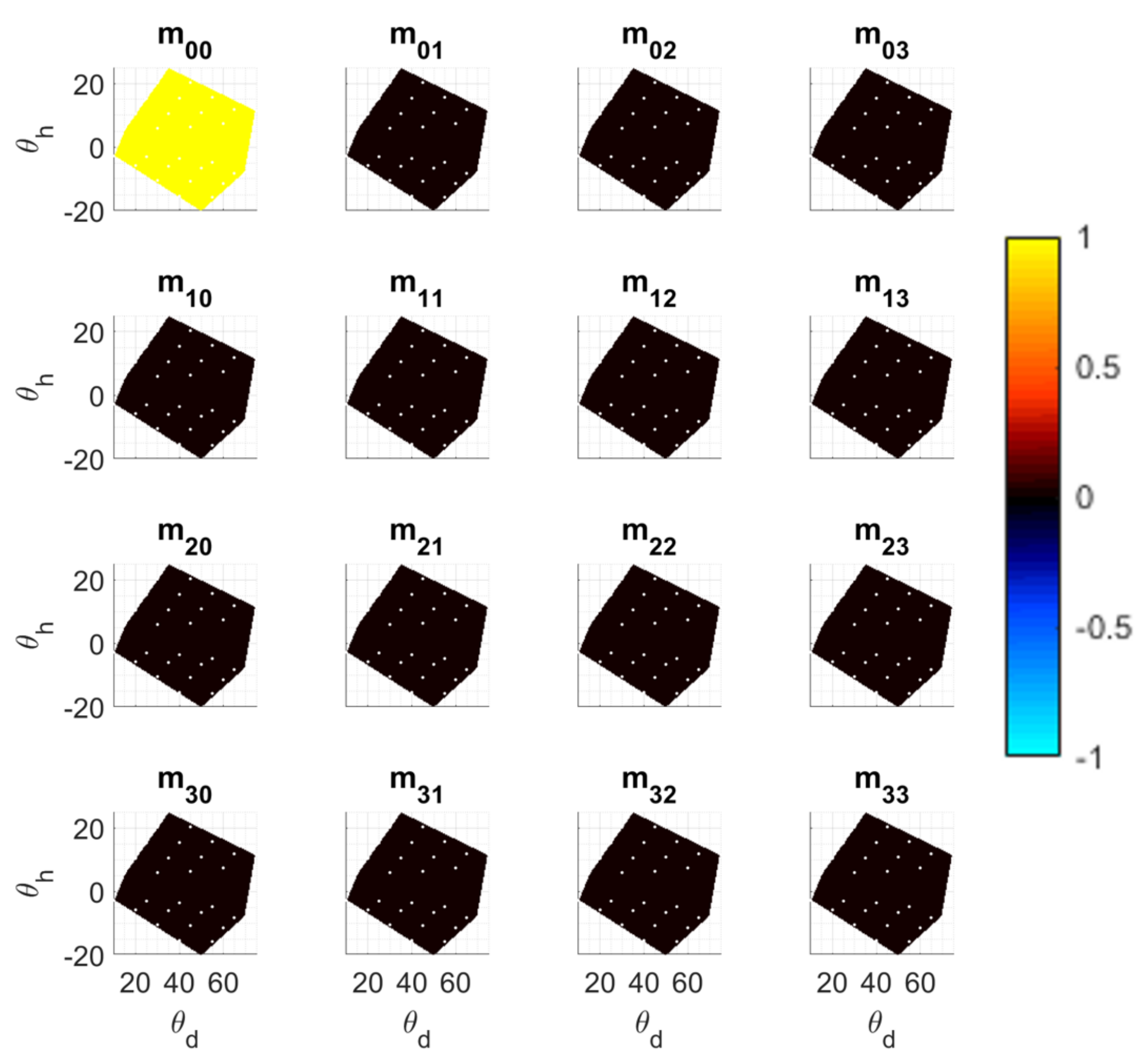}}\end{subfigure}
    \begin{subfigure}[$\widetilde{\mathbf{f}}^R_{1,1.54}(\widehat{\boldsymbol{\omega}}_i,\widehat{\boldsymbol{\omega}}_o)$;Eq. \ref{eq:RotFresnel}]{\includegraphics[width=0.49\textwidth]{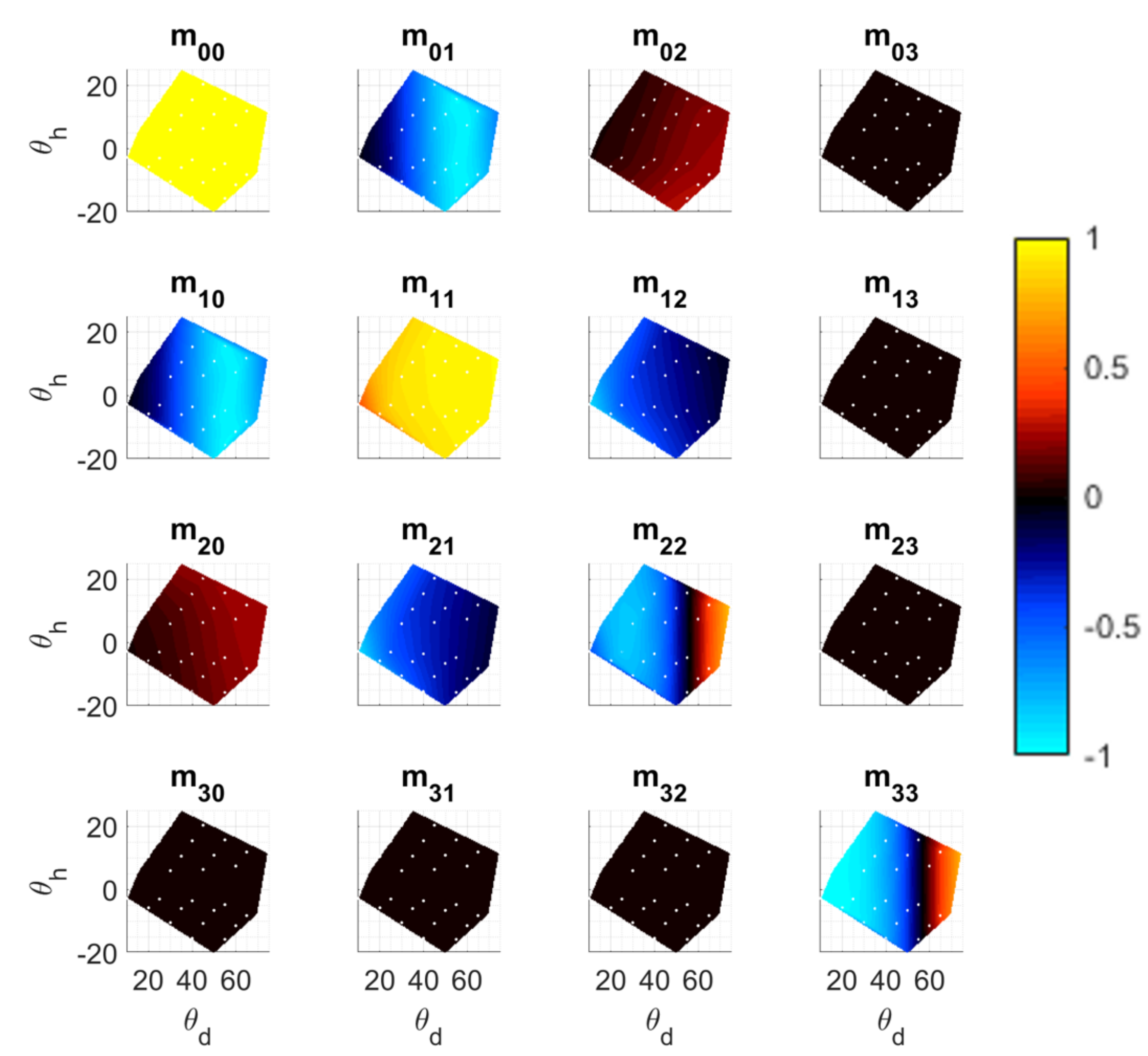}}\end{subfigure}
    \begin{subfigure}[$\mathbf{p}^{d}(\widehat{\boldsymbol{\omega}}_i,\widehat{\boldsymbol{\omega}}_o,\widehat{\mathbf{n}};\frac{1}{3})$; Eq. \ref{eq:bulkscatterHd}]{\includegraphics[width=0.49\textwidth]{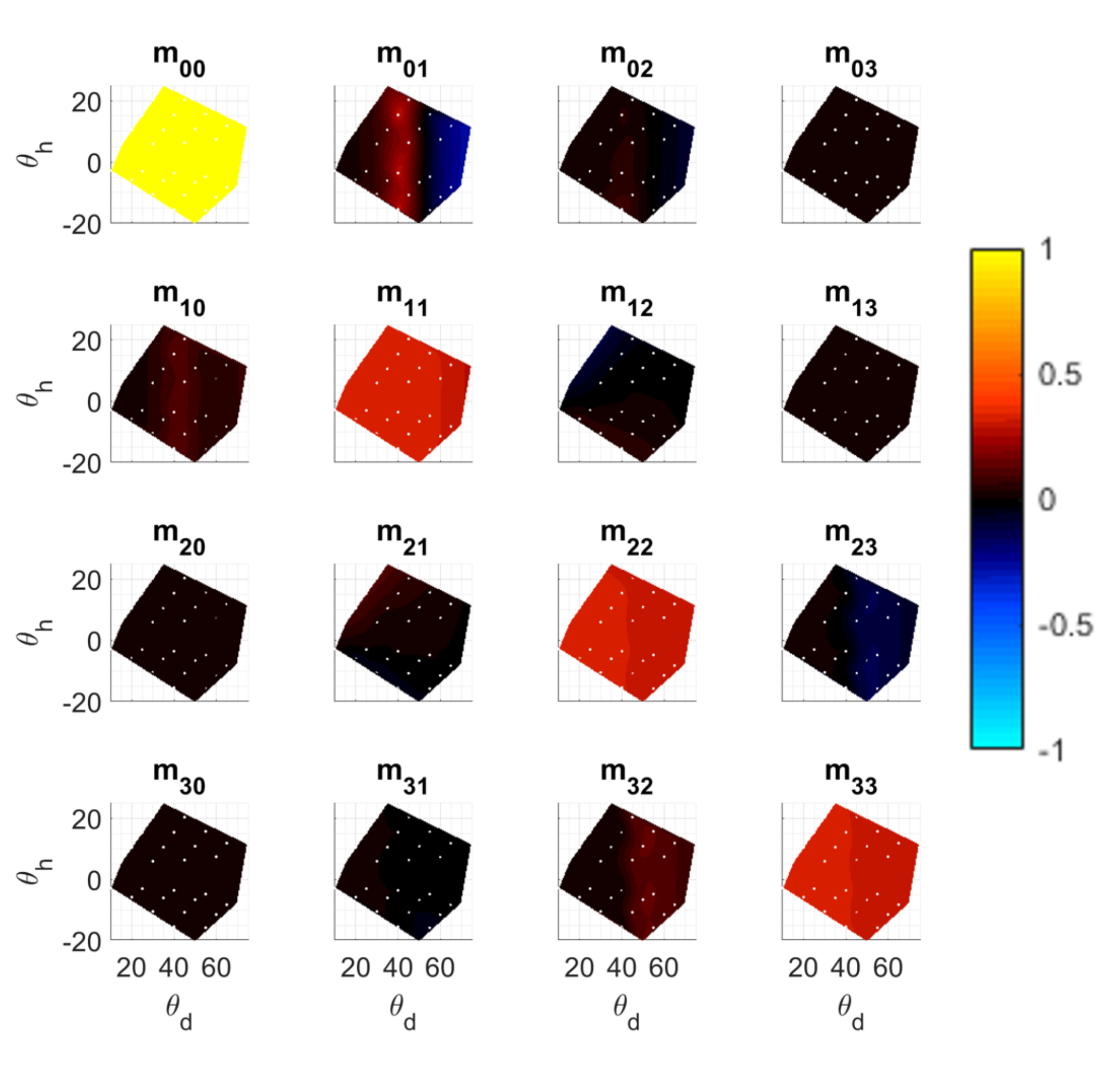}} \end{subfigure}
    \begin{subfigure}[$\mathbf{q}(\widetilde{\mathbf{f}}^{R}_{1,1.54}(\widehat{\boldsymbol{\omega}}_i,\widehat{\boldsymbol{\omega}}_o,\widehat{\mathbf{n}}))$; Eq. \ref{eq:comp}]{\includegraphics[width=0.49\textwidth]{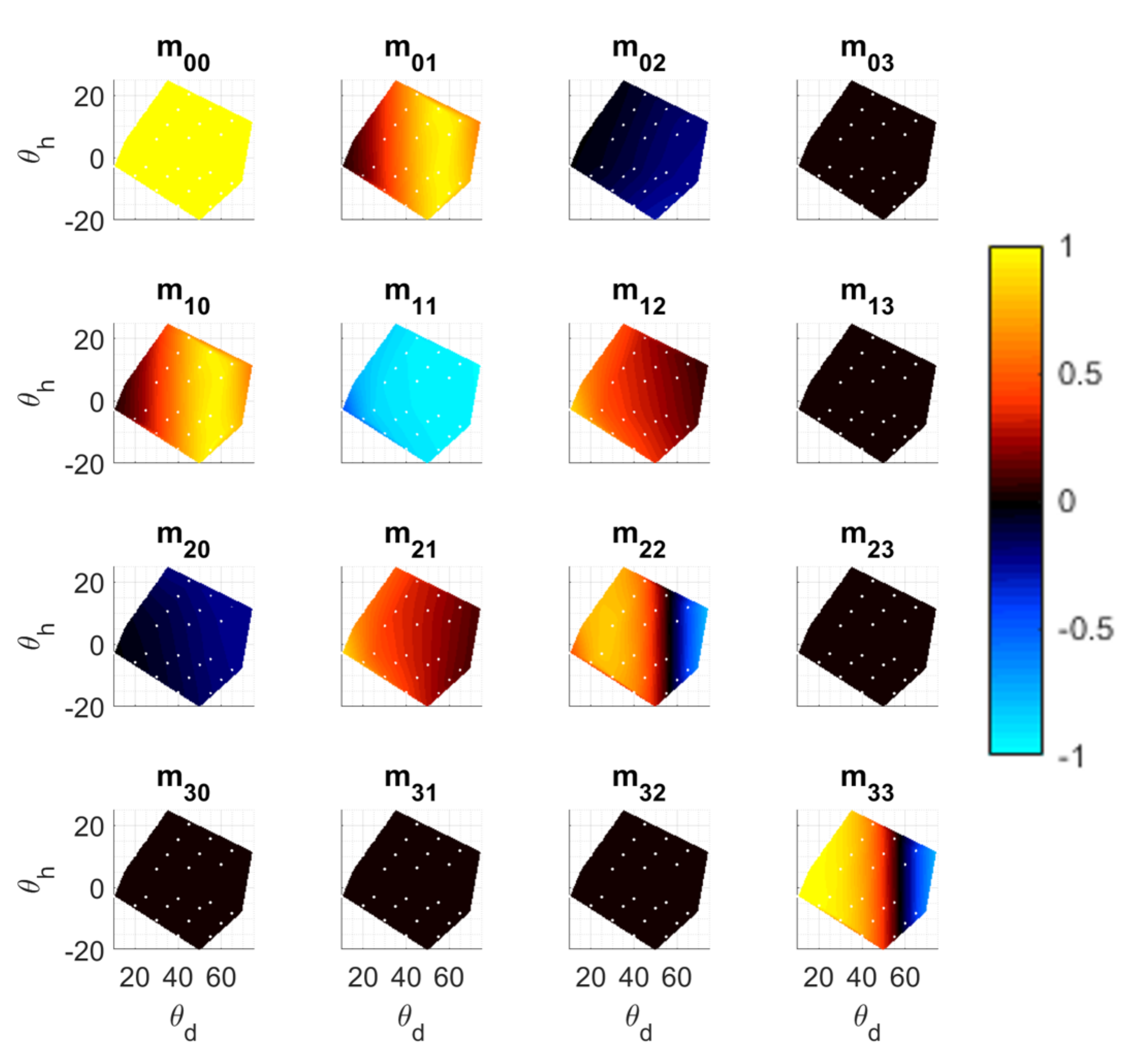}}\end{subfigure}
    \caption{These normalized Mueller matrices are the components of pBSDF models in (a) ideal depolarizer $\mathbf{D}(0)$ and (b) Fresnel reflection matrix $\widetilde{\mathbf{f}}^R_{1,1.54}(\widehat{\boldsymbol{\omega}}_i,\widehat{\boldsymbol{\omega}}_o,\widehat{\mathbf{n}})$, (c) the isotropic depolarizer $\mathbf{D}(d)$ with  $d=\frac{1}{3}$, and (d) the normalized complementary Mueller matrix $\mathbf{q}(\widetilde{\mathbf{f}}^{R}_{n_0,n_1}(\widehat{\boldsymbol{\omega}}_i,\widehat{\boldsymbol{\omega}}_o,\widehat{\mathbf{n}})$ to the Fresnel reflection are plotted in signed $\theta_h$ versus $\theta_d$ space.     
    The (a) ideal depolarizer and (b) Fresnel reflection matrix are the component Mueller matrices of the $\mathbf{p}^{(1)}$ model.
    The $\mathbf{p}^{(2)}$ model uses the same component Mueller matrices as $\mathbf{p}^{(1)}$ with an additional (c) diffuse component $\mathbf{p}^{d}(\widehat{\boldsymbol{\omega}}_i,\widehat{\boldsymbol{\omega}}_o,\widehat{\mathbf{n}},d)$ where $d$ is treated as a fit parameter.
    The (b) Fresnel reflection matrix and (d) the complementary matrix are the component Mueller matrices of the $\mathbf{p}^{(0)}$ model.
    }
\label{fig:ComponentMMs}
\end{figure}

Conventionally, a Mueller pBSDF model's components describe the potential ray paths light may experience in a light-matter interaction. The base model and bulk model both include Mueller matrices which are direct descriptions of ray paths. Fresnel reflection from a microfacet is a rotated Fresnel reflection matrix designated as $\widetilde{\mathbf{F}}^R$. The $\widetilde{.}$ indicates a rotation to adjust for the frame of reference changes between the incident plane and exitant plane. Appendix \ref{app:RotM} explains frame of reference rotation. The dependence of rotated Fresnel reflection on acquisition geometry could be made explicit by denoting $\widetilde{\mathbf{F}}^{R}_{n_0,n_1}(\widehat{\boldsymbol{\omega}}_i,\widehat{\boldsymbol{\omega}}_o,\widehat{\mathbf{n}})$, but this dependence is assumed in this section. For brevity, the matrix is written as $\widetilde{\mathbf{F}}^{R}_{n_0,n_1}$.

The base model analyzed in this work is a normalized polarimetric interpretation of the original Cook and Torrence model \cite{cooktorrance}. This polarimetric interpretation separates the Mueller matrix into a non-depolarizing and a fully depolarizing component, which is an alternate decomposition of a Mueller matrix from a spectral decomposition \cite{kostinski1992depolarization}. Another potential ray path is transmission into a material, depolarizing bulk scatter inside the material, and transmission out of a material; denoted $\widetilde{\mathbf{F}}^T_{n_1,n_0} \mathbf{D}(d) \widetilde{\mathbf{F}}^T_{n_0,n_1} $. This ray path is considered as a third term in the bulk scattering model $\mathbf{p}^{(2)}$, which is a modified form of the Baek\cite{Baek2018} and Kondo models \cite{kondo2020accurate}. The complementary model $\mathbf{p}^{(0)}$ introduced in this work uses a direct ray path description with the Fresnel reflection $\widetilde{\mathbf{F}}^R$ term, but the second term is a complementary matrix $\mathbf{Q}(\widetilde{\mathbf{f}}^{R}_{n_0,n_1})$ which comes from analysis of the coherency matrix; see Sec.\ref{sec:DepReps}. 

The GGX microfacet distribution\cite{Trowbridge:75} is applied to every model, which adds the fit parameter $\sigma$ related to surface roughness. The GGX distribution is a state-of-the-art microfacet distribution used in both the Baek et al. model and the Kondo et al. to describe surface texture effects. Appendix \ref{sec:MicroDistFuncs} describes the GGX distribution. 

Each pBSDF model type is fit over measurements in $\theta_h$ vs. $\theta_d$ space using least-squares fitting. The least-squares fitting routine aims to minimize the mean squared error of simulated irradiances computed from the polarimetric measurement matrix $\mathbf{W}$ over all measurement geometries, $\bar{\Delta}(\mathbf{m},\mathbf{p}|\mathbf{W})$ (see Eq. \ref{eq:FOM_GGX}).

\subsubsection{The Complementary Model} \label{sec:CompModels} 
The component Mueller matrices of the complementary model $\mathbf{p}^{(0)}$ are the normalized rotated Fresnel reflection matrix $\widetilde{\mathbf{f}}^{R}_{n_0,n_1}$ with the $\mathbf{m}_{00}$ element set to zero and its complementary Mueller matrix $\mathbf{Q}\left(\widetilde{\mathbf{f}}^{R}_{n_0,n_1}\right)$

\begin{eqnarray}
    \mathbf{p}^{(0)}(\widehat{\boldsymbol{\omega}}_i,\widehat{\boldsymbol{\omega}}_o,\widehat{\mathbf{n}};\sigma) =  \frac{4}{3}\left[
    \frac{z^R_\lambda\left[{\mathbf{F}}_{n_0,n_1}^R
    \right]_{00}}{\gamma(\widehat{\boldsymbol{\omega}}_i,\widehat{\boldsymbol{\omega}}_o, \widehat{\mathbf{n}};\sigma)} \left(\widetilde{\mathbf{f}}^R_{n_0,n_1}-\mathbf{D}(0)\right) +\mathbf{Q}(\widetilde{\mathbf{f}}^{R}_{n_0,n_1})\right] \label{eq:CompModel} 
\end{eqnarray}
and
\begin{equation}
    \gamma(\widehat{\boldsymbol{\omega}}_i,\widehat{\boldsymbol{\omega}}_o, \widehat{\mathbf{n}};\sigma)=\frac{p(\widehat{\boldsymbol{\omega}}_i,\widehat{\boldsymbol{\omega}}_o,\widehat{\mathbf{n}};\sigma) G(\widehat{\boldsymbol{\omega}}_i,\widehat{\boldsymbol{\omega}}_o,\widehat{\mathbf{n}};\sigma)}{4(-\widehat{\boldsymbol{\omega}}_i\cdot\widehat{\mathbf{n}})(\widehat{\boldsymbol{\omega}}_o\cdot\widehat{\mathbf{n}})(\widehat{\mathbf{h}}\cdot\widehat{\mathbf{n}})}.
    \label{eq:gamma}
\end{equation}
Here $\gamma(\widehat{\boldsymbol{\omega}}_i,\widehat{\boldsymbol{\omega}}_o, \widehat{\mathbf{n}};\sigma)$ is a scalar-valued function which combines the microfacet distribution function $p(\widehat{\boldsymbol{\omega}}_i,\widehat{\boldsymbol{\omega}}_o,\widehat{\mathbf{n}};\sigma)$, the associated shadowing-masking function $G(\widehat{\boldsymbol{\omega}}_i,\widehat{\boldsymbol{\omega}}_o,\sigma)$, and other established geometrical factors\cite{cooktorrance,priest2002polarimetric}. The fit parameter $z^R_\lambda$ is albedo-dependent, which is denoted by the $\lambda$ subscript. Use of the optional fit parameter $\sigma$ is dependent on the selected microfacet distribution function. If a microfacet distribution with a fit parameter $\sigma$ is applied, then $\mathbf{p}^{(0)}$ is a two parameter model.

The Mueller matrix complementary to Fresnel reflection is denoted $\mathbf{Q}(\widetilde{\mathbf{f}}^{R}_{n_0,n_1})$; see Eq. \ref{eq:comp} for the complementary Mueller matrix definition. The complementary model is a simplification of the coherency matrix assuming a triply degenerate eigenspectrum; see Section \ref{sec:DepReps}. The most significant Mueller-Jones matrix $\widehat{\mathbf{m}}_0$ is assumed to be Fresnel reflection. Figures \ref{fig:ComponentMMs} (a), (b), and (d) show the Mueller matrices $\mathbf{D}(0)$, $\widetilde{\mathbf{f}}^{R}_{n_0,n_1}(\widehat{\boldsymbol{\omega}}_i,\widehat{\boldsymbol{\omega}}_o,\widehat{\mathbf{n}})$, and $\mathbf{Q}(\widetilde{\mathbf{f}}^{R}_{n_0,n_1}(\widehat{\boldsymbol{\omega}}_i,\widehat{\boldsymbol{\omega}}_o,\widehat{\mathbf{n}}))$ plotted in signed $\theta_h$ versus $\theta_d$ space.  
 
\subsubsection{The Base Model} \label{sec:BaseModel}
The base model is a normalized polarimetric interpretation of the original Cook and Torrence model \cite{cooktorrance}. This model consists of an ideal depolarizer $\mathbf{D}(0)$ and a rotated Fresnel reflection matrix $\widetilde{\mathbf{F}}_{n_0,n_1}^R$
\begin{equation} \label{eq:BaseModel} 
\mathbf{p}^{(1)}(\widehat{\boldsymbol{\omega}}_i,\widehat{\boldsymbol{\omega}}_o, \widehat{\mathbf{n}};\sigma)= \frac{\mathbf{D}(0)  + z^s_\lambda \gamma(\widehat{\boldsymbol{\omega}}_i,\widehat{\boldsymbol{\omega}}_o, \widehat{\mathbf{n}};\sigma) \widetilde{\mathbf{F}}^{R}_{n_0,n_1}}{1+z^s_\lambda \gamma(\widehat{\boldsymbol{\omega}}_i,\widehat{\boldsymbol{\omega}}_o, \widehat{\mathbf{n}};\sigma)\left[{\widetilde{\mathbf{F}}}_{n_0,n_1}^R\right]_{00}}.
\end{equation}
The Fresnel reflection term $\widetilde{\mathbf{F}}^R_{n_0,n_1}$ from a specularly oriented microfacet is weighted using a term which depends on measurement geometry and up to two fit parameters. Fit parameter $z^s_\lambda$ is albedo-dependent, denoted by the $\lambda$ subscript. The fit parameter $\sigma$ may be included if the microfacet distribution function selected for $p(\widehat{\boldsymbol{\omega}}_i,\widehat{\boldsymbol{\omega}}_o,\widehat{\mathbf{n}};\sigma)$ includes a $\sigma$ fit parameter. Figure \ref{fig:ComponentMMs} (a) and (b) are the $\mathbf{D}(0)$ and $\widetilde{\mathbf{f}}^{R}_{n_0,n_1}(\widehat{\boldsymbol{\omega}}_i,\widehat{\boldsymbol{\omega}}_o,\widehat{\mathbf{n}})$ component Mueller matrices plotted in signed $\theta_h$ versus $\theta_d$ space, respectively. A relative-reflectance implementation of this model uses a $a_\lambda$ fit parameter to weight the $\mathbf{D}(0)$ component as seen in prior work, Li et al. 2020\cite{Lietal2020}. However, since the models in this work are evaluated in a normalized form, the $a_\lambda$ and $z^s_\lambda$ components can be consolidated into one parameter. 

\subsubsection{The Bulk Model} \label{sec:BulkModel}
The bulk model is a transmission-inclusive pBSDF model which incorporates the assumption that multiple scattering inside the media is depolarizing. This model adds a partially-depolarizing bulk scattering component $\mathbf{P}^d(\widehat{\boldsymbol{\omega}}_i,\widehat{\boldsymbol{\omega}}_o,\widehat{\mathbf{n}},d) $ to the $\mathbf{p}^{(1)}$ model. 
The normalized bulk model is
\begin{equation}
    \mathbf{p}^{(2)}(\widehat{\boldsymbol{\omega}}_i,\widehat{\boldsymbol{\omega}}_o,\widehat{\mathbf{n}},d;\sigma) = \frac{\mathbf{D}(0)  +  z^s_\lambda \frac{\widetilde{\mathbf{F}}^{R}_{n_0,n_1}(\widehat{\boldsymbol{\omega}}_i,\widehat{\boldsymbol{\omega}}_o,\widehat{\mathbf{n}})}{\gamma(\widehat{\boldsymbol{\omega}}_i,\widehat{\boldsymbol{\omega}}_o, \widehat{\mathbf{n}};\sigma)} +z^d_\lambda\mathbf{P}^d(\widehat{\boldsymbol{\omega}}_i,\widehat{\boldsymbol{\omega}}_o,\widehat{\mathbf{n}},d)}{[\mathbf{P}^{(2)}(\widehat{\boldsymbol{\omega}}_i,\widehat{\boldsymbol{\omega}}_o,\widehat{\mathbf{n}},d;\sigma)]_{00}}\label{eq:BulkModel}
\end{equation}
where the normalization factor is
\begin{equation}
    [\mathbf{P}^{(2)}(\widehat{\boldsymbol{\omega}}_i,\widehat{\boldsymbol{\omega}}_o,\widehat{\mathbf{n}},d;\sigma)]_{00} = 1+z^s_\lambda\frac{\left[{\widetilde{\mathbf{F}}}_{n_0,n_1}^R\right]_{00}}{\gamma(\widehat{\boldsymbol{\omega}}_i,\widehat{\boldsymbol{\omega}}_o, \widehat{\mathbf{n}};\sigma)}  + z^d_\lambda\left[\mathbf{P}^d(\widehat{\boldsymbol{\omega}}_i,\widehat{\boldsymbol{\omega}}_o,\widehat{\mathbf{n}},d)\right]_{00}. 
\end{equation}
The bulk scattering component 
\begin{equation}
\mathbf{P}^d(\widehat{\boldsymbol{\omega}}_i,\widehat{\boldsymbol{\omega}}_o,\widehat{\mathbf{n}},d) 
= z_\lambda^d (-\widehat{\boldsymbol{\omega}}_i\cdot\widehat{\mathbf{n}}) \mathbf{R}(\alpha_o) \mathbf{F}^T_{n_1,n_0}(\theta_o') \hspace{.1cm} \mathbf{D}(d) \hspace{.1cm} \mathbf{F}^T_{n_0,n_1}(\theta_i) \mathbf{R}(-\alpha_i)
\label{eq:bulkscatterHd}
\end{equation}
traces a path of transmission from air into the media, bulk scattering within the media, and propagation from media back to air.  The $\mathbf{p}^{(2)}$ model includes three fit parameters: $z^s_\lambda$, $z^d_\lambda$, and $d$. Parameters denoted with a $\lambda$ subscript are wavelength-dependent. 

This bulk scattering component $\mathbf{P}^d$ is an extension of the diffuse component of the partial Mueller matrix pBSDF model proposed by Baek et al. \cite{Baek2018} and used in Kondo et al. \cite{kondo2020accurate}. Instead of assuming the bulk scattering inside the material's volume is completely depolarizing, partial isotropic depolarization is allowed. The ratio of partial isotropic depolarization $d$ is a fit parameter. As $d$ approaches 0, more scattering events take place inside the volume of the material. A value of $d=1$ is equivalent to the $4\times4$ identity matrix and indicates no scattering events occur. In the case that $d=0$, the $\mathbf{p}^{(2)}$ model is a normalized implementation of the Kondo et al. model with a delay factor equal to zero. In the case that no $\mathbf{D}(0)$ term is included and $d=0$, the $\mathbf{p}^{(2)}$ model is a normalized implementation of the Baek model\cite{Baek2018}.

\subsection{Measurement Agreement}\label{sec:MeasAgree}
Two Mueller matrices can be compared by the mean squared error (MSE) of simulated irradiance values from a given polarimetric measurement matrix $\mathbf{W}$
\begin{equation}
    \Delta(\mathbf{m},\mathbf{p}|\mathbf{W})=\frac{1}{L}\sum^L_{l=1}\left|\left[\mathbf{i}(\mathbf{m})-\mathbf{i}(\mathbf{h})\right]_l\right|^2=\frac{1}{L}\sum^L_{l=1}\left|\left[\mathbf{W}^t(\vec{\mathbf{m}}-\vec{\mathbf{h}})\right]_l\right|^2\label{eq:FOM}
\end{equation}
where $\left[\cdot\right]_l$ is the $l^{th}$ irradiance value. In this work, $L=40$ and the polarimetric measurement matrix is computed from the PSA/PSG pairs of the RGB950 imaging Mueller Matrix polarimeter \cite{RGB950}. For fitting pBSDF Mueller matrix models, the distance metric in Eq. \ref{eq:FOM} is averaged over $K$ measurement geometries 
\begin{eqnarray}
    \bar{\Delta}(\mathbf{m}_\lambda(\small{\widehat{\boldsymbol{\omega}}_{i}},\small{\widehat{\boldsymbol{\omega}}_{o}}),\mathbf{p}_\lambda(\small{\widehat{\boldsymbol{\omega}}_{i}},\small{\widehat{\boldsymbol{\omega}}_{o}})|\mathbf{W})&=&
    \frac{1}{K}\sum_{k=1}^{K}{\Delta}(\mathbf{m}_\lambda([\small{\widehat{\boldsymbol{\omega}}_{i}}]_k,[\small{\widehat{\boldsymbol{\omega}}_{o}}]_k),\mathbf{p}_\lambda([\small{\widehat{\boldsymbol{\omega}}_{i}}]_k,[\small{\widehat{\boldsymbol{\omega}}_{o}}]_k)|\mathbf{W})\label{eq:FOM_GGX}\\
    &=&\frac{1}{LK}\sum_{k=1}^{K}\sum^L_{l=1}\left|\left[\mathbf{W}^t\left(\mathbf{m}_\lambda([\small{\widehat{\boldsymbol{\omega}}_{i}}]_k,[\small{\widehat{\boldsymbol{\omega}}_{o}}]_k))-\mathbf{p}_\lambda([\small{\widehat{\boldsymbol{\omega}}_{i}}]_k,[\small{\widehat{\boldsymbol{\omega}}_{o}}]_k)\right)\right]_l\right|^2\nonumber
\end{eqnarray}
where the incident and exitant propogation directions of the $k^{th}$ measurement geometry are $[\small{\widehat{\boldsymbol{\omega}}_{i}}]_k$ and $[\small{\widehat{\boldsymbol{\omega}}_{o}}]_k$, respectively. In this work, $K=30$ and measurement geometries are reported in Table \ref{tab:MeasGeos} in Appendix \ref{app:MeasGeo}.

\section{Results} \label{sec:Results}
\subsection{pBSDF Model Fitting Results} \label{sec:pBSDFResults} 
\begin{figure}[ht!]
    \centering
    \begin{subfigure}[pBSDF model ${\mathbf{p}}^{(0)}_\lambda$ (Eq. \ref{eq:CompModel})]{\includegraphics[width=\textwidth]{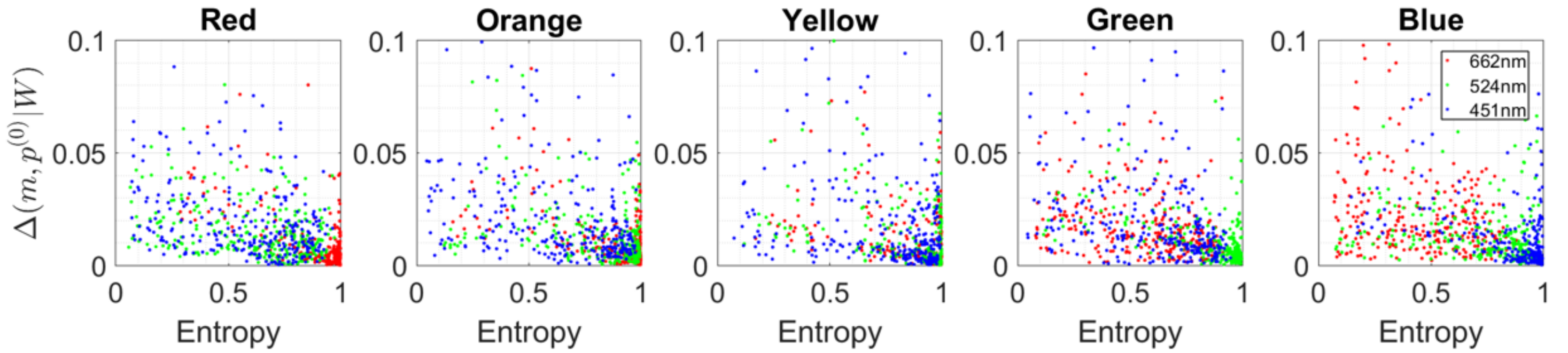}} \end{subfigure}
    \begin{subfigure}[pBSDF model ${\mathbf{p}}^{(1)}_\lambda$ (Eq. \ref{eq:BaseModel})]{\includegraphics[width=\textwidth]{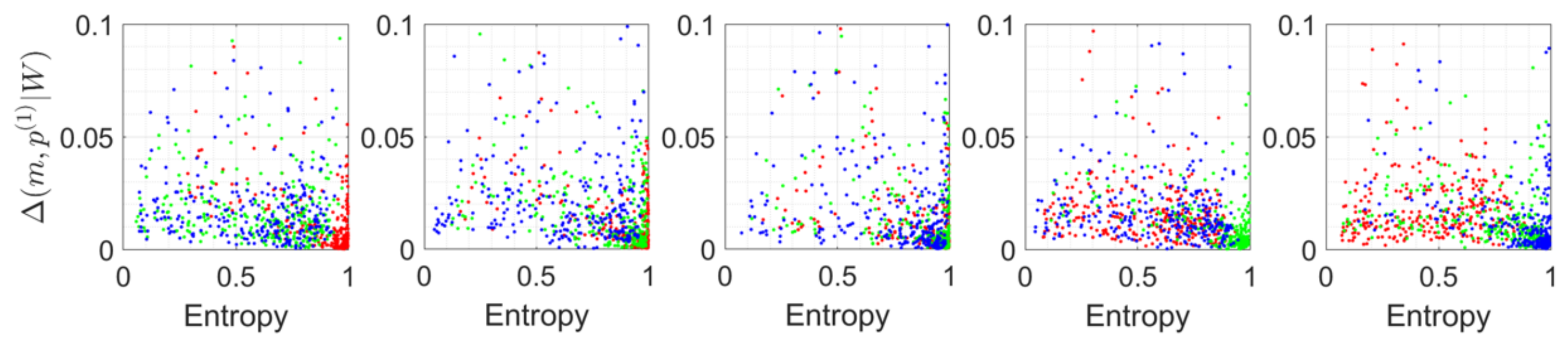}}
    \end{subfigure}
    \begin{subfigure}[pBSDF model $\mathbf{p}^{(2)}_\lambda$ (Eq. \ref{eq:BulkModel})]{\includegraphics[width=\textwidth]{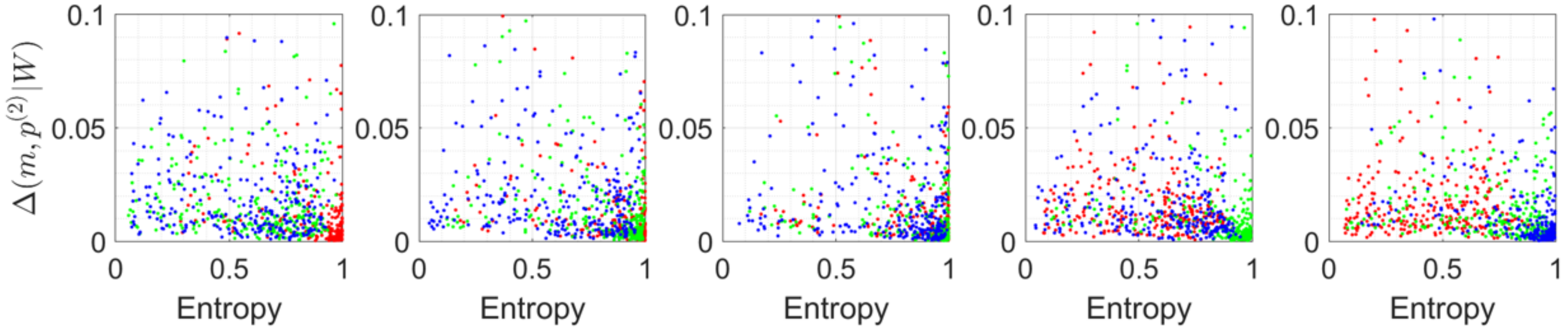}} \end{subfigure}
    \caption{For illumination 662 nm (red), 524 nm (green), and 451 nm (blue) irradiance agreement ${\Delta}(\mathbf{m},\mathbf{p}|\mathbf{W})$ (Eq.\ref{eq:FOM_GGX}) versus polarization entropy $E(\mathbf{M})$ (Eq. \ref{eq:PolEnt}) for the pBSDF models: (a) $\mathbf{p}^{(0)}$, (b) $\mathbf{p}^{(1)}$, and (c) $\mathbf{p}^{(2)}$. The GGX microfacet distribution function Eq. \ref{eq:GGX} is applied to each model and the merit function is $\bar{\Delta}(\mathbf{m},\mathbf{p}|\mathbf{W})$ (Eq. \ref{eq:FOM_GGX}). The average MSEs for each model over all entropies are $\left<\bar{\Delta}(\mathbf{m},\mathbf{p}^{(0)}|\mathbf{W})\right>_{E}=0.0163$, $\left<\bar{\Delta}(\mathbf{m},\mathbf{p}^{(1)}|\mathbf{W})\right>_{E}=0.0165$, and  $\left<\bar{\Delta}(\mathbf{m},\mathbf{p}^{(2)}|\mathbf{W})\right>_{E}=0.0165$. 
    Fewer than 15 data points out of 810 in each plot are truncated by setting the max $\Delta=0.1$.
    \label{fig:DeltavEnt}}
\end{figure}

Models $\mathbf{p}^{(0)}$, $\mathbf{p}^{(1)}$, and $\mathbf{p}^{(2)}$ are evaluated through comparing $\left<\bar{\Delta}(\mathbf{m},\mathbf{p}|\mathbf{W})\right>_{E}$, which is the MSE defined in Eq. \ref{eq:FOM_GGX} averaged over all entropies (E). The average MSE for each model over all entropies, over only low entropies, and over only high entropies are given in Tab. \ref{tab:MSEResults}. Measurement agreement over all 4050 measurements is reported in the $\left<\bar{\Delta}\right>_{E}$ column. The $\left<\bar{\Delta}(\mathbf{m},\mathbf{p}|\mathbf{W})\right>_{E}$ are similar within $\pm0.002$ for the ${\mathbf{p}}^{(0)}$, ${\mathbf{p}}^{(1)}$, and ${\mathbf{p}}^{(2)}$ models. Subsequent columns in Tab. \ref{tab:MSEResults} report $\left<\bar{\Delta}\right>_{E}$ over four quadrants:  $E<0.565, 0.565\leq E<0.816, 0.816\leq E<0.954$, and  $E\geq0.954$. For the lowest entropy measurements where $E<0.565$, the $\mathbf{p}^{(2)}$ model produces the lowest measurement agreement $\left<\bar{\Delta}(\mathbf{m},\mathbf{p}^{(2)}|\mathbf{W})\right>_{E<0.565}=0.0250$, compared to  $\left<\bar{\Delta}(\mathbf{m},\mathbf{p}^{(0)}|\mathbf{W})\right>_{E<0.565}=0.0295$ and  $\left<\bar{\Delta}(\mathbf{m},\mathbf{p}^{(1)}|\mathbf{W})\right>_{E<0.565}=0.0267$. For all other entropy values $\geq 0.565$, the $\mathbf{p}^{(0)}$ model produces the lowest measurement agreement $\left<\bar{\Delta}(\mathbf{m},\mathbf{p}^{(0)}|\mathbf{W})\right>_{E\geq0.565}=0.0119$, compared to  $\left<\bar{\Delta}(\mathbf{m},\mathbf{p}^{(1)}|\mathbf{W})\right>_{E\geq0.565}=0.0131$ and  $\left<\bar{\Delta}(\mathbf{m},\mathbf{p}^{(2)}|\mathbf{W})\right>_{E\geq0.565}=0.0138$. 
 
The addition of more fit parameters and a potential ray path in the $\mathbf{p}^{(2)}$ bulk model does not increase polarimetric accuracy for measurements where $E\geq0.565$. Larger $\xi_0$ values means that differences in $\xi_1, \xi_2, \xi_3$ are relatively smaller in magnitude, but Tab. \ref{tab:MSEResults} shows that $\left<\Delta\right>_{E<0.565}$ is greater than $\left<\Delta\right>_{E\geq0.565}$ over all tested models. This is attributable to the relative magnitude elements in a normalized Mueller matrix. For the purpose of analysis in this work, half of all measurements are categorized as high albedo and the other half are categorized as low albedo. Polarization entropies $E\geq0.816$ are considered high entropy, with a total of 2029 out of 4050 measurements falling into this category. Low entropy measurements have larger magnitude in the normalized Mueller matrix elements (see: Fig. \ref{fig:HighLowAlbedo}) that produce a larger mean simulated irradiance, $\left<\bar{\Delta}\right>_{E}$. 

Measurements with low entropy include low albedo measurements taken at on-specular and near-specular geometries ($\theta_h<10^\circ$) over all $\theta_d$, high albedo measurements of smoother textures (\emph{e.g.} T1, T2, T3, T4) taken at on-specular and near-specular geometries, and high albedo measurements taken of rougher textures  (\emph{e.g.} T5, T6, T7, T8, T9) at on-specular and near-specular geometries when $\theta_d>65^\circ$. All other measurements are categorized as high entropy measurements.

\begin{table}[ht]
    \centering
    \begin{tabular}{c|c|c|c}
    \hline
     & $\mathbf{p}^{(0)}$ &  $\mathbf{p}^{(1)}$ & $\mathbf{p}^{(2)}$ 
    \\ \hline 
    $\left<\bar{\Delta}\right>_{E}$ & $0.0163\pm0.0118$ & $0.0165\pm0.0108$ & $0.0165\pm0.0116$ \\
    $\left<\bar{\Delta}\right>_{E<0.565}$ & $0.0295\pm0.0218$ & $0.0267\pm0.0162$ & $0.0250\pm0.0168$ \\
    $\left<\bar{\Delta}\right>_{0.565\leq E<0.816}$ & $0.0181\pm0.0124$ & $0.0184\pm0.0113$ & $0.0188\pm0.0112$ \\
    $\left<\bar{\Delta}\right>_{0.816\leq E<0.954}$ & $0.0096\pm0.0107$ & $0.0116\pm0.0114$ & $0.0124\pm0.0120$ \\
    $\left<\bar{\Delta}\right>_{E\geq0.954}$ & $0.0080\pm0.0093$ & $0.0093\pm0.0097$ & $0.0100\pm0.0104$ \\
    \hline
    \end{tabular}
    \caption{The average MSE (Eq. \ref{eq:FOM_GGX}) over all measurements and over measurements divided into quartiles of entropy ranges. The average MSE are given for the complementary $\mathbf{p}^{(0)}$, base $\mathbf{p}^{(1)}$, and bulk $\mathbf{p}^{(2)}$ models. The standard errors for MSE over all measurements and over each quartile are indicated by the $\pm$ values. Over all entropies, the complementary, base, and bulk models return approximately similar performance. The bulk model performs best in the lower entropy measurements ($E<0.565$), while the complementary models perform best for higher entropy samples ($E\geq0.565$).} \label{tab:MSEResults}
\end{table}
\begin{figure}[ht!]
    \centering
    \begin{subfigure}[$\xi_0$ calculated by Eq. \ref{eq:CohVecs} from Mueller measurements (left-right): R9:662nm,  B1:451nm, R9:451nm, and B1:662nm.]{\includegraphics[width=\textwidth]{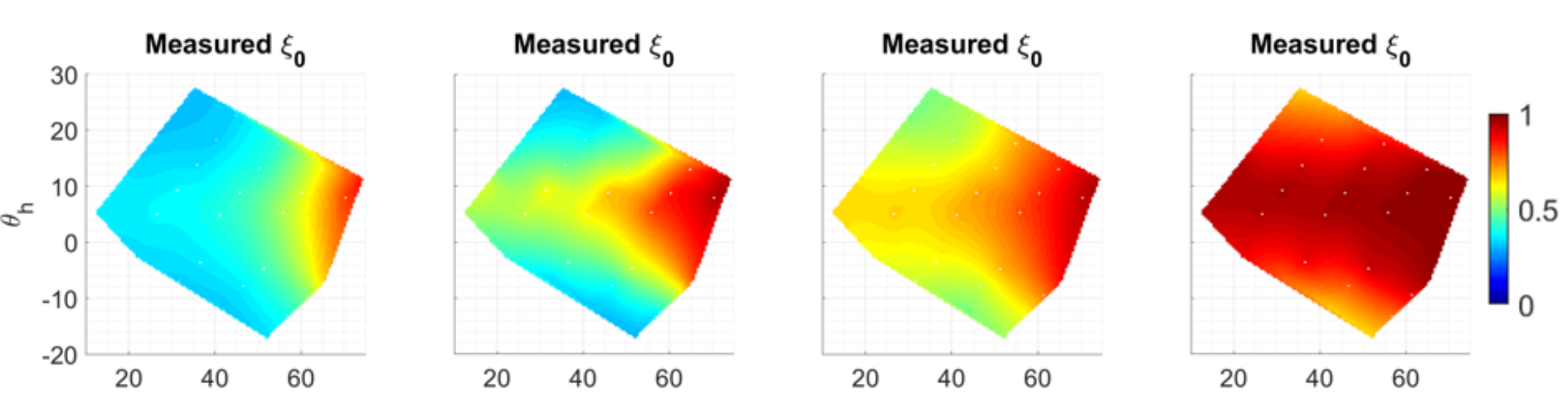}}\end{subfigure}
    \begin{subfigure}[$\xi_0$ estimated from $\mathbf{p}^{(0)}$ model by fitting Eq. \ref{eq:FOM_GGX} (left-right): R9:662nm,  B1:451nm, R9:451nm, and B1:662nm.]{\includegraphics[width=.98\textwidth]{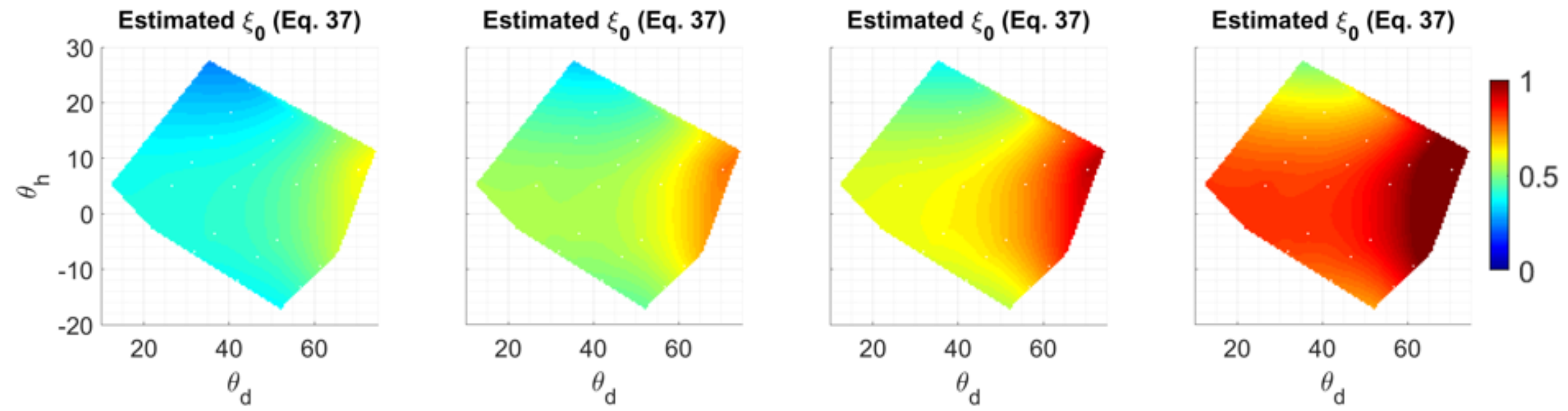}} \end{subfigure}
    \begin{subfigure}[$\xi_0$ estimated from $\mathbf{p}^{(0)}$ model by fitting Eq. \ref{eq:DeltaXi} (left-right): R9:662nm,  B1:451nm, R9:451nm, and B1:662nm.]{\includegraphics[width=.98\textwidth]{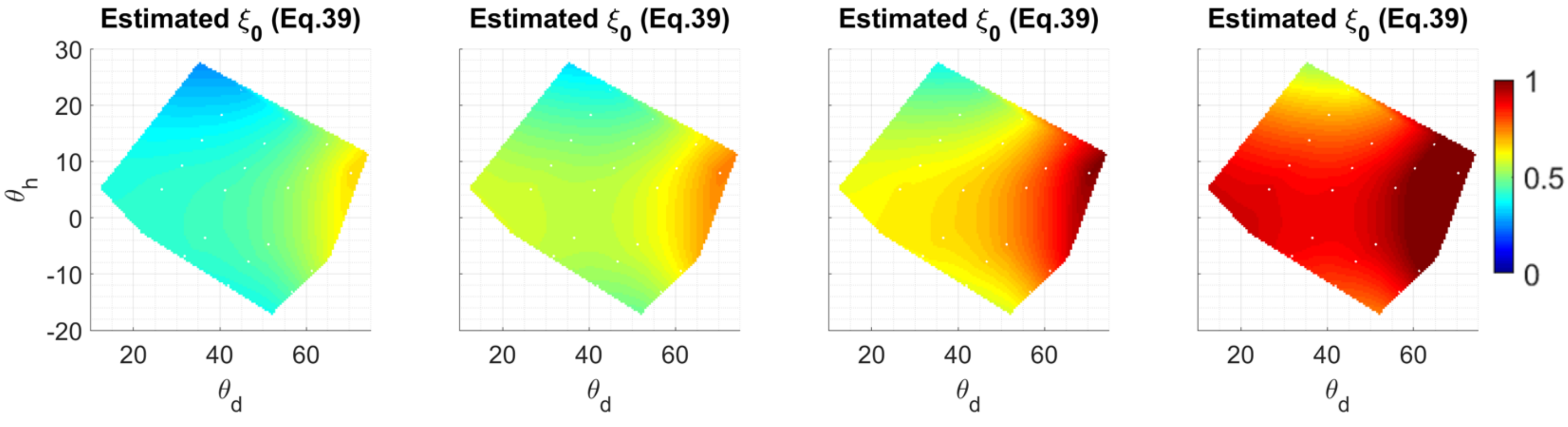}} \end{subfigure}
    \caption{In signed halfway angle $\theta_h$ and difference angle $\theta_d$ space the largest normalized eigenvalue $\xi_0$ in (a) calculated from Mueller matrix measurements, (b) estimated from the $\mathbf{p}^{(0)}$ model using Eq. \ref{eq:FOM_GGX}, and (c) estimated from Eq. \ref{eq:DeltaXi}. In (a) $\xi_0$ has a larger magnitude along the $\theta_h=5^\circ$ axis than both estimates in (b,c). The same bricks, B1 and R9, are also used to show $\widehat{\mathbf{m}}_0$ in Fig. \ref{fig:BasisMJMs}.}
    \label{fig:Estxi0_thhthd}
\end{figure}
Figure \ref{fig:DeltavEnt} compares the $\Delta(\mathbf{m},\mathbf{p}|\mathbf{W})$ versus entropy of: \ref{fig:DeltavEnt}(a) $\mathbf{p}^{(0)}$, \ref{fig:DeltavEnt}(b) $\mathbf{p}^{(1)}$, and \ref{fig:DeltavEnt}(c) $\mathbf{p}^{(2)}$ models at individual measurements. The $\Delta(\mathbf{m},\mathbf{p}|\mathbf{W})$ values plotted are not the merit function, but instead are individual geometries produced from a fit which minimizes $\bar{\Delta}(\mathbf{m},\mathbf{p}|\mathbf{W})$ for each combination of brick colors and surface roughness (T1-T9). 

Over all measurements, the $\mathbf{p}^{(0)}$ complementary model produces better measurement agreement than the $\mathbf{p}^{(1)}$ base model by $\left<\Delta(\mathbf{m},\mathbf{p}^{(1)}|\mathbf{W})\right>_{E\geq0.954}-\left<\Delta(\mathbf{m},\mathbf{p}^{(0)}|\mathbf{W})\right>_{E\geq0.954}=0.0013$. 
The $\mathbf{p}^{(0)}$ complementary model produces better measurement agreement than the $\mathbf{p}^{(1)}$ base model by $\left<\Delta(\mathbf{m},\mathbf{p}^{(1)}|\mathbf{W})\right>_{0.816\leq E\leq0.954}-\left<\Delta(\mathbf{m},\mathbf{p}^{(0)}|\mathbf{W})\right>_{0.816\leq E\leq0.954}=0.002$.  
An additional fit parameter and component Mueller matrix in the $\mathbf{p}^{(2)}$ bulk model does not increase total polarimetric accuracy, except for low entropy measurements where $E<0.565$. Low entropy measurements correspond to larger $\xi_0$ magnitudes. While larger $\xi_0$ means that differences in $\xi_{1,2,3}$ are relatively smaller in magnitude, the large $\left<\Delta\right>_{E<0.565}$ values for all models in Tab. \ref{tab:MSEResults} is attributable to the relative magnitude elements in a normalized Mueller matrix.
Low entropy measurements have larger magnitude in the normalized Mueller matrix elements; see Fig \ref{fig:HighLowAlbedo}.

Numerical fit results using the complementary model are provided in Appendix \ref{app:FitResults}

\subsection{Estimating $\xi_0$} \label{sec:EstXi0}
\begin{figure}[ht!]
    \centering
    \begin{subfigure}[662nm illumination ]{\includegraphics[width=.49\textwidth]{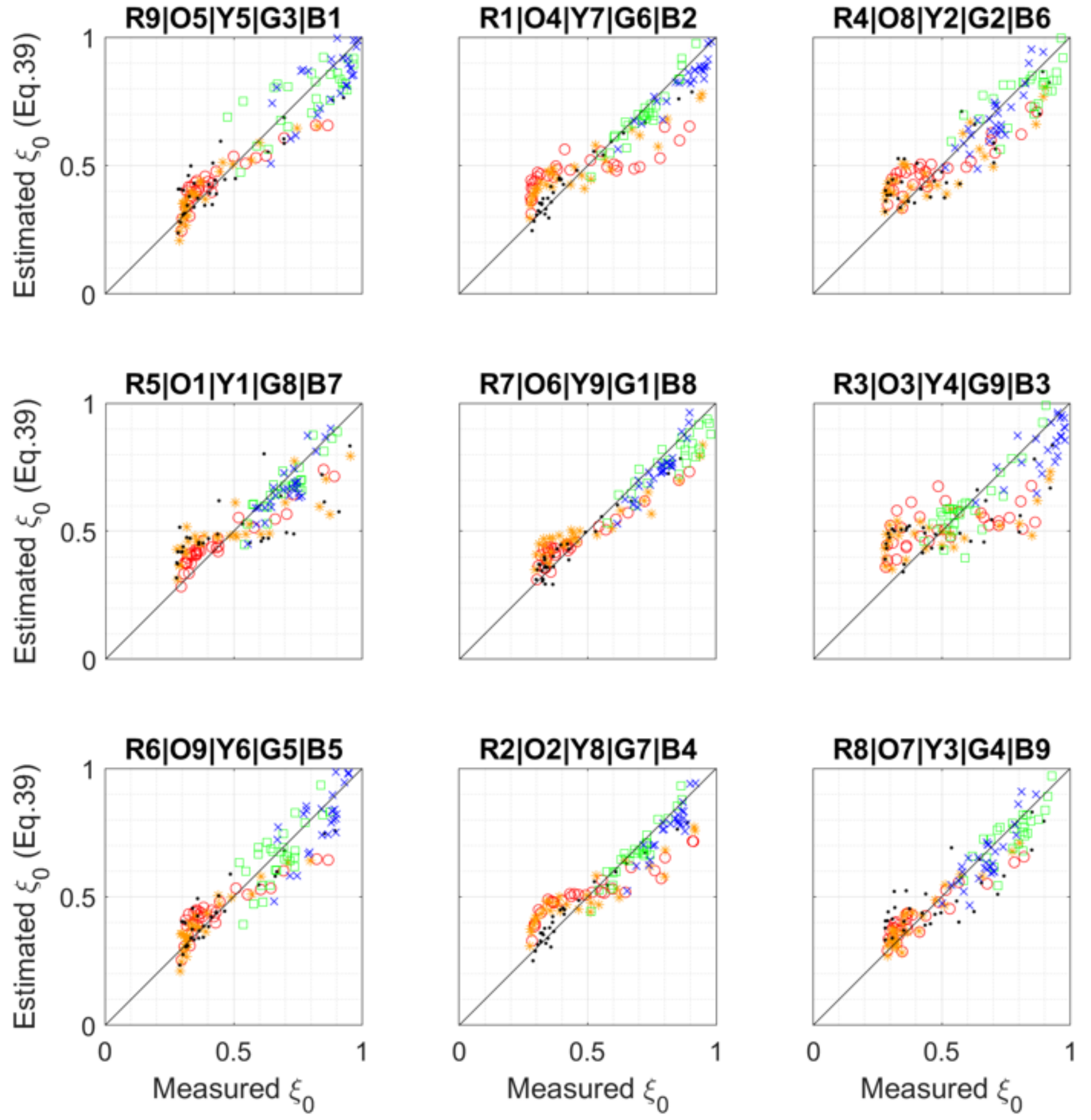}}\end{subfigure}
    \begin{subfigure}[524nm  illumination ]{\includegraphics[width=.49\textwidth]{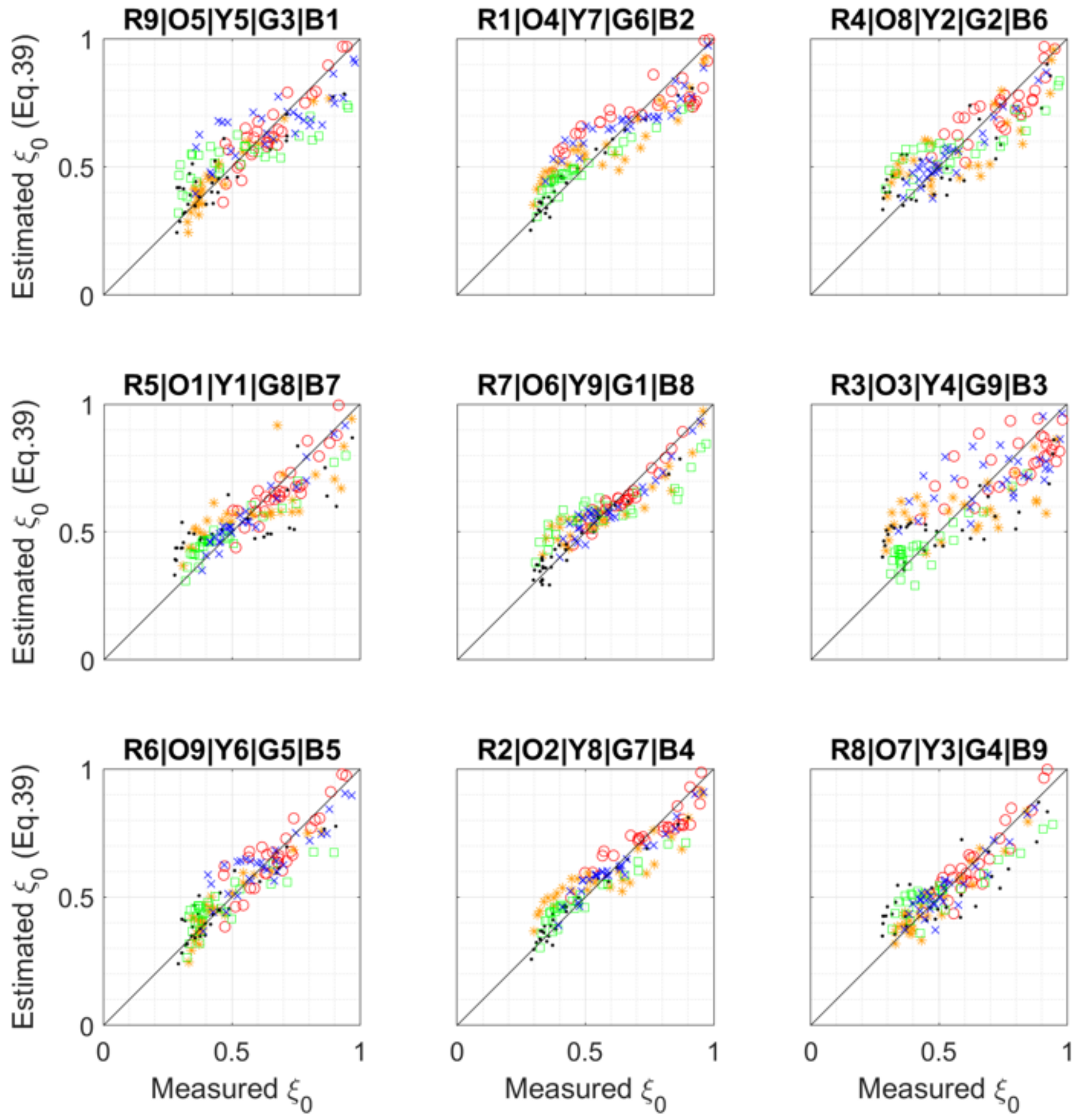}}\end{subfigure}
    \begin{subfigure}[451nm  illumination ]{\includegraphics[width=.6\textwidth]{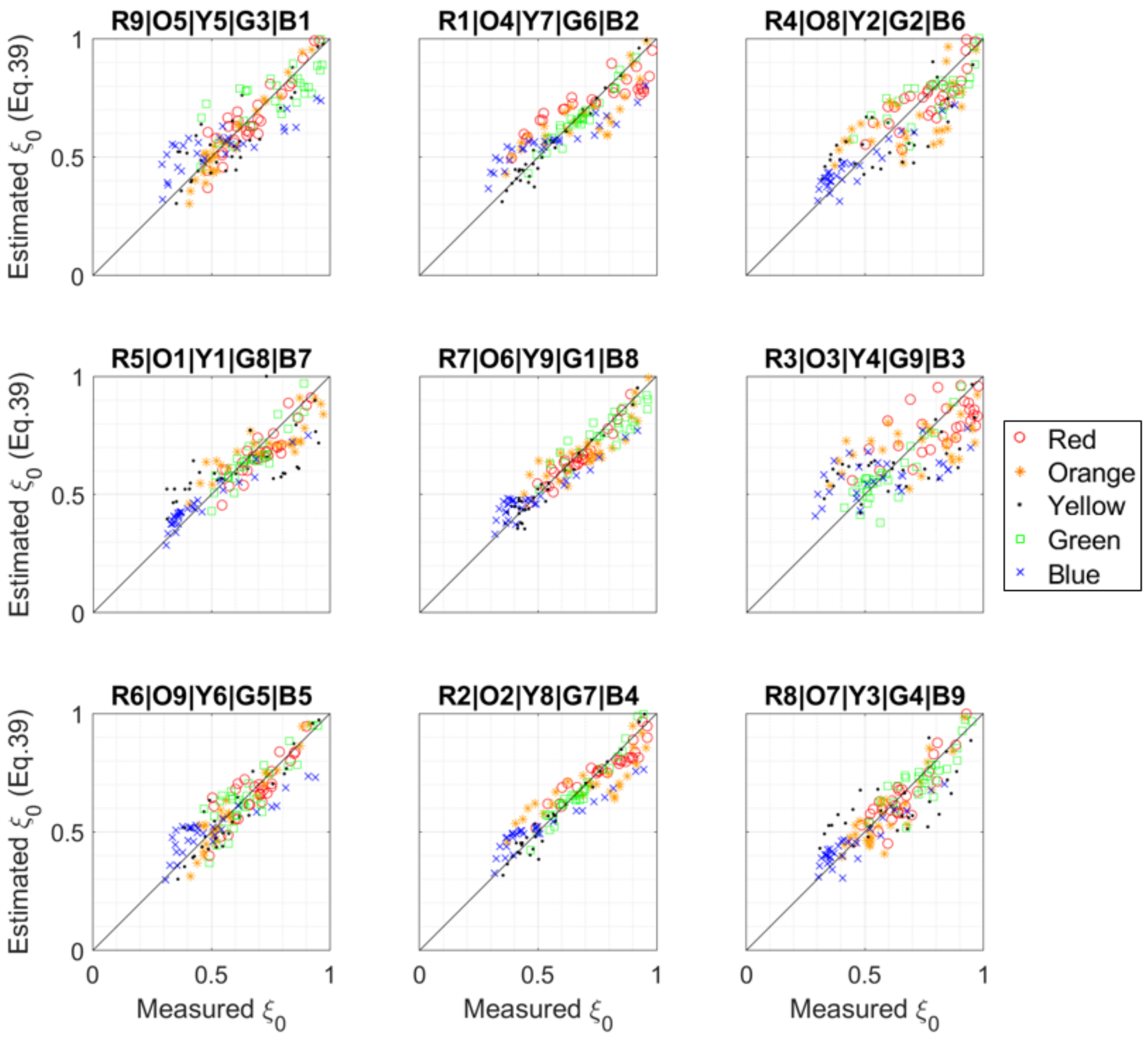}}\end{subfigure}
\caption{Estimated values for normalized $\xi_0$ using Eq.\ref{eq:EstXi0} are plotted versus measured values of $\xi_0$. Equation \ref{eq:DeltaXi} is used as a merit function. The individual scatter plots' positions in this $3\times3$ subplot orientation correspond to the measured brick's position in the experimental setup. Each subplot's title indicates which sample is in the position by using the first letter of the brick color and the corresponding texture label (1-9) as a key. The legend indicates the color of the brick under observation.}  \label{fig:Dirxi0_BrixWav}
\end{figure}

If the exact measured value for $\xi_0$ is plugged into the complementary model (Eq. \ref{eq:CompModel}) in place of the existing $z^R_\lambda\left[\mathbf{F}^R_{n_0,n_1}\right]_{00}/\gamma(\widehat{\boldsymbol{\omega}}_i,\widehat{\boldsymbol{\omega}}_o,\widehat{\mathbf{n}};\sigma)$ term, then the resulting $\left<\Delta(\mathbf{m},\mathbf{p}^{(0)}|\mathbf{W})\right>_{E}$ is 0.0102. Therefore, this measurement agreement is the best possible performance achievable from assuming triple degeneracy and perfect estimation of $\xi_0$. Using the exact measured $\xi_0$ value in Eq.\ref{eq:1par_base}, which is an alternate parameterization of the base model $\mathbf{p}^{(1)}$, produces the same measurement agreement. The $\mathbf{p}^{(2)}$ model is not parameterized by $\xi_0$. 

The $z^R_\lambda\left[\mathbf{F}^R_{n_0,n_1}\right]_{00}/\gamma(\widehat{\boldsymbol{\omega}}_i,\widehat{\boldsymbol{\omega}}_o,\widehat{\mathbf{n}};\sigma)$ term in the complementary model $\mathbf{p}^{(0)}$ can be used as a distribution function to estimate $\xi_0$ value 
\begin{eqnarray}
    \xi_0(\widehat{\boldsymbol{\omega}}_i,\widehat{\boldsymbol{\omega}}_o, \widehat{\mathbf{n}};\sigma) &=&  \frac{z^R_\lambda}{\gamma(\widehat{\boldsymbol{\omega}}_i,\widehat{\boldsymbol{\omega}}_o, \widehat{\mathbf{n}};\sigma)}\left[{\mathbf{F}}_{n_0,n_1}^R(\widehat{\boldsymbol{\omega}}_i,\widehat{\boldsymbol{\omega}}_o,\widehat{\mathbf{n}})\right]_{00} \label{eq:EstXi0} \\
    &=& z^R_\lambda\left[{\mathbf{F}}_{n_0,n_1}^R(\widehat{\boldsymbol{\omega}}_i,\widehat{\boldsymbol{\omega}}_o,\widehat{\mathbf{n}})\right]_{00} \frac{4 (\widehat{\mathbf{h}}\cdot\widehat{\mathbf{n}})(-\widehat{\boldsymbol{\omega}}_i\cdot\widehat{\mathbf{n}})(\widehat{\boldsymbol{\omega}}_o\cdot\widehat{\mathbf{n}})}
    {p(\widehat{\boldsymbol{\omega}}_i,\widehat{\boldsymbol{\omega}}_o,\widehat{\mathbf{n}};\sigma) G(\widehat{\boldsymbol{\omega}}_i,\widehat{\boldsymbol{\omega}}_o,\widehat{\mathbf{n}};\sigma)}. \nonumber
\end{eqnarray}
Figure \ref{fig:Estxi0_thhthd} compares this estimated $\xi_0$ to the measured $\xi_0$. The measured eigenvalue $\xi_0$ has a larger magnitude along the $\theta_h=0$ axis than the $\xi_0$ value calculated from $\mathbf{p}^{(0)}(\widehat{\boldsymbol{\omega}}_i,\widehat{\boldsymbol{\omega}}_o,\widehat{\mathbf{n}})$. By fitting $\xi_0$, the overall $\left<\bar{\Delta}\right>_{E}$ is increased to 0.0172. The increase in $\left<\bar{\Delta}\right>_{E}$ is not surprising because $\Delta$ is not the merit function used in fitting $\xi_0$. Fitting a distribution function to the normalized largest eigenvalue minimizes $\Delta\xi_0$ as the merit function. The mean square error between the measured and estimated $\xi_0$ is
\begin{equation}
    \Delta\xi_0=
    \frac{1}{K}\sum_{k=1}^{K}\left([\xi_0]_k-\frac{z^R_\lambda   }{\gamma([\small{\widehat{\boldsymbol{\omega}}_i}]_k,[\small{\widehat{\boldsymbol{\omega}}_o}]_k,\widehat{\mathbf{n}})} \left[{\mathbf{F}}_{n_0,n_1}^R([\small{\widehat{\boldsymbol{\omega}}_i}]_k,[\small{\widehat{\boldsymbol{\omega}}_o}]_k,\widehat{\mathbf{n}})\right]_{00} \right)^2
    \label{eq:DeltaXi}
\end{equation}
where $[\xi_0]_k$ is the normalized largest eigenvalue of a Mueller matrix at the $k^{th}$ geometry. This eigenvalue can be computed from the spectral decomposition of a Mueller matrix, see Eq.\ref{eq:Mn2M}, or estimated from a reduced number of measurements as described in Sec.\ref{sec:MeasAgree}. Compared to the figure of merit for $L$ simulated irradiance values at varying measurement geometries in Eq. \ref{eq:FOM_GGX}, the figure of merit for the normalized largest eigenvalue in Eq. \ref{eq:DeltaXi} contains just the single summation over $K$ measurement geometries. For an estimated $\xi_0$ using the $\mathbf{p}^{(0)}$ model, $\left<\Delta\xi_0)\right>_{E}=0.0101$. A distribution function that more closely describes $\xi_0$ may be better suited to modeling opaque plastics than the GGX distribution. 

Figure \ref{fig:Dirxi0_BrixWav} demonstrates measurement agreement to the $\xi_0$ value when using Eq.\ref{eq:EstXi0} as a model and Eq.\ref{eq:DeltaXi} as a merit function. The plots are arranged in a $3\times3$ grid which indicates the bricks' position in the sample plane. The title of each plot is a key that indicates which color and texture brick is positioned in each location during measurement.

\section{Conclusion} \label{sec:Conclusion}

This work applies the spectral decomposition of normalized Mueller matrix measurements to pBSDF modeling. The object ensemble consists of red, orange, yellow, green, and blue bricks roughened using nine textures of sandpaper listed in Tab. \ref{tab:LegoTextures}; see Sec.\ref{sec:Dataset}. The Mueller matrix data set consists of measurements at thirty different incident illumination and exitant observation positions (Tab. \ref{tab:MeasGeos}) under 662$\pm22$, 524$\pm35$, and 451$\pm20$ nm illumination for a total of 4050 Mueller matrices; see Sec.\ref{sec:AcqGeo}.  

Spectral decomposition \cite{cloude1986group} analysis on these 4050 Mueller matrix measurements reveals an approximate triple degeneracy in the smallest three eigenvalues $\xi_1, \xi_2,$ and $\xi_3$; see Sec.\ref{sec:3DegEig}. When the spectral decomposition is triply degenerate, the pBSDF model's depolarization is completely parameterized by the largest normalized eigenvalue, denoted $\xi_0$, which represents the fractional weight of the most significant Mueller-Jones matrix. The most significant Mueller-Jones matrix for these back-scattering measurements is well-approximated by the Fresnel reflection matrix. Figure \ref{fig:BasisMJMs} exemplifies how this Fresnel reflection approximation is applicable over varying albedo and texture. Polarization entropy (Eq. \ref{eq:PolEnt}), which has an inverse relationship with the largest coherency eigenvalue $\xi_0$, increases as surface texture becomes rougher; see Fig. \ref{fig:PolEntropyvsRa}. The magnitude of $\xi_0$ decreases as acquisition geometry moves away from specular orientations, scattering angle approaches normal incidence, and/or albedo increases; see Fig. \ref{fig:HiLoT1T9EntEig}. Triple degeneracy in the eigenspectrum allows the most popular implementation of a back-scattering pBSDF model, the weighted addition of an ideal depolarizer $\mathbf{D}(0)$ and Fresnel reflection $\mathbf{F}^{R}_{n_0,n_1}$ to be represented using only a single, measurable parameter: $\xi_0$. The capability to formulate a pBSDF model from a smaller quantity of measurements than required to formulate a Mueller matrix is a way to utilize a triply degenerate assumption by computing the largest normalized eigenvalue from Eq. \ref{eq:2Meas}.

This work assesses three normalized models: the $\mathbf{p}^{(0)}$ complementary model (Eq.\ref{eq:CompModel}), the $\mathbf{p}^{(1)}$ base model (Eq.\ref{eq:BaseModel}), and the $\mathbf{p}^{(2)}$ bulk model (Eq.\ref{eq:BulkModel}). 
Normalized Mueller matrix measurements $\mathbf{m}$, normalized Mueller pBSDF models $\mathbf{p}$, and the respective polarimetric measurement matrices $\mathbf{W}$ are used to evaluate the polarimetric accuracy of a model. Normalized Mueller matrices set the relative irradiance equal to one by dividing each element by the $\mathrm{M}_{00}$ element. Comparisons using normalized Mueller matrices assess polarimetric accuracy in the non-$\mathrm{M}_{00}$ elements independently from relative irradiance. 

The proposed $\mathbf{p}^{(0)}$ complementary model produces similar measurement agreement $\bar{\Delta}(\mathbf{m},\mathbf{p}|\mathbf{W})$ as the existing pBSDF models $\mathbf{p}^{(1)}$ and $\mathbf{p}^{(2)}$ when assessed using the measurement agreement of the simulated normalized irradiance; see Eq.\ref{eq:FOM_GGX} for $\bar{\Delta}(\mathbf{m},\mathbf{p}|\mathbf{W})$ in Sec.\ref{sec:MeasAgree}.
$\bar{\Delta}(\mathbf{m},\mathbf{p}^{(0)}|\mathbf{W})$ is similar to $\bar{\Delta}(\mathbf{m},\mathbf{p}^{(1)}|\mathbf{W})$ and $\bar{\Delta}(\mathbf{m},\mathbf{p}^{(2)}|\mathbf{W})$ within $\bar{\Delta}(\mathbf{m},\mathbf{p}|\mathbf{W})\pm0.0002$.
Table \ref{tab:MSEResults} lists the average measurement agreement $\bar{\Delta}_{E}$ over different quadrants of data separated by polarization entropy.
For measurements where entropy is $\geq0.0565$, the $\mathbf{p}^{(0)}$ proposed model produces the closest measurement agreement by a $\bar{\Delta}_{E}$ improvement between 0.0003 and 0.0020.
For measurements where entropy is $<0.0565$, the $\mathbf{p}^{(2)}$ bulk model produces the closest measurement agreement by an improvement to $\bar{\Delta}_{E}$ of 0.0045. 

One advantage of the proposed complementary model $\mathbf{p}^{(0)}$ is the relationship to the measurable quantity $\xi_0$. Section \ref{sec:MeasAgree} describes a method for measuring $\xi_0$. Another advantage of the complementary model $\mathbf{p}^{(0)}$ is a normalized Mueller matrix output which does not require a division by $\mathrm{M}_{00}$ step. Any normalized Mueller pBSDF model in this work can be easily adapted for applications sensitive to relative irradiance using Eq.\ref{eq:BRDFReflectance}, Eq.\ref{eq:norm_n}, and any of the many existing non-polarized, scalar BSDF models. 

The spectral decomposition analysis that produced the $\mathbf{p}^{(0)}$ model also leads to a novel method of fitting a Mueller pBSDF model, described in Sec. \ref{sec:EstXi0}. The second merit function introduced in this work is $\Delta\xi_0$, the measurement agreement for the largest coherency eigenvalue, $\xi_0$; see Sec.\ref{sec:EstXi0}. Equation \ref{eq:gamma} is one potential distribution function to describe the profile of $\xi_0$ over varying $\widehat{\boldsymbol{\omega}}_i$, $\widehat{\boldsymbol{\omega}}_o$, and $\widehat{\mathbf{n}}$. The $\xi_0$ parameter is useful for Mueller pBSDF modeling because it can either be computed from a full Mueller matrix or directly measured. A full Mueller matrix requires at least 16 polarized measurements, but a direct measurement of $\xi_0$ requires at least two polarized measurements. Section \ref{sec:MethodMeasXi0} describes a novel method of measuring $\xi_0$, assuming the coherency matrix for the light-matter interaction under observation has a triply degenerate eigenspectrum. 

For Mueller matrix measurements that are not triply degenerate more than two component Mueller matrices are appropriate. If the most significant Mueller-Jones matrix is not known \emph{a priori}, then more fit parameters need to be added to the pBSDF model. In this work, these two conditions have been used to demonstrate simplified merit functions for pSBDF fitting and simplified measurements schemes for pBSDF characterization. 

%Authors may also include Supplemental Documents: (PDF documents with expanded descriptions or methods) with the primary manuscript. At this time, supplemental PDF files are not accepted for partner titles, JOCN and Photonics Research. To reference the supplementary document, the statement ``See Supplement 1 for supporting content.'' should appear at the bottom of the manuscript (above the References heading).

%After proofreading the manuscript, compress your .tex manuscript file and all figures (which should be in EPS or PDF format) in a ZIP, TAR or TAR-GZIP package. All files must be referenced at the root level (e.g., file \texttt{figure-1.eps}, not \texttt{/myfigs/figure-1.eps}). If there are supplementary materials, the associated files should not be included in your manuscript archive but be uploaded separately through the Prism interface.

\section{Acknowledgements}
The authors would like to thank Jace Malm for data collection and Khalid Omer and Quinn Jarecki for helpful discussions. Construction and continued technical support of the RGB950 instrument is provided by Axometrics in Huntsville, Alabama. 
\section{Disclosures}
The authors have no disclosures to make.

\bibliography{report.bib}

\begin{thebibliography}{10}
\newcommand{\enquote}[1]{``#1''}

\bibitem{CloudeDepolSynthesis}
S.~R. Cloude, \enquote{Depolarization synthesis: understanding the optics of
  {M}ueller matrix depolarization,} {\protect\JournalTitle{JOSA A}}
  \textbf{30}, 691--700 (2013).

\bibitem{cloude1986group}
S.~R. Cloude, \enquote{Group theory and polarisation algebra,}
  {\protect\JournalTitle{Optik (Stuttgart)}} \textbf{75}, 26--36 (1986).

\bibitem{chipman2005metrics}
R.~A. Chipman, \enquote{Metrics for depolarization,} in \emph{Polarization
  Science and Remote Sensing II,}  vol. 5888 (International Society for Optics
  and Photonics, 2005), p. 58880L.

\bibitem{cloudepottier}
S.~R. Cloude and E.~Pottier, \enquote{Concept of polarization entropy in
  optical scattering,} {\protect\JournalTitle{Opt. Eng.}} \textbf{36} (1995).

\bibitem{bickel1988mueller}
W.~S. Bickel, J.-Y. Hsu, S.-C. Chiao, D.~Abromson, and V.~Iafelice,
  \enquote{The mueller matrix-stokes vector representation of surface
  scattering,} in \emph{Polarization Considerations for Optical Systems,}  vol.
  891 (International Society for Optics and Photonics, 1988), pp. 32--41.

\bibitem{torrance1967theory}
K.~E. Torrance and E.~M. Sparrow, \enquote{Theory for off-specular reflection
  from roughened surfaces,} {\protect\JournalTitle{J. Opt. Soc. Am.}}
  \textbf{57}, 1105--1114 (1967).

\bibitem{cooktorrance}
R.~L. Cook and K.~E. Torrance, \enquote{A reflectance model for computer
  graphics,} in \emph{Computer Computer Graphics SIGGRAPH,}  vol.~15 (1981),
  pp. 301--316.

\bibitem{Walter2007}
B.~Walter, S.~R. Marschner, H.~Li, and K.~E. Torrance, \enquote{Microfacet
  models for refraction through rough surfaces.}
  {\protect\JournalTitle{Rendering techniques}} \textbf{2007}, 18th (2007).

\bibitem{MSPIpBRDF}
D.~J. Diner, F.~Xu, J.~V. Martonchik, B.~E. Rheingans, S.~Geier, V.~M.
  Jovamovic, A.~Davis, R.~A. Chipman, and S.~C. McClain, \enquote{Exploration
  of a polarized surface bidirectional reflectance model using the ground-based
  multiangle spectropolarimetric imager,} {\protect\JournalTitle{Atmosphere}}
  \textbf{3}, 591--619 (2012).

\bibitem{Baek2018}
S.~Baek, D.~S. Jeon, X.~Tong, and M.~H. Kim, \enquote{Simultaneous acquisition
  of polarimetric {SVBRDF} and normals,} {\protect\JournalTitle{ACM Trans.
  Graph.}} \textbf{37}, 268--1 (2018).

\bibitem{Baek2020Image}
S.-H. Baek, T.~Zeltner, H.~J. Ku, I.~Hwang, X.~Tong, W.~Jakob, and M.~H. Kim,
  \enquote{Image-based acquisition and modeling of polarimetric reflectance,}
  {\protect\JournalTitle{Transactions on Graphics (Proceedings of SIGGRAPH)}}
  \textbf{39} (2020).

\bibitem{kondo2020accurate}
Y.~Kondo, T.~Ono, L.~Sun, Y.~Hirasawa, and J.~Murayama, \enquote{Accurate
  polarimetric {BRDF} for real polarization scene rendering,} in \emph{European
  Conference on Computer Vision,}  (Springer, 2020), pp. 220--236.

\bibitem{TomiYama}
S.~Tominaga and T.~Yamamoto, \enquote{Metal-dielectric object classification by
  polarization degree map,} in \emph{19th International Conference on Pattern
  Recognition,}  (2008), pp. 1--4.

\bibitem{polmultiviewstereo}
Z.~{Cui}, J.~{Gu}, B.~{Shi}, P.~{Tan}, and J.~{Kautz}, \enquote{Polarimetric
  multi-view stereo,} in \emph{2017 IEEE Conference on Computer Vision and
  Pattern Recognition (CVPR),}  (2017), pp. 369--378.

\bibitem{breon2017brdf}
F.-M. Breon and F.~Maignan, \enquote{A brdf--bpdf database for the analysis of
  earth target reflectances,} {\protect\JournalTitle{Earth System Science
  Data}} \textbf{9}, 31--45 (2017).

\bibitem{Lietal2020}
L.~Li, R.~Chipman, and M.~Kupinski, \enquote{Effects of surface roughness and
  albedo on depolarization in {M}ueller matrices,} in \emph{Polarization:
  Measurement, Analysis, and Remote Sensing XIV,}  vol. 11412 of \emph{Proc.
  SPIE} (2020).

\bibitem{Torrance67}
K.~E. Torrance and E.~M. Sparrow, \enquote{Theory for off-specular reflection
  from roughened surfaces,} {\protect\JournalTitle{J. Opt. Soc. Am.}}
  \textbf{57}, 1105--1114 (1967).

\bibitem{Breon}
F.~Bréon, \enquote{An analytical model for the cloud-free atmosphere/ocean
  system reflectance,} {\protect\JournalTitle{Remote sensing of environment}}
  \textbf{43}, 179--192 (1993).

\bibitem{ashikmin2000microfacet}
M.~Ashikmin, S.~Premo{\v{z}}e, and P.~Shirley, \enquote{A microfacet-based brdf
  generator,} in \emph{Proceedings of the 27th annual conference on Computer
  graphics and interactive techniques,}  (2000), pp. 65--74.

\bibitem{RGB950}
J.~M. López-Téllez, R.~A. Chipman, L.~W. Li, S.~C. McEldowney, and M.~H.
  Smith, \enquote{Broadband extended source imaging {M}ueller-matrix
  polarimeter,} {\protect\JournalTitle{Opt. Lett.}} \textbf{44}, 1522--1547
  (2019).

\bibitem{germer1999polarization}
T.~A. Germer and C.~C. Asmail, \enquote{Polarization of light scattered by
  microrough surfaces and subsurface defects,} {\protect\JournalTitle{JOSA A}}
  \textbf{16}, 1326--1332 (1999).

\bibitem{Rusinkiewicz:1998:ANC}
S.~Rusinkiewicz, \enquote{A new change of variables for efficient {BRDF}
  representation,} in \emph{Rendering Techniques (Proc. Eurographics Workshop
  on Rendering),}  (1998).

\bibitem{PriestGermer2000}
R.~G. Priest and T.~A. Germer, \enquote{Polarimetric {BRDF} in the microfacet
  model: Theory and measurements,} in \emph{Proceedings of the Military Sensing
  Symposia (MSS) Specialty Group Meeting on Passive Sensors,}  (2000), Military
  Sensing Symposia (MSS).

\bibitem{Chipman1}
R.~A. Chipman, W.~T. Lam, and G.~Young, \emph{Polarized Light and Optical
  Systems} (CRC Press, Boca Raton, Florida, 2019).

\bibitem{depind}
J.~J. Gil and E.~Bernabeu, \enquote{Depolarization and polarization indices of
  an optical system,} {\protect\JournalTitle{Optica Acta: International Journal
  of Optics}} \textbf{33}, 185--189 (1986).

\bibitem{ossikovski2010alternative}
R.~Ossikovski, \enquote{Alternative depolarization criteria for mueller
  matrices,} {\protect\JournalTitle{JOSA A}} \textbf{27}, 808--814 (2010).

\bibitem{kostinski1992depolarization}
A.~B. Kostinski, \enquote{Depolarization criterion for incoherent scattering,}
  {\protect\JournalTitle{Applied optics}} \textbf{31}, 3506--3508 (1992).

\bibitem{le1996optical}
F.~Le~Roy-Br{\'e}honnet, B.~Le~Jeune, P.~Elies, J.~Cariou, and J.~Lotrian,
  \enquote{Optical media and target characterization by mueller matrix
  decomposition,} {\protect\JournalTitle{Journal of Physics D: Applied
  Physics}} \textbf{29}, 34 (1996).

\bibitem{aiello2005physical}
A.~Aiello and J.~Woerdman, \enquote{Physical bounds to the
  entropy-depolarization relation in random light scattering,}
  {\protect\JournalTitle{Physical review letters}} \textbf{94}, 090406 (2005).

\bibitem{pires2008statistics}
H.~D.~L. Pires and C.~Monken, \enquote{On the statistics of the
  entropy-depolarization relation in random light scattering,}
  {\protect\JournalTitle{Optics Express}} \textbf{16}, 21059--21068 (2008).

\bibitem{physreal90}
S.~R. Cloude, \enquote{Conditions for the physical realisability of matrix
  operators in polarimetry,} in \emph{Polarization Considerations for Optical
  Systems II,}  vol. 1166 R.~A. Chipman, ed., International Society for Optics
  and Photonics (SPIE, 1990), pp. 177--187.

\bibitem{physreal93}
C.~R. Givens and A.~B. Kostinski, \enquote{A simple necessary and sufficient
  condition on physically realizable {M}ueller matrices,}
  {\protect\JournalTitle{J. Mod. Opt.}} \textbf{40}, 471–481 (1993).

\bibitem{Trowbridge:75}
T.~S. Trowbridge and K.~P. Reitz, \enquote{Average irregularity representation
  of a rough surface for ray reflection,} {\protect\JournalTitle{J. Opt. Soc.
  Am.}} \textbf{65}, 531--536 (1975).

\bibitem{priest2002polarimetric}
R.~G. Priest and S.~R. Meier, \enquote{Polarimetric microfacet scattering
  theory with applications to absorptive and reflective surfaces,}
  {\protect\JournalTitle{Optical Engineering}} \textbf{41}, 988--993 (2002).

\bibitem{Guarnera2016brdf}
D.~Guarnera, G.~C. Guarnera, A.~Ghosh, C.~Denk, and M.~Glencross,
  \enquote{{BRDF} representation and acquisition,} in \emph{Computer Graphics
  Forum,}  vol.~35 (Wiley Online Library, 2016), pp. 625--650.

\end{thebibliography}
\appendix

\section{Geometries} \label{app:MeasGeo} 
In computer graphics literature, the use of a halfway vector $\widehat{\mathbf{h}}$ is common in the implementation of microfacet BSDF models \cite{Guarnera2016brdf,Rusinkiewicz:1998:ANC,Walter2007,Baek2018,Baek2020Image}. For backscattering events, the halfway vector bisects the incident and reflected ray \cite{Rusinkiewicz:1998:ANC}.  An example of the halfway vector is depicted in Figure \ref{fig:Coordinates}. 
The vectors $\widehat{\boldsymbol{\omega}}_i$ and $\widehat{\boldsymbol{\omega}}_o$ point in the direction of light travel; the halfway vector for reflection is defined as
\begin{equation}
    {\mathbf{h}}=-\widehat{\boldsymbol{\omega}}_i+\widehat{\boldsymbol{\omega}}_o.
    \label{eq:hr}
\end{equation}
In a system where the surface of the subject under observation is a defined microfacet profile and not a statistical model, a reflection only occurs if the micronormal is parallel to the halfway vector $\widehat{\mathbf{m}}\|\widehat{\mathbf{h}}$ \cite{Walter2007}. In rendered images, the microfaceted surface profile would be created using a statistical model. 

The halfway angle $\theta_h$ describes the deviation from a specular microfacet from the surface normal calculated as the angle between $\widehat{\mathbf{h}}$ and $\widehat{\mathbf{n}}$. The cosine of the halfway angle is therefore equivalent to as $\cos(\theta_h) = \widehat{\mathbf{h}} \cdot \widehat{\mathbf{n}}$. In the special case where $\widehat{\boldsymbol{\omega}}_i = \widehat{\boldsymbol{\omega}}_o$, the halfway vector is equivalent to $\widehat{\boldsymbol{\omega}}_o$.

\begin{figure}[!t]
    \centering
    \begin{subfigure}[macro-incident angle $\theta_i$]{\includegraphics[width=0.4\textwidth]{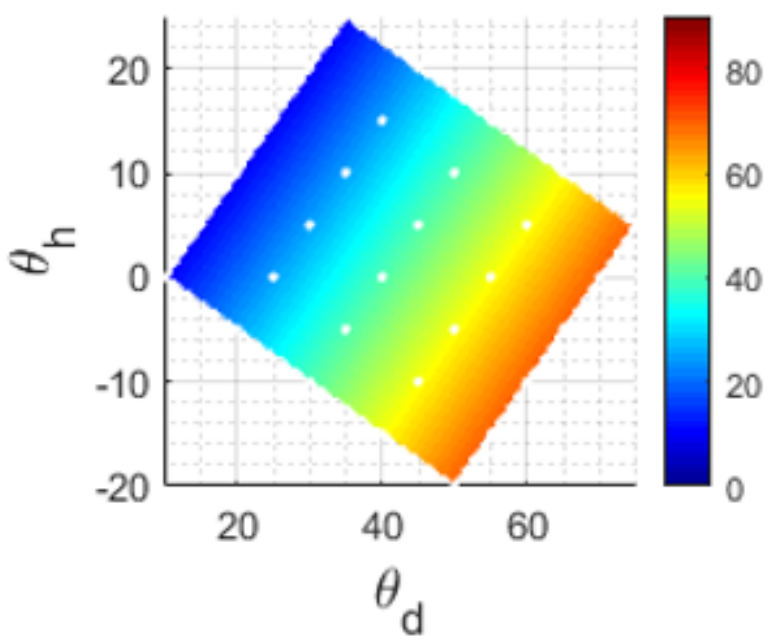}} \end{subfigure}
    \begin{subfigure}[macro-exitant angle $\theta_o$]{\includegraphics[width=0.4\textwidth]{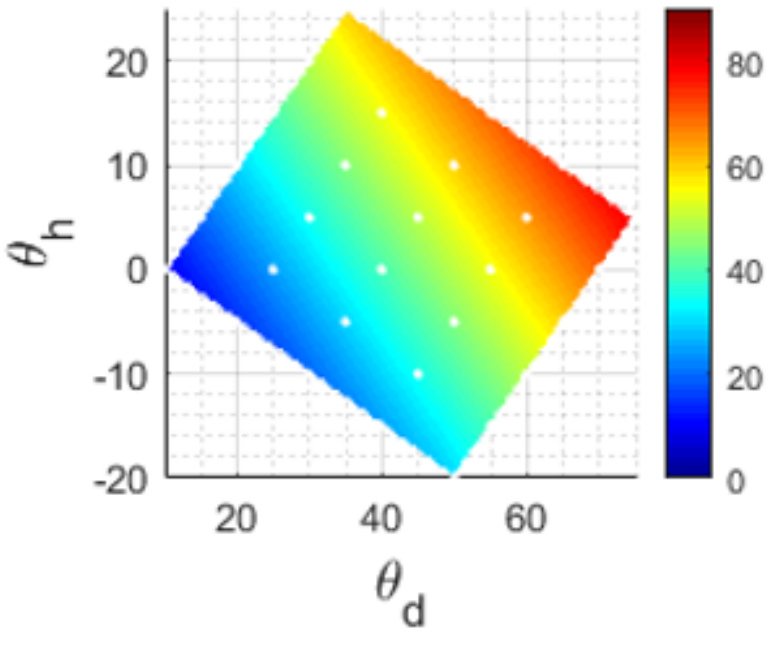}} \end{subfigure}
    \begin{subfigure}[halfway angle $\theta_h$]{\includegraphics[width=0.4\textwidth]{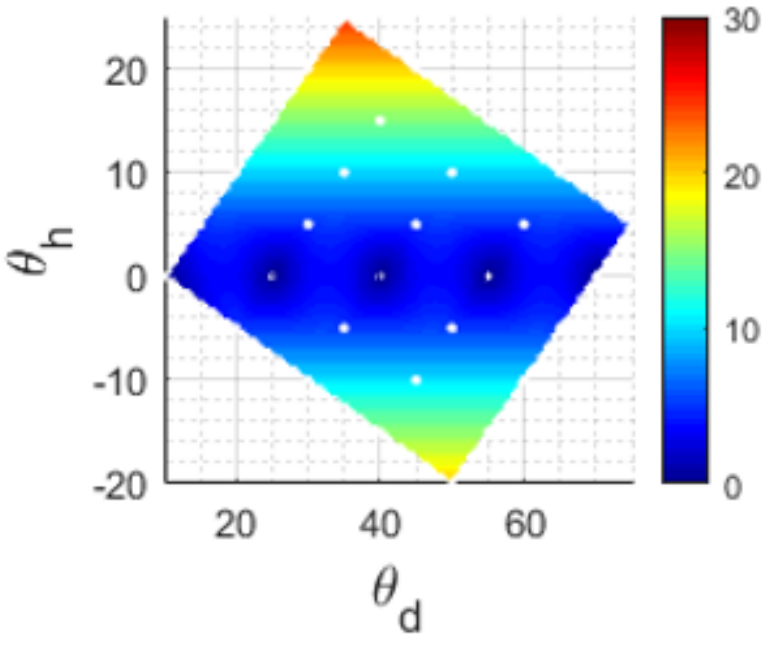}} \end{subfigure}
    \begin{subfigure}[halfway azimuth $\phi_h$]{\includegraphics[width=0.4\textwidth]{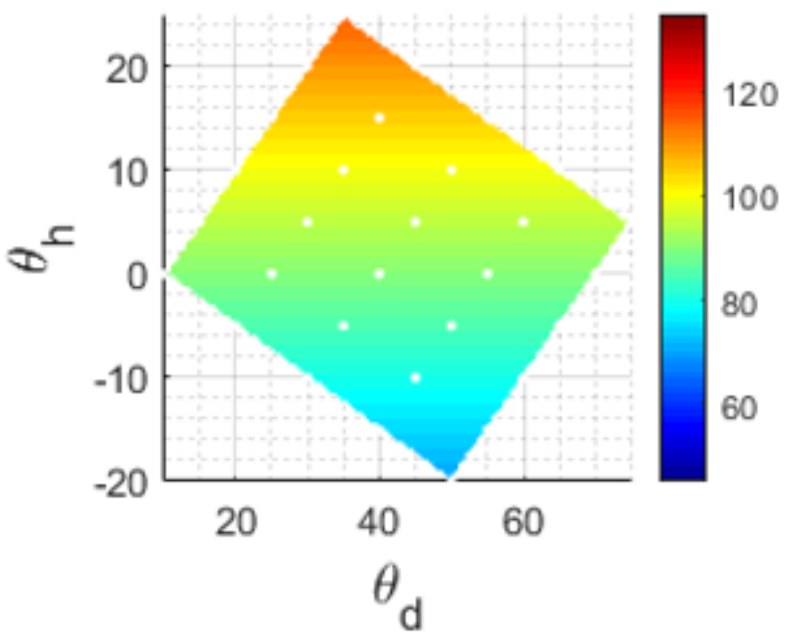}} \end{subfigure}
    \caption{Over a range of acquisition geometries the plots are: (a) the macro-incident angle $\theta_i$, (b) the macro-exitant angle $\theta_o$, (c) the halfway angle $\theta_h$, and (d) the halfway azimuth $\phi_h$. Every angle is plotted in the signed $\theta_h$ vs $\theta_d$ space in units of degrees. These plots correspond to a brick positioned in the center of the $3\times3$ tower. The halfway angle is typically constrained on the interval $[0^\circ,90^\circ]$. A signed version of the halfway angle is used to unfold the $\theta_h$ versus $\theta_d$ space where positive-value indicates $\phi_h>90^\circ$ and a negative-value indicates $\phi_h<90^\circ$. }
    \label{fig:thetaplots}
\end{figure}

\section{Fresnel Matrices} \label{app:FresnelM}
\subsection{Jones Matrix Form}
The Fresnel matrices are Muller-Jones matrices which describe the polarization effects at a material interface derived from the Fresnel equations. The Fresnel Mueller matrices can be calculated from the Fresnel Jones matrices using 
\begin{equation}
\mathbf{M} = \mathbf{U} (\mathbf{J}^* \otimes \mathbf{J}) \mathbf{U}^{-1}.
\label{eq:J2M}
\end{equation}
The matrix $\mathbf{U}$ is the unitary matrix from Eq. \ref{eq:UMatrix}. The Jones matrix for Fresnel interactions is typically written as
\begin{equation}
  \begin{bmatrix}
  s & 0\\
  0 & p\\
  \end{bmatrix}
\label{eq:JonesFresnel}    
\end{equation}
where $s$ and $p$ are Fresnel amplitude coefficients perpendicular and parallel to the plane of incidence, respectively. These coefficients are expressed as $r_s$ or $r_p$ for reflection and $t_s$ or $t_p$ for transmission. The plane of incidence contains the incident propagation vector $\widehat{\boldsymbol{\omega}}_i$ and the surface normal $\widehat{\mathbf{n}}$ of the material. A rotation into the plane of incidence which contains $\widehat{\boldsymbol{\omega}}_i$ and the halfway vector $\widehat{\mathbf{h}}$ is needed to described the Fresnel reflection from a microfacet.

In systems where the x-axis correspond to s-polarized light, Eq. \ref{eq:JonesFresnel} is used. In systems where the y-axis corresponds to s-polarized light, the positions of the s and p polarization in Eq. \ref{eq:JonesFresnel} are swapped. The measurements in this work result in a global coordinate definition where the vertical y-axis corresponds to the orientation of s-polarized light; this geometrical difference is accounted for in model implementation to ensure a geometrical match with the measurement setup.
\subsection{Mueller Matrix Form}
The polarization effects of a specular reflection from a microfacet are described by the non-depolarizing Fresnel reflection matrix
\begin{equation}
\label{eq:FmatrixR}
\mathbf{F}^R(n_0, n_1, \theta_d)=
\frac{1}{2}\begin{bmatrix}
 (|r_s^2| + |r_p^2|) & (|r_s^2| - |r_p^2|)& 0 & 0 \\
 (|r_s^2| - |r_p^2|)& (|r_s^2| + |r_p^2|)& 0 & 0 \\
 0 & 0 & (|r_p r_s^*| + |r_p^* r_s|)& i (|r_p r_s^*| - |r_p^* r_s|)\\
 0 & 0 & i (|r_p r_s^*| - |r_p^* r_s|)& (|r_p r_s^*| + |r_p^* r_s|)\\
\end{bmatrix}.
\end{equation}
An asterisk indicates a complex conjugate and Eq. \ref{eq:JonesFresnel} has been converted from a Jones to a Mueller matrix. The Fresnel reflection coefficients $r_s$ and $r_p$ are
\begin{equation}
r_s(n_0, n_1, \theta_d)= \frac{n_0\cos(\theta_d)-n_1\cos(\theta_d')}{n_0\cos(\theta_d)+n_1\cos(\theta_d')}
\label{eq:rs}
\end{equation}
and
\begin{equation}
r_p(n_0, n_1, \theta_d)= \frac{n_1\cos(\theta_d)-n_0\cos(\theta_d')}{n_1\cos(\theta_d)+n_0\cos(\theta_d')},
\label{eq:rp}
\end{equation}
where $n_0$ is the starting refractive index, $n_1$ is the ending refractive index, $\theta_d$ is the incident angle upon the material microfacet, and $\theta_d'$ is calculated using Snell's law: $n_0\sin{(\theta_d)}=n_1\sin{(\theta_d')}$. For a material which is illuminated by light which propagates through air first, $n_0=1$ while $n_1>1$.

The polarization effects of light passing from inside the media back to air toward the camera are described by the non-depolarizing Fresnel transmission matrix
\begin{equation}
\label{eq:FmatrixT}
\mathbf{F}^T(n_0, n_1, \theta_d)=
\frac{1}{2}\begin{bmatrix}
 (|t_s^2| + |t_p^2|) & (|t_s^2| - |t_p^2|)& 0 & 0 \\
 (|t_s^2| - |t_p^2|)& (|t_s^2| + |t_p^2|)& 0 & 0 \\
 0 & 0 & (|t_p t_s^*| + |t_p^* t_s|)& i (|t_p t_s^*| - |t_p^* t_s|)\\
 0 & 0 & i (|t_p t_s^*| - |t_p^* t_s|)& (|t_p t_s^*| + |t_p^* t_s|)\\
\end{bmatrix}.
\end{equation} 
The Fresnel transmission coefficients $t_s$ and $t_p$ are calculated using Eq. \ref{eq:ts} and Eq. \ref{eq:tp}.
\begin{equation} \label{eq:ts}
t_s(n_0, n_1, \theta_d)= \frac{2 n_0\cos(\theta_d)}{n_0\cos(\theta_d)+n_1\cos(\theta_d')}
\end{equation}
\begin{equation} \label{eq:tp}
t_p(n_0, n_1, \theta_d)= \frac{2 n_0\cos(\theta_d)}{n_1\cos(\theta_d)+n_0\cos(\theta_d')}
\end{equation}
In Eq. \ref{eq:ts} and \ref{eq:tp}, $n_0$ is the ambient material's refractive index (e.g. air) and $n_1$ is the observed material refractive index. The angle $\theta_d$ is the incident angle upon the boundary between media from the $n_0$ side, and $\theta_d' = \arcsin(n_0\sin(\theta_d)/n_1)$ as calculated from Snell's Law. For transmission from air into a non-air material, $n_0=1$ and $n_1>1$. For transmission from a non-air material out to air, $n_0>1$ and $n_1=1$.
\subsection{Mueller Matrix Rotation} \label{app:RotM}
A polarization ellipse is specified by two orthogonal directions in the transverse plane \cite{Chipman1}. A rotation of this coordinate system, which preserves the transverse plane but changes the local coordinate system is described by Eq. \ref{eq:RotM}.
\begin{equation}
\mathbf{R}(\alpha)=
\begin{bmatrix}
1 & 0 & 0 & 0 \\
0 & \cos(2\alpha) & -\sin(2\alpha) & 0 \\
0 & \sin(2\alpha) & \cos(2\alpha) & 0 \\
0 & 0 & 0 & 1 \\
\end{bmatrix},
\label{eq:RotM}
\end{equation}
where $\alpha$ is the angle between two coordinate systems. This matrix rotates positive angles in the counterclockwise direction about the propagation direction. 

The local coordinate system of a Mueller matrix can be specified using the unit vectors $\widehat{\boldsymbol{\Xi}}_{i,o}$ and $\widehat{\boldsymbol{\Sigma}}_{i,o}$ which are adapted from Priest \& Germer \cite{PriestGermer2000}. $\widehat{\boldsymbol{\Xi}}_{i,o}$ corresponds to the macrosurface while $\widehat{\boldsymbol{\Sigma}}_{i,o}$ corresponds to the microsurface. The subscripts $i$ or $o$ indicate whether the vectors correspond to incident or exitant directions of travel for a light-matter interaction.

The unit vector $\widehat{\boldsymbol{\Sigma}}_i = (\widehat{\boldsymbol{\omega}}_i\times\widehat{\mathbf{h}})/|\widehat{\boldsymbol{\omega}}_i\times\widehat{\mathbf{h}}|$ is perpendicular to the photon travel direction $\widehat{\boldsymbol{\omega}}_i$ and the halfway vector $\widehat{\mathbf{h}}$; which is to say $\widehat{\boldsymbol{\Sigma}}_i$ is the unit vector perpendicular to the micro-incident plane formed by $\widehat{\boldsymbol{\omega}}_i$ and $\widehat{\mathbf{h}}$.
The unit vector $\widehat{\boldsymbol{\Xi}}_i = (\widehat{\boldsymbol{\omega}}_i\times\widehat{\mathbf{n}})/|\widehat{\boldsymbol{\omega}}_i\times\widehat{\mathbf{n}}|$ is perpendicular to a photon travel direction $\widehat{\boldsymbol{\omega}}_i$ and the macronormal $\widehat{\mathbf{n}}$. The vector $\widehat{\boldsymbol{\Xi}}_i$ is perpendicular to the macro-incident plane formed by $\widehat{\boldsymbol{\omega}}_i$ and $\widehat{\mathbf{n}}$.
The angle from $\widehat{\boldsymbol{\Xi}}_i$ to  $\widehat{\boldsymbol{\Sigma}}_i$ rotates the coordinate system from the macro-incident plane to the plane of incidence for a Fresnel microfacet, $\emph{i.e.}$ the micro-incident plane \cite{PriestGermer2000}. The four-quadrant tangent function is used to calculate the incident rotation angle\cite{MSPIpBRDF}
\begin{equation}
\tan{(\alpha_i)} 
=\frac{(\widehat{\boldsymbol{\omega}}_i\times\widehat{\mathbf{n}})\cdot\widehat{\mathbf{h}}}{(\widehat{\mathbf{h}}\cdot\widehat{\mathbf{n}})-(\widehat{\boldsymbol{\omega}}_i\cdot\widehat{\mathbf{h}})(\widehat{\boldsymbol{\omega}}_i\cdot\widehat{\mathbf{n}})}
\label{eq:alphai}
\end{equation}
where identities $(\mathbf{a}\times\mathbf{b})\cdot(\mathbf{c}\times\mathbf{d})=(\mathbf{a}\cdot\mathbf{c})(\mathbf{b}\cdot\mathbf{d})-(\mathbf{a}\cdot\mathbf{d})(\mathbf{b}\cdot\mathbf{c})$ and $(\mathbf{a}\times\mathbf{b})\times(\mathbf{a}\times\mathbf{c})=((\mathbf{a}\times\mathbf{b})\cdot\mathbf{c})\mathbf{a}$ have been used. In principle, Eq. \ref{eq:alphai} should be equal to $\frac{|\widehat{\boldsymbol{\Xi}}_i\times\widehat{\boldsymbol{\Sigma}}_i|}{\widehat{\boldsymbol{\Xi}}_i\cdot\widehat{\boldsymbol{\Sigma}}_i}=  \frac{\sin{(\alpha_i)}}{\cos{(\alpha_i)}}$, but in practice this has led to $\alpha$ rotation angles constrained between $0^\circ$ and $90^\circ$. 

A similar rotation is needed for the exitant light.
The unit vector $\widehat{\boldsymbol{\Xi}}_o = -(|\widehat{\boldsymbol{\omega}}_o\times\widehat{\mathbf{n}})/|\widehat{\boldsymbol{\omega}}_o\times\widehat{\mathbf{n}}|$ is perpendicular to a photon travel direction $\widehat{\boldsymbol{\omega}}_o$ and the macronormal $\mathbf{n}$; the $\widehat{\boldsymbol{\Xi}}_o$ vector is perpendicular to the macro-exitant plane.
The unit vector $\widehat{\boldsymbol{\Sigma}}_o = -(\widehat{\boldsymbol{\omega}}_o\times\widehat{\mathbf{h}})/|\widehat{\boldsymbol{\omega}}_o\times\widehat{\mathbf{h}}|$ is perpendicular to the photon travel direction $\widehat{\boldsymbol{\omega}}_o$ and the halfway vector $\widehat{\mathbf{h}}$; the $\widehat{\boldsymbol{\Sigma}}_o$ vector is perpendicular to the macro-exitant plane \cite{PriestGermer2000}.
The angle $\alpha_o$ is
\begin{equation}
\tan{(\alpha_o)} 
=\frac{(\widehat{\boldsymbol{\omega}}_o\times\widehat{\mathbf{n}})\cdot\widehat{\mathbf{h}}}{(\widehat{\mathbf{h}}\cdot\widehat{\mathbf{n}})-(\widehat{\boldsymbol{\omega}}_o\cdot\widehat{\mathbf{h}})(\widehat{\boldsymbol{\omega}}_o\cdot\widehat{\mathbf{n}})}.
\label{eq:alphao}
\end{equation}
\subsection{Rotated Fresnel Matrices} \label{app:RotFR}
The Fresnel matrices in Eq. \ref{eq:FmatrixR} and Eq. \ref{eq:FmatrixT} are applied to light-material interactions in the plane of incidence defined by the incident propagation vector and the micronormal, here called the micro-incident plane. The Fresnel matrix $\mathbf{F}$ is rotated from the macro-incident plane by $\alpha_i$ to the micro-incident plane, where the Fresnel matrix is applied, and then to the macro-exitant plane by $\alpha_o$
\begin{equation}
  \widetilde{\mathbf{F}}^{R,T}_{n_0,n_1}(\widehat{\boldsymbol{\omega}}_i,\widehat{\boldsymbol{\omega}}_o,\widehat{\mathbf{n}})=\mathbf{R}({\alpha_o)}\mathbf{F}^{R,T}(n_0,n_1,\theta_d)\mathbf{R}({-\alpha_i})
  \label{eq:RotFresnel}
\end{equation}
where $\alpha_i$ is the angle from the macro-incident plane to the micro-incident plane and $\alpha_o$ is the angle from micro-incident plane to the macro-exitant plane.  
Eq. \ref{eq:alphai} and Eq. \ref{eq:alphao} in Appendix \ref{app:MeasGeo} are used to calculate the transverse rotation angles.
 
\section{GGX Microfacet Distribution Function} \label{sec:MicroDistFuncs} 
Microfacet distribution functions $p(\widehat{\boldsymbol{\omega}}_i,\widehat{\boldsymbol{\omega}}_o,\widehat{\mathbf{n}};\sigma)$ are probability distribution functions used to describe how much light scatters after a material interaction from a given source position $\widehat{\boldsymbol{\omega}}_i$, exitant propagation direction $\widehat{\boldsymbol{\omega}}_o$, and (optionally) surface roughness $\sigma$. Shadowing-masking functions $G(\widehat{\boldsymbol{\omega}}_i,\widehat{\boldsymbol{\omega}}_o,\widehat{\mathbf{n}};\sigma)$ are optionally applied to BSDFs to adjust for the effects of steep microfacet orientations that shadow neighboring microfacets. This work applies the GGX microfacet distribution function to the Fresnel reflection matrix. 

Microfacet distribution functions $p(\widehat{\boldsymbol{\omega}}_i,\widehat{\boldsymbol{\omega}}_o,\widehat{\mathbf{n}};\sigma)$ are distinct from microfacet response functions $D(\widehat{\boldsymbol{\omega}}_i,\widehat{\boldsymbol{\omega}}_o,\widehat{\mathbf{n}};\sigma)$ which are sometimes used in other works to describe a statistical distribution of surface normals over a surface \cite{Walter2007,Baek2018}. Microfacet distribution functions $p(\widehat{\boldsymbol{\omega}}_i,\widehat{\boldsymbol{\omega}}_o,\widehat{\mathbf{n}};\sigma)$ are preferrentially used in this work for being probability distribution functions which integrate to 1. They are generally related through the relationship $p(\widehat{\boldsymbol{\omega}}_i,\widehat{\boldsymbol{\omega}}_o,\widehat{\mathbf{n}};\sigma)  = (\widehat{\mathbf{h}} \cdot \widehat{\mathbf{n}}) D(\widehat{\boldsymbol{\omega}}_i,\widehat{\boldsymbol{\omega}}_o,\widehat{\mathbf{n}};\sigma)$.

The GGX microfacet function \cite{Walter2007,Trowbridge:75} is designed to be applied to both reflection and transmission events with a consideration for conservation of radiance included for the case of transmission. The GGX distribution was developed by fitting reflection measurements from and transmission measurements through different finishes of roughened glass \cite{Walter2007}. Eq. \ref{eq:GGX} and Eq. \ref{eq:GGXshadowB} describe the GGX distribution function and its associated shadowing-masking function, respectively. The GGX distribution includes an $\sigma$ fit parameter which tunes the shape of the microfacet distribution curve. Not all microfacet distribution functions include a $\sigma$ parameter.

\begin{equation}
p(\theta_h;\sigma) =
\frac{\sigma^2}{\pi(\widehat{\mathbf{h}}\cdot\widehat{\mathbf{n}})^3 \left(\sigma^2+\tan^2{(\theta_h)}\right)^2}
\label{eq:GGX}
\end{equation}
\begin{equation}
G(\theta_i,\theta_o;\sigma) = \left(\frac{2}{1+\sqrt{1+\sigma^2\tan^2{\theta_i}}}\right)\left(\frac{2}{1+\sqrt{1+\sigma^2\tan^2{\theta_o}}}\right)
\label{eq:GGXshadowB}
\end{equation}

\section{Dataset Details} \label{app:Dataset}
\subsection{Measurement Geometry}
Table \ref{tab:MeasGeos} reports the position of the illumination source and camera as $\theta_i$ and $\theta_o$, respectively. The center of the illumination source and camera's optical axis are the same height as the center of the sample plane. The angles of incidence $\theta_i$ and angles of observation $\theta_o$ quoted in Tab. \ref{tab:MeasGeos} are in reference the the center brick in the 3x3 LEGO tower arrangement (Fig. \ref{fig:3x3Towers}). Exact regions of interest in the other eight off-center bricks will have $\theta_i$ and $\theta_o$ values which vary up to $\pm8^\circ$ depending on the measurement geometry under observation.
\begin{table}[ht]
    \centering
    \begin{tabular}{c|c|c|c|c|c}
         \hline
         $\theta_i$& -10 & -25 & -40 & -55 &-70  \\
         \hline
         $\theta_{o,1}$& 10 & 15 & 20 & 25 & 30\\
         $\theta_{o,2}$& 20 & 25 & 30 & 35 & 40\\
         $\theta_{o,3}$& 30 & 35 & 40 & 45 & 50\\
         $\theta_{o,4}$& 40 & 45 & 55 & 60 & 65\\
         $\theta_{o,5}$& 50 & 55 &  60 & 65 & 70\\
         $\theta_{o,6}$& 60 & 65 & 70 & 75 & 80\\
         \hline
    \end{tabular}
    \caption{The nominal measurements for five different source positions and six different camera positions each are presented in units of degrees $[^\circ]$. The angles of incidence $\theta_i$ and angles of observation $\theta_o$ quoted here are in reference the the center brick in the 3x3 LEGO tower arrangement.}
    \label{tab:MeasGeos}
\end{table}
\begin{table}[ht]
    \centering
    \begin{tabular}{c|c|c|c|c|c|c}
    \hline
         Texture & Grit & Red & Orange & Yellow & Green & Blue \\
         \hline
         1 & 3000 & 0.2124 & 0.2534 & 0.2278 & 0.5578 & 0.4899\\ 
         2 & 2000 & 0.2849 & 0.2593 & 0.4093 & 0.6589 & 0.5579\\ 
         3 & 1500 & 0.3710 & 0.5417 & 0.3443 & 0.3537 & 0.3054\\ 
         4 & 1000 & 0.8945 & 0.8492 & 0.9741 & 1.2183 & 1.2636\\ 
         5 & 800  & 1.3268 & 1.6097 & 1.3267 & 2.2167 & 1.6839\\ 
         6 & 400  & 3.0182 & 2.6990 & 1.4738 & 2.7064 & 3.3533\\ 
         7 & 240  & 2.9624 & 3.4901 & 3.1755 & 3.0469 & 3.5590\\ 
         8 & 180  & 3.0252 & 3.5997 & 3.2248 & 2.7321 & 2.6173\\ 
         9 & 80   & 6.2977 & 6.0830 & 7.0237 & 8.7648 & 6.3219\\ 
         \hline
    \end{tabular}
    \caption{The leftmost column indicates the texture label of 1-9 and the second column indicates the grit of the sandpaper used for surface roughening treatment. Each subsequent row reports the reports arithmetic mean surface roughness (Ra) in microns for each LEGO brick color. These surface profiles are measured by a white light interferometer. }
    \label{tab:LegoTextures}
\end{table}

\section{Fit Results} \label{app:FitResults}
\begin{table}[ht]
  \centering 
    \begin{tabular}{c|c|c|c|c|c}
    \hline
    Color/Texture & Ra [$\mu$m] & $z^R_{662}$ & $z^R_{524}$ & $z^R_{451}$ & $\sigma$\\
    \hline
    R1&0.21&0.24&0.41&0.43&1.81\\
    R2&0.28&0.25&0.42&0.44&1.85\\
    R3&0.37&0.93&1.46&1.51&1.11\\
    R4&0.89&0.18&0.33&0.35&2.03\\
    R5&1.33&0.13&0.31&0.33&2.30\\
    R6&3.02&0.21&0.37&0.39&1.86\\
    R7&2.96&0.23&0.40&0.42&1.87\\
    R8&3.03&0.14&0.28&0.30&2.09\\
    R9&6.30&0.18&0.31&0.32&1.96\\
    \hline
    O1&0.25&0.18&0.25&0.34&2.14\\
    O2&0.26&0.16&0.21&0.28&2.13\\
    O3&0.54&0.57&0.70&0.91&1.35\\
    O4&0.85&0.13&0.18&0.24&2.23\\
    O5&1.31&0.11&0.14&0.19&2.17\\
    O6&2.70&0.19&0.26&0.35&2.10\\
    O7&3.49&0.09&0.12&0.16&2.34\\
    O8&3.60&0.10&0.15&0.20&2.41\\
    O9&6.08&0.11&0.15&0.19&2.16\\
    \hline
    Y1&0.23&0.10&0.11&0.14&2.67\\
    Y2&0.41&0.08&0.10&0.13&2.61\\
    Y3&0.34&0.09&0.10&0.14&2.49\\
    Y4&0.97&0.30&0.32&0.41&1.77\\
    Y5&1.33&0.09&0.09&0.12&2.47\\
    Y6&1.47&0.09&0.09&0.13&2.46\\
    Y7&3.18&0.07&0.08&0.11&2.63\\
    Y8&3.22&0.07&0.08&0.14&2.63\\
    Y9&7.02&0.07&0.07&0.10&2.90\\
    \hline
  \end{tabular}
  \caption{Fit parameters for the complementary model using a GGX distribution for different plastic brick colors and microfacet distribution functions under 662nm, 524nm, and 451nm illumination. Values are given for the textures T1, T5, and T9.}
  \label{tab:CompFits1}
\end{table} 
\begin{table}[ht]
\centering 
\begin{tabular}{c|c|c|c|c|c}
    \hline
    Color/Texture & Ra [$\mu$m] &  $z^R_{662}$ & $z^R_{524}$ & $z^R_{451}$ & $\sigma$\\
    \hline
    G1&0.56&0.64&0.42&0.63&1.58\\
    G2&0.66&0.31&0.20&0.31&2.01\\
    G3&0.35&0.44&0.28&0.43&1.77\\
    G4&1.22&0.27&0.14&0.26&2.06\\
    G5&2.22&0.31&0.19&0.30&1.87\\
    G6&2.71&0.30&0.18&0.28&1.94\\
    G7&3.05&0.30&0.18&0.29&1.91\\
    G8&2.73&0.23&0.12&0.22&2.22\\
    G9&8.76&0.34&0.24&0.35&1.71\\
    \hline
    B1&0.49&0.44&0.33&0.26&1.81\\
    B2&0.56&0.39&0.29&0.23&1.91\\
    B3&0.31&1.23&0.99&0.80&1.23\\
    B4&1.26&0.30&0.21&0.17&2.02\\
    B5&1.68&0.35&0.25&0.20&1.93\\
    B6&3.35&0.21&0.15&0.12&2.13\\
    B7&3.56&0.18&0.13&0.10&2.39\\
    B8&2.62&0.37&0.26&0.21&1.90\\
    B9&6.32&0.20&0.14&0.11&2.13\\
    \hline
  \end{tabular}
  \caption{Fit parameters for the complementary model using a GGX distribution for different plastic brick colors and microfacet distribution functions under 662nm, 524nm, and 451nm illumination.}
  \label{tab:CompFits2}
\end{table} 

\end{document}